\documentclass[nofootinbib,preprint,11pt]{revtex4-1}

\usepackage{tikz}

\usepackage[section]{placeins}
\usepackage{fancyhdr}
\usepackage{array}
\usepackage{mathtools}
\usepackage{amssymb}
\usepackage{graphicx}
\usepackage{epsfig}
\usepackage{wrapfig}
\usepackage{graphics}
\usepackage{slashed}	
\usepackage{enumitem}
\usepackage{color}
\definecolor{SeaBlue}{rgb}{0.1,0.4,0.85}
\definecolor{DarkBlue}{rgb}{0.1,0.3,0.65}
\definecolor{Maroon}{rgb}{0.6,0.2,0.2}
\definecolor{SeaGreen}{rgb}{0.2,0.4,0.2}
\definecolor{Purple}{rgb}{0.7,0.1,0.6}
\definecolor{Red}{rgb}{0.8,0.2,0.2}
\definecolor{Black}{rgb}{0.0,0.0,0.0}
\usepackage[colorlinks]{hyperref}
\hypersetup{
     colorlinks   = true,
     citecolor    = red,
	linkcolor=SeaBlue
}
\usepackage[T1]{fontenc}
\usepackage{titlesec}
\titleformat{\section}
  { \color{Maroon} \normalfont \scshape}
  {\thesection}{1em}{\bf \MakeUppercase}
\titleformat{\subsection} { \color{DarkBlue}  \normalfont\scshape}{\thesubsection}{1em}{\bf }{}
\titleformat{\subsubsection}
  { \color{Purple}  \normalfont\itshape}{\thesubsubsection }{1em}{}{}

\usetikzlibrary{decorations.pathmorphing}	
\usetikzlibrary{decorations.markings}
\tikzset{
    vector/.style={decorate, decoration={snake}, draw},
	provector/.style={decorate, decoration={snake,amplitude=2.5pt}, draw},
	antivector/.style={decorate, decoration={snake,amplitude=-2.5pt}, draw},
    fermion/.style={draw=black, postaction={decorate},
        decoration={markings,mark=at position .55 with {\arrow[draw=black]{>}}}},
    fermionbar/.style={draw=black, postaction={decorate},
        decoration={markings,mark=at position .55 with {\arrow[draw=black]{<}}}},
    fermionnoarrow/.style={draw=black},
    gluon/.style={decorate, draw=black,
        decoration={coil,amplitude=4pt, segment length=5pt}},
    scalar/.style={dashed,draw=black, postaction={decorate},
        decoration={markings,mark=at position .55 with {\arrow[draw=black]{>}}}},
    scalarbar/.style={dashed,draw=black, postaction={decorate},
        decoration={markings,mark=at position .55 with {\arrow[draw=black]{<}}}},
    scalarnoarrow/.style={dashed,draw=black},
    electron/.style={draw=black, postaction={decorate},
        decoration={markings,mark=at position .55 with {\arrow[draw=black]{>}}}},
	bigvector/.style={decorate, decoration={snake,amplitude=4pt}, draw},
}

\tikzstyle{block} = [draw, rectangle, 
    minimum height=3em, minimum width=6em]

\newcommand{\bfx}{ {\bf x} }
\newcommand{\bfu}{ {\bf u} }
\newcommand{\bfv}{ {\bf v} }
\newcommand{\bfr}{ {\bf r} }
\newcommand{\bfk}{ {\bf k} }
\newcommand{\bfp}{ {\bf p} }

\newcommand{\bfq}{ {\bf q} }
\newcommand{\qmod}{ | {\bf q} | }
\newcommand{\eV}{ \textrm{eV} }
\newcommand{\keV}{ \textrm{keV} }
\newcommand{\MeV}{ \textrm{MeV} }
\newcommand{\GeV}{ \textrm{GeV} }

\newcommand{\mDM}{m_{\rm DM}}
\newcommand{\kpc}{ \textrm{kpc} }

\newcommand{\Yfo}{Y_{\rm fo} }
\newcommand{\Mpl}{M_{\rm pl} }
\newcommand{\sigv}{ \langle \sigma v \rangle }

\newcommand{\Neff}{N_\textrm{eff}}

\definecolor{cerulean}{rgb}{0., 0.52,0.65}

\def\exercise#1{\begin{itemize}[label=$\star$]
	\item {\emph{Exercise:}} {#1}\end{itemize} }

\newcolumntype{P}[1]{>{\raggedright\arraybackslash}p{#1}}

\usepackage{aas_macros}

\makeatletter
\def\l@subsubsection#1#2{}
\makeatother

\begin{document}

\title{\Large TASI lectures on dark matter models and direct detection}
\author{\large Tongyan Lin}
\affiliation{\large Department of Physics, University of California, San Diego, CA 92093, USA }
\date{\today}
\begin{abstract} 
\normalsize These lectures provide an introduction to models and direct detection of dark matter. We summarize the general features and motivations for candidates in the full dark matter mass range, and then restrict to the $\sim$keV--TeV mass window. Candidates in this window can be produced by thermal mechanisms in the standard cosmology, and are an important target for experimental searches. We then turn to sub-GeV dark matter (light dark matter) and dark sectors, an area where many new models and experiments are currently being proposed. We discuss the cosmology of dark sectors, specific portal realizations, and some of the prospects for detection. The final parts of these lectures focus on the theory for direct detection, both reviewing the fundamentals for nuclear recoils of WIMPs and describing new directions for sub-GeV candidates. A version of these lectures was originally presented at the TASI 2018 summer school on ``Theory in an Era of Data''.
\end{abstract}
\maketitle
\tableofcontents

\newpage

{\emph{Note on intended audience:}} These lectures are aimed at graduate students getting started in research, or more advanced researchers seeking an introduction to the field. The notes assume some familiarity with field theory and the fundamentals of cosmology, such as FRW metrics and the history of the universe starting with Big Bang Nucleosynthesis (BBN). Those who would appreciate a reminder of cosmology basics will find a brief review in Lecture 2. In addition, Table~\ref{tab:units} in Appendix~\ref{sec:convention} may be helpful for some, since we will use a mix of units from astrophysics (when quoting results in that community) and natural units (where $\hbar=c=1$, more comfortable for particle physicists). A handful of exercises are included as part of the lectures, particularly in parts where a calculation/justification of a statement would take us too far from the main narrative.

\section{Overview of dark matter candidates \label{sec:models} }

The depth of knowledge about our cosmological history has grown immensely over the past few decades. The data have revealed the presence of dark matter and dark energy as the dominant contributions to the average energy density in the universe, while their underlying particle physics description remains unknown.

Any dark matter candidate must be consistent with a broad range of observations on astrophysical and cosmological scales~\cite{Bertone:2016nfn}, while also satisfying laboratory bounds. We begin these lectures with a summary of observational facts and constraints from astrophysics and cosmology. Guided by these facts, we will then sketch out a few broad classes of dark matter candidates. Our discussion will focus on the bare minimum of ingredients for such models, and we will not spend much time at all on ``top-down'' motivations for candidates. Other reviews, such as Ref.~\cite{Feng:2010gw}, describe in more detail the potential links between specific DM candidates and extensions of the Standard Model (SM), such as supersymmetry, models of neutrino mass, and so on.

The foundation of cosmology is the $\Lambda$CDM paradigm. In a nutshell, this refers to a cosmology (usually in a spatially flat spacetime) containing a number of effective fluids, including baryons, photons, and dark matter. Specifically, the dark matter fluid has an equation of state with pressure $p = 0$ and only gravitational interactions with the SM matter. Perturbations in these fluids are seeded  initially by nearly scale-invariant metric fluctuations, and their evolution gives rise to the structures we see today. This framework makes predictions for two important observables that inform our understanding of DM: the first is the power spectrum for the cosmic microwave background (CMB) photons, and the second is the {\it matter power spectrum} $P(k)$, the spectrum of density fluctuations in matter. Since dark matter is the dominant form of matter in the universe, $P(k)$ gives a good measure of the spectrum of DM density fluctuations.

We begin by focusing on observations of the matter power spectrum, rather than the CMB. This is because $P(k)$ directly encodes the clustering of dark matter\footnote{It is difficult to directly measure $P(k)$, and it is more often the situation that measurements of galaxies (for example) are compared to the prediction in a specific model. }. The possible behavior of dark matter is made manifest. The matter power spectrum is the variance in density perturbations on the length scale $2 \pi /k$, where $k$ is comoving wavenumber. One can also define a dimensionless power spectrum: 
\begin{align}
	\Delta^2(k) \equiv 4 \pi \left( \frac{k}{2\pi} \right)^3 P(k),
\end{align}
where $\Delta^2(k) \sim 1$ corresponds to $O(1)$ density fluctuations. For large density fluctuations, gravitational collapse into halos will occur. Nevertheless, it is common to present the {\it linear} matter power spectrum, since it doesn't involve the modeling of halo formation. 

\begin{figure}[h!]
\includegraphics[width=0.7\textwidth]{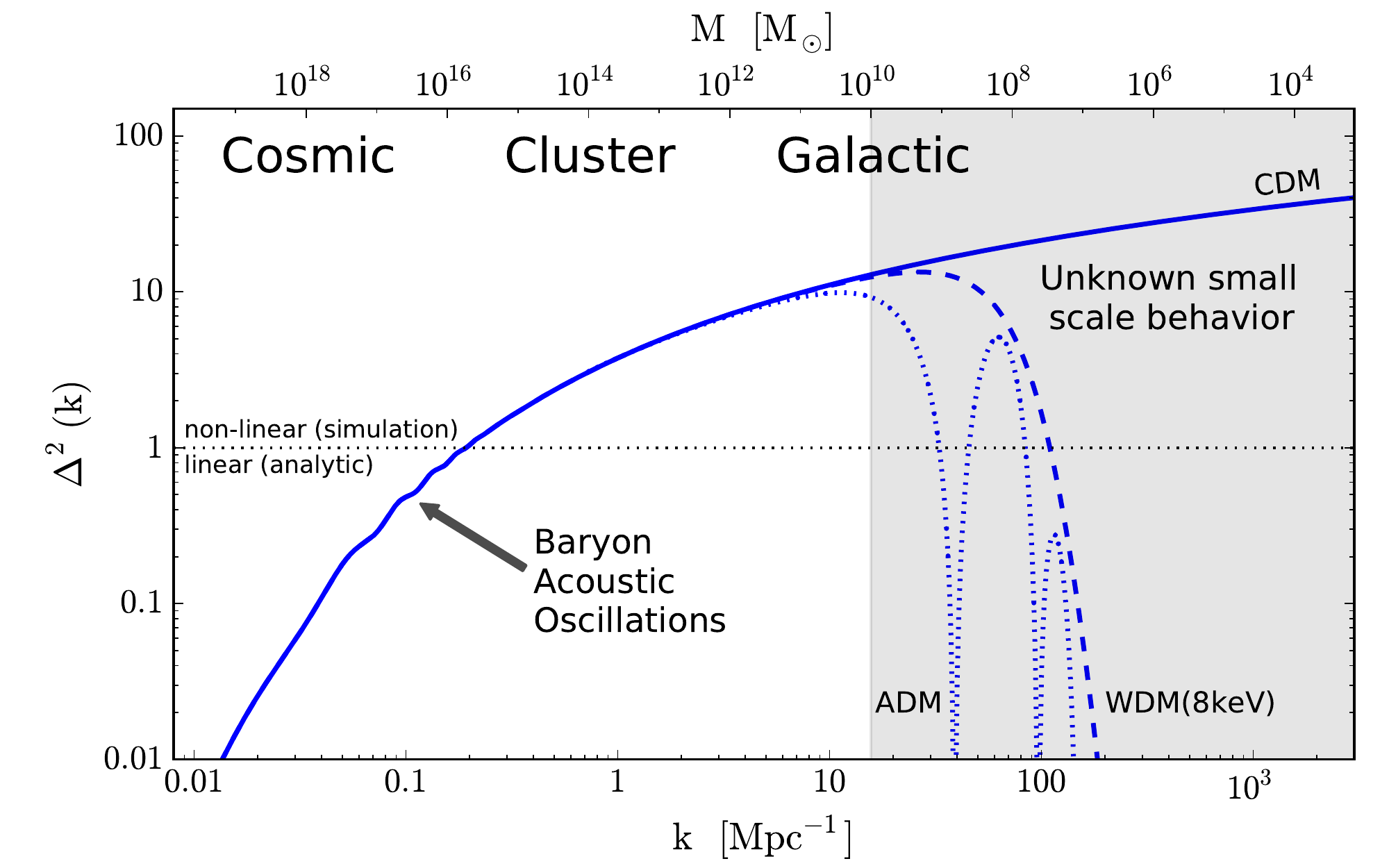}\hspace{1cm}
 \caption{Reproduced from Ref.~\cite{Kuhlen:2012ft}.  The solid blue line is the linear matter power spectrum of density perturbations in $\Lambda$CDM at $z=0$, where $\Delta^2(k) = 4 \pi (k/2\pi)^3 P(k)$.  The other lines show the power spectrum in models of atomic dark matter (ADM, dotted line) and warm dark matter (WDM, dashed line) with a mass of 8 keV. \label{fig:powerspectrum}}
\vspace{0.6cm}
\hspace{-0.2cm}\includegraphics[width=0.8\textwidth]{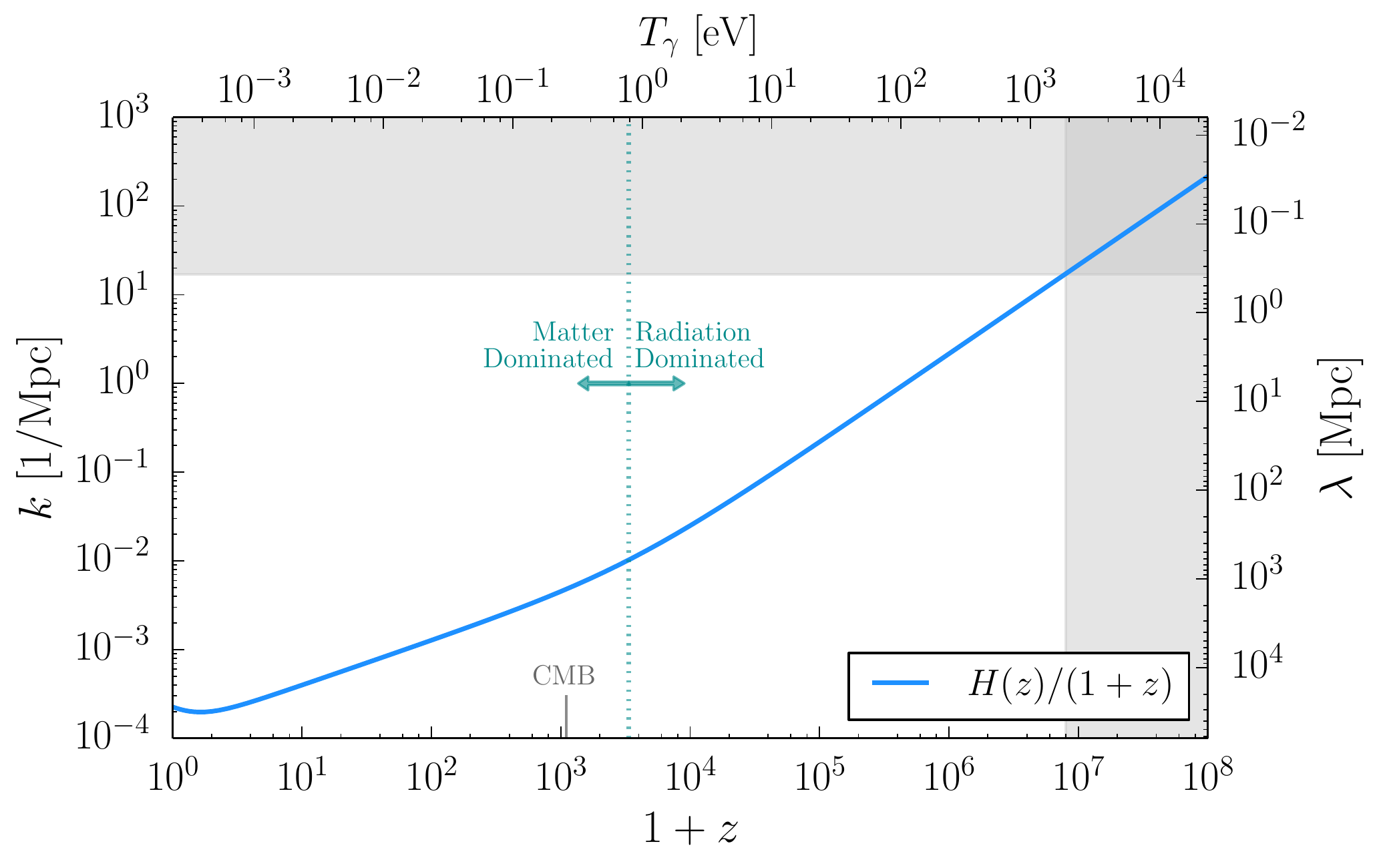}
 \caption{Perturbation modes with comoving wavenumber $k$ (or wavelength $\lambda$ at $z=0$) entered the horizon when $k = aH = H/(1+z)$. Small-scale modes entered the horizon at high redshift, where the matter power spectrum is less constrained; this region is shaded light gray. The top scale shows the photon temperature at that redshift.  \label{fig:kmode}}
\end{figure}

The matter power spectrum is a redshift-dependent quantity, and there are numerous observational probes of $P(k)$ for $k \sim 10^{-3} - 10$ Mpc$^{-1}$ and over a range of redshifts $z \lesssim 3-4$. (For a summary plot including a number of measurements, see for example Fig 5 in Ref.~\cite{Hlozek:2011pc}.) So far, these agree beautifully with the $\Lambda$CDM prediction, which is shown in Fig.~\ref{fig:powerspectrum} at $z=0$. The shaded region of $k$ values indicates where we currently lack reliable measurements  (where the boundary of this region is fuzzy, of course).  This tells us the modes where we can be reasonably confident of the evolution of DM, namely $k \lesssim 10-20$ Mpc$^{-1}$.  By integrating the mass enclosed within a radius $2 \pi /k$, we can also associate a mass scale with a given mode, with $M \gtrsim 10^{10} M_\odot$ for scales where the power spectrum has been well-measured.

Gaining information about the power spectrum at larger $k$ (smaller scales) informs us about the behavior of DM at earlier times. This is because the condition for when a specific $k$  mode started evolving is when $k \sim H(z)/(1+z)$, where $z$ is the redshift: this is when the physical mode size $2\pi a(t) / k = 2\pi/[k(1+z)] $ becomes smaller than the horizon size $H^{-1}$. (Note that here we are assuming Newtonian gauge, where super-horizon density modes are constant in time, following Ref.~\cite{Dodelson:2003ft}.) A comparison of $H(z)/(1+z)$ with $k$ is shown in Fig.~\ref{fig:kmode}. Here we see that larger $k$ modes started evolving earlier, at higher redshifts. A density perturbation in cold dark matter can start growing by gravitational interactions as soon as it is contained within a horizon size.  This explains the trend in Fig.~\ref{fig:powerspectrum}, where smaller scales have larger $\Delta^2(k)$. Changes to the $\Lambda$CDM assumptions could have delayed (or enhanced) the growth of DM perturbations at early times, leading to a $k$-dependent suppression (or enhancement) in the matter power spectrum relative to the $\Lambda$CDM prediction.

The shaded regions of Fig.~\ref{fig:kmode} illustrate the relationship between our cosmological history and the $k$-modes where the power spectrum is currently not reliably measured. Those modes would have started evolving at $z \sim 10^7$, or when the photon temperature of the universe was around $T_\gamma \sim$ keV. Particle interactions of DM at temperatures above $\sim $keV are therefore relatively unconstrained by the matter power spectrum. This epoch represents an exciting frontier in dark matter physics, connecting small-scale structure in dark matter with possible particle physics models.

Let us now combine our knowledge of $P(k)$ with a variety of other astrophysical measurements, and make some general and quantitative statements about properties of DM:
\begin{itemize}
	\item {\bf Dark with respect to SM interactions} -- This statement means several different things: first, that DM is not luminous in galaxies or clusters observed today. One of the commonly used images that illustrates this spectacularly is the merging Bullet Cluster. Second, from our discussion above, observations of the matter power spectrum and CMB require the presence of a matter component that has only gravitational interactions.
	Interactions of DM with SM particles at early times would suppress the power spectrum, since the radiation pressure of the baryons and photons would prevent DM density perturbations from growing. This resulting $P(k)$ would be similar to the line labeled `ADM' (although that model is slightly different, as we'll come to in a moment).  Specifically, DM interactions with the SM should be tiny after $z \lesssim 10^7$ (unshaded region in Fig.~\ref{fig:kmode}).
	\item {\bf Cold (nonrelativistic)} -- A second conclusion following from Fig.~\ref{fig:powerspectrum} is that DM is sufficiently nonrelativistic by $z \sim 10^7$. If DM is relativistic, then perturbations within a horizon can become washed out due to the motion of the DM. As a result, there is a relative suppression in the matter power spectrum for modes which enter the horizon while the DM is still relativistic. This effect can be seen in the warm dark matter (WDM) line in the figure. Note that this depends on the assumed temperature of the DM.
	\item {\bf Collisionless within the DM sector on large scales} -- Similar to the first point, non-gravitational interactions within the dark matter sector could modify the matter power spectrum. For example, the presence of a ``dark radiation'' bath which interacts with a component of the nonrelativistic dark matter would delay growth of density perturbations and lead to the presence of ``dark acoustic oscillations''. The `ADM' line shown in Fig.~\ref{fig:powerspectrum} illustrates this with a model of atomic dark matter. There are also observational constraints on DM self-scattering at late times (i.e., today) on  galaxy and cluster scales.
	\item {\bf (Close to) stable} -- At a minimum, we know the lifetime of DM should be much greater than the age of the universe. This, too, can be constrained with the matter power spectrum~\cite{Poulin:2016nat,Bringmann:2018jpr}. In specific scenarios where DM decays to SM particles, much stronger statements can be made: a late decay to SM particles can be constrained by indirect detection or 21cm cosmology.
	\item {\bf Preserves successes of BBN} -- This requirement can partly be viewed as a restatement of the first bullet point. The light elements were formed during BBN, and their primordial abundances depend on the total energy density through the expansion rate, which puts a constraint on the energy density in dark sector particles at the time of BBN.  Looking back in time, we have a good sense of the most important baryonic physics as far back as around $T \sim $ few MeV, but so far there is a huge zoo of possibilities for our cosmological history prior to that. 
	Hence, in most models, the work of making the DM is  done well before BBN.
\end{itemize}

\begin{figure}[t]
\includegraphics[width=0.9\textwidth]{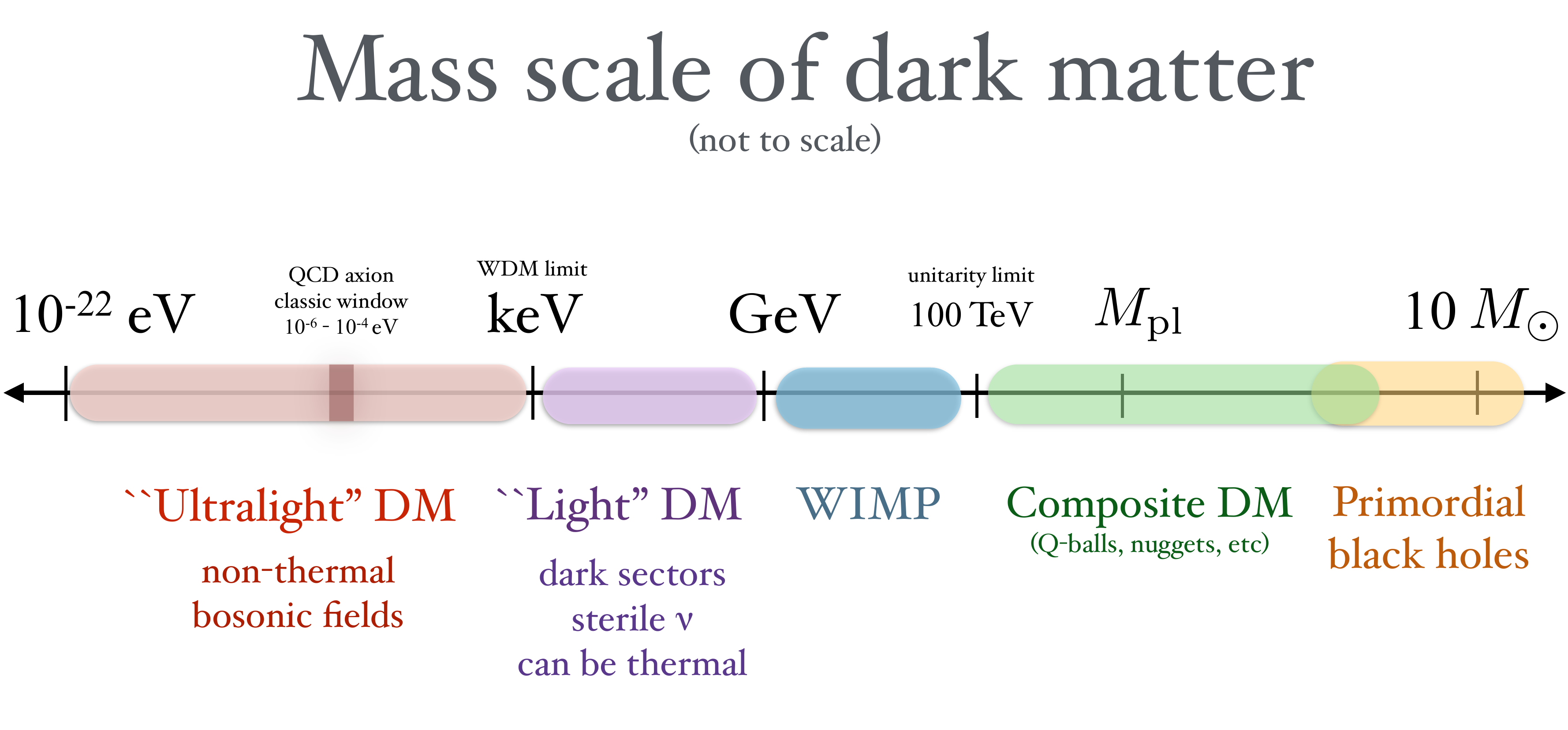}
 \caption{The mass range of allowed DM candidates, comprising both particle candidates and primordial black holes. Mass ranges are only approximate (in order of magnitude), and meant to indicate general considerations.\label{fig:mass_scale}}
\end{figure}

What are the possibilities for a DM candidate? Whether the DM around us is present in the form of fundamental particles, or as composite objects, we would like to know a few basic properties, such as the mass and spin. And we would of course like to characterize its interactions with SM particles or other new states. In this lecture, we will start by classifying candidates to the extent possible by mass and spin. Fig.~\ref{fig:mass_scale} gives a compact summary of the landscape and the main tourist spots - we will visit each below.

{\it A brief aside on MOND.}  --- MOdified Newtonian Dynamics (MOND) is a framework for modified gravity on galactic scales~\cite{1983ApJ...270..365M}, originally put forth as an alternative to dark matter. A specific relativistic theory is needed to obtain predictions during the early universe. Assuming no additional matter content, popular candidates such as TeVeS~\cite{Bekenstein:2004ne} give a notably worse fit to CMB and large scale structure data  compared to $\Lambda$CDM~\cite{Skordis:2005xk,Zuntz:2010jp}. A recent analysis of Milky Way rotation curve and stellar kinematics data  is also in tension with MOND~\cite{Lisanti:2018qam}.

\subsection{Bosons vs. fermions and the WDM limit \label{sec:WDMlimit} }

The keV mass scale is a special scale which, roughly speaking, demarcates thermally-produced DM (either a fermion or boson)  from nonthermally-produced bosonic DM. There are two separate arguments here: first, a fermion DM candidate must have mass greater than $O(\keV)$ in order to be consistent with observations of galaxies, and second, DM that is thermally produced from the SM bath must also have mass greater than $O(\keV)$ to be consistent with observations of large scale structure.

Using observations of the kinematics of stars in galaxies, a general statement can be made about the spin of a potential DM candidate. Galaxies reside inside dark matter halos, gravitationally bound overdensities that extend well beyond the typical radius for the stellar component of the galaxy. As a simple example, we can model this halo as an object that underwent gravitational collapse and is now {\it virialized}. Except close to the baryonic component, the gravitational potential is dominated by the mass in the dark matter halo. This means we can neglect the effect of the baryons on the dark matter halo, and use the kinematics of stars as a tracer for the dark matter density and   gravitational potential\footnote{There were quite a few white lies in this paragraph. The reality is that dark matter halos are undergoing merger processes and not in equilibrium, baryonic physics can have significant effects on the dark matter, and the mapping from stellar kinematics to dark matter properties is an area of active research in the era of {\it Gaia} data.}.

Since the gravitational potential of a galaxy can be inferred from data, this sets an upper bound on the possible velocity of dark matter particles in the halo. Combining this argument with the Pauli exclusion principle leads to a maximum density of dark matter. Specifically, for fermions there is an upper bound on the phase space density $f(\bfx,\bfp)$:
\begin{align}
	f(\bfx, \bfp) \le g \hbar^3
\end{align} 
where $g$ is the number of spin and flavor states. Let us take as a toy model a degenerate, nonrelativistic fermion gas in three dimensions. 

The local DM number density is given by
\begin{align}
	n(\bfx) = \int \frac{d^3 \bfp}{(2 \pi)^3}  f(\bfx, \bfp) \le  \frac{g}{8 \pi^3} \frac{4 \pi}{3} p_{max}^3.
\end{align}
This is an extremely conservative assumption! In the above equation, $p_{max}$ is the maximum possible momentum, and physically must be cut off at $\mDM v_{esc}$, where the  local escape velocity
\begin{align}
	v_{esc}(\bfx) =  \left[ - 2 \Phi(\bfx) \right]^{1/2} \simeq \sqrt{ \frac{ 2 G M_{vir} }{ R_{vir} } }.
\end{align}
With the assumption of equilibrium for a virialized halo, a particle with velocity greater $v_{esc}$ cannot be bound to the galaxy.  (In a typical galaxy, there are also unvirialized DM components that are unbound, but the fraction of the density in this population is small.) As an estimate of the bound, we will just use the average density of the halo, $n \sim  (3 M_{vir})/(4 \pi R_{vir}^3 \mDM)$. Requiring that this average density is below the maximum number density above, we obtain
\begin{align}
	\mDM^4 \gtrsim \frac{5}{g} G^{-3/2} M_{vir}^{-1/2} R_{vir}^{-3/2} \rightarrow \mDM \gtrsim 5\ \eV
\end{align}
if we use the mass ($10^{12} M_\odot$) and virial radius ($0.3$ Mpc) of the Milky Way. Applied more carefully to some of the dwarf satellites (e.g. with $M \sim 10^9 M_\odot$), this bound can be improved to a few hundred eV~\cite{Boyarsky:2008ju}.

This argument is related to the Tremaine-Gunn bound~\cite{Tremaine:1979we}, which specifically addresses the situation where the DM was {\emph{ever}} in thermal equilibrium. By applying Liouville's theorem which states that the phase space density at the microscopic level is conserved, they obtain a slightly stronger bound, since the phase space density of a fermion in equilibrium is given by $f(p) = g/(\exp(E_p/T) +1) \le g/2$. In addition, the argument can be applied to both fermions and scalars since their starting point is that the phase space distribution is initially thermal with the same temperature as the photons. With additional assumptions on the possible phase space distribution, these bounds could be even stronger.  Setting aside these details, we find that fermions well below the $\sim$ keV mass scale are {\it not} plausible candidates to be all of the DM.

{\it The warm dark matter bound.}  ---  Another general, qualitative statement can be made about DM candidates with mass below keV. Often referred to as the warm dark matter (WDM) bound, the idea is that there is a suppression in the matter power spectrum for sufficiently low mass DM, see the example in Fig.~\ref{fig:powerspectrum}. Currently, the strongest bounds are from observations of the Lyman-$\alpha$ forest, which is a tracer for the matter power spectrum (see Refs.~\cite{Baur:2017stq,Irsic:2017ixq,Yeche:2017upn,Murgia:2018now} for recent constraints on WDM). Turning to Fig.~\ref{fig:kmode}, the smallest measured scales for the power spectrum correspond to $k \sim 10-20/$Mpc, modes which entered the horizon and started growing at $z \sim 10^7$.  At this time, the photon temperature was $T_\gamma (1+z) \sim$~keV. Therefore, if dark matter was in thermal equilibrium and had similar temperature as the photons, its mass should satisfy $\mDM \gtrsim$~keV -- otherwise, it would be relativistic and lead to damping of the power spectrum. Of course, this is not a hard boundary and specific models can fit observational data depending on the actual velocity of the DM in the early universe. 

\subsection{Ultralight bosonic dark matter}

We will refer to the entire span of candidates below $\sim$ keV as ultralight bosonic dark matter.  The very low mass end of DM candidates is usually quoted as around $\mDM \approx 10^{-22}$ eV. First of all, what happens when DM is this light? It behaves as a coherent field. Let's look at the number of DM particles within a volume given by the de Broglie wavelength:
\begin{align}
	\lambda_{\rm dB} &= \frac{2\pi}{\mDM v} \approx 0.4\ \kpc \left( \frac{10^{-22} \eV}{\mDM} \right) \\
	N &=  \frac{\rho_{\rm DM}}{\mDM} (\lambda_{\rm dB})^3 \approx  10^{94} \left( \frac{10^{-22} \eV}{\mDM} \right)^4 = 75 \left( \frac{10 \eV}{\mDM} \right)^4
\end{align}
Here we used $v \sim 10^{-3}$, as in the Milky Way, and $\rho_{\rm DM} = 0.4 $ GeV/cm$^3$ as the average DM density near the Sun. $N$ is the occupation number, and when $N \gg 1$, then we expect that we can describe the DM as a classical field. In the solar neighborhood, we can describe the DM as a scalar field $\phi$:
\begin{align}
	\phi = \phi_0 \cos (\bfk \cdot \bfx -  \omega_k t) \approx \phi_0 \cos (\bfk \cdot \bfx -  m_\phi t)
\end{align}
where $|\bfk| \simeq 10^{-3} m_\phi$. The magnitude and direction of the vector $\bfk$ is random and fluctuates over length scales $\sim 1/|\bfk| v_0$, where $v_0 \sim 10^{-3}$ is the DM velocity dispersion. Ignoring the gradient energy of the field, the local energy density is
\begin{align}
	E \approx \frac{1}{2} \dot \phi^2 + V(\phi) = \frac{1}{2} \dot \phi^2 + \frac{1}{2} m_\phi^2 \phi^2 = \frac{1}{2} m_\phi^2 \phi_0^2 \, .
\end{align}
We have also dropped any quartic terms in $V(\phi)$. Therefore, the local field value is $\phi_0 = \sqrt{ 2 \rho_{\rm DM}}/m_\phi$.

{\it Formation of cores.}  ---  Neglecting the Hubble expansion, $\phi$ satisfies the classical wave equation $\Box \, \phi +m_\phi^2\, \phi = 0$. In the presence of a gravitational potential, and taking the non-relativistic limit, this then becomes the Schrodinger-Poisson equation. Solving this equation reveals that an ultralight scalar field has a Jeans scale, below which growth of perturbations is suppressed. The result is a suppression of the matter power spectrum, and DM halo profiles which are more cored. In galaxies, the core scale is just $\lambda_{\rm dB}$, since one cannot compress a collection of particles with momentum $mv$ any more than that. On the other hand, the scale of 1 kpc corresponds to the half-light radius of dwarf spheroidal galaxies, and as such the kinematics of stars in these galaxies can be used to constrain ultralight DM. Studies of the Lyman-$\alpha$ forest and dwarf galaxies generally require $m_\phi \gtrsim 10^{-22}$ eV. The regime within an order of magnitude or so of $10^{-22}$ eV is known as Fuzzy DM~\cite{Hu:2000ke,Hui:2016ltb}.

\exercise{Estimate on what scales bosonic DM could undergo gravitational collapse. Use the fact that for a perturbation of size $r = \lambda_{\rm dB}$, we can estimate the velocity of particles inside the perturbation as $v \sim \sqrt{G M/r}$. Assuming an $O(1)$ overdensity, then $v \sim \sqrt{ G \bar \rho_{\rm dm} } r $, where $\bar \rho_{\rm dm}$ is the average DM density.}

{\it Scalar fields as cold dark matter.}  ---  Since the DM is so light, it would be highly relativistic if it had been produced in thermal equilibrium. The abundance here is instead set by misalignment, which is just a generic term for a mechanism that leads to $\phi$ being away from the origin. Suppose at some time $t_i$ an initial value $\phi_i$ was set (within one comoving horizon). One of the defining characteristics of cold dark matter is that the energy drops as the cube of the scale factor $a(t)$, appropriate for a non-relativistic species. We check that this is the case for a scalar field. 

\begin{figure}[t]
\includegraphics[width=0.65\textwidth]{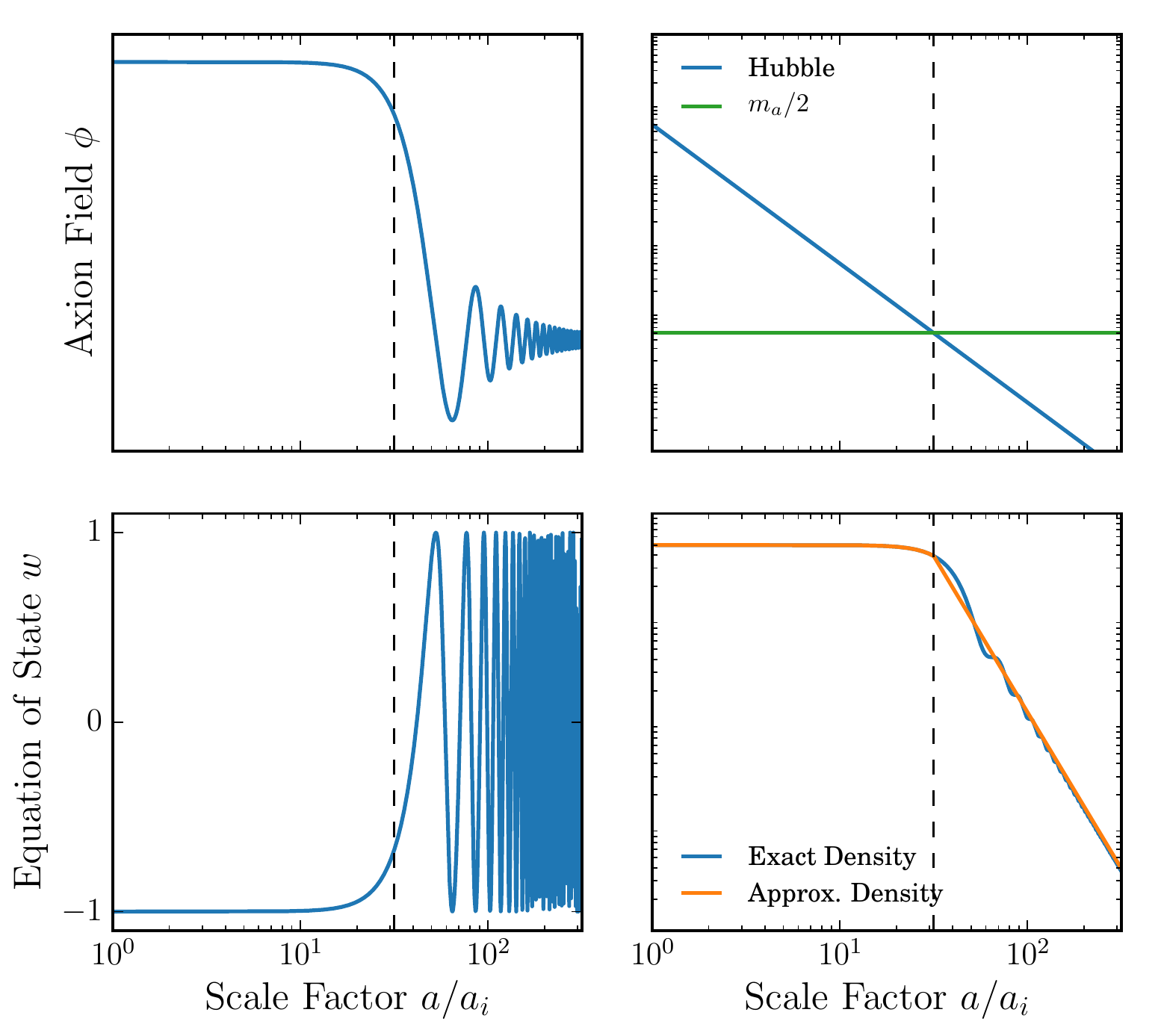}
 \caption{Reproduced from Ref.~\cite{Marsh:2015xka}, the evolution of a scalar field in a radiation dominated universe. The scalar field has mass $m_a$ and the dashed line indicates when $H = m_a/2$, which is approximately when the field starts oscillating.  The lower left panel shows the evolution of the equation of state, where $w = -1$ at early times. At late times, the equation of state oscillates rapidly between $-1$ and 1 but on cosmological time scales of $\sim H^{-1}$ we can approximate $\langle w \rangle \approx 0$, which describes cold dark matter. The lower right panel shows the evolution of the energy density. The orange line is the evolution of the density assuming cold dark matter for $H < m_a/2$, and we see there are some residual oscillations in the exact energy density as we transition to the oscillatory phase. \label{fig:axion_evolution}}
\end{figure}

Considering just the homogenous part of the scalar field, the field equation is 
\begin{align}
	\ddot \phi + 3 H \dot \phi + V'(\phi) = 0
\end{align}
where dots are time derivatives, and $V'(\phi) = dV/d\phi = m_\phi^2\, \phi$ for a free scalar. The second term accounts for the expanding universe through the Hubble expansion $H(t)$, and is known as a Hubble friction term. The solution for $\phi(t)$ can be obtained in two limits:
\begin{align}
	\phi(t) \approx \begin{cases}
		\phi_0  & H \gg m_\phi \\
		\frac{\phi_1}{a(t)^{3/2} } \sin (m_\phi t)  &   H \ll m_\phi
	\end{cases}
\end{align}
where the relations between the constants $\phi_0$ and $\phi_1$ must be determined by matching the numerical solution across the time $t_{\rm osc}$ where $H(t_{\rm osc}) \approx m_\phi$. An exact solution is shown in the top left panel of Fig.~\ref{fig:axion_evolution}. As long as the energy density due to the momentum of $\phi$ is negligible, the total energy density $\frac{1}{2} m_\phi^2 \phi^2$ redshifts as $1/a^3$ when $ H \ll m_\phi$.   Since $H$ is decreasing, this scalar field behaves as a cold dark matter component in the late universe. The evolution of the energy density and equation of state are also shown in Fig.~\ref{fig:axion_evolution}.

A free scalar field having no interactions with the SM, and with an average value set by the misalignment mechanism, is indeed a perfectly valid DM candidate that is extremely difficult to conclusively test. However, if even a tiny coupling is present, there are a variety of interesting experimental proposals to search for this kind of DM, as we will discuss later (briefly).  In addition, the QCD axion is a specific example of bosonic field dark matter where there {\emph{are}} couplings to the SM, which gives a compelling reason to search for weakly-coupled bosons.

{\it The QCD axion.}  ---  In Fig.~\ref{fig:mass_scale}, there is a small but important sliver marked around $10^{-5}$ eV. This is the ``classic'' window for the QCD axion to be all of the dark matter, but note that this is {\emph{not}} the only allowed window. The QCD axion is a pseudo-Nambu-Goldstone boson of an approximate $U(1)$ called a Peccei-Quinn (PQ) symmetry, proposed to solve the strong CP problem. The classic window corresponds to when the PQ symmetry is unbroken during inflation; applying various astrophysical constraints and requiring that the axion comprises all of the DM leads to a remaining narrow mass range. We limit the discussion here, as there are many reviews on axions. An introduction to the axion as a way to solve the strong CP problem can be found within the same TASI proceedings as these~\cite{Hook:2018dlk}, while an earlier extensive review on particle physics models for axions can be found in Ref.~\cite{Kim:2008hd}. These reviews also give an introduction to detection techniques for axions. Refs.~\cite{Sikivie:2006ni,Marsh:2015xka} provide more extended discussion of axion cosmology.

\subsection{Superheavy dark matter and primordial black holes}

What is the largest possible mass of a DM candidate?  The least-massive known galaxies reside in DM halos with mass as low as $\sim 10^5-10^6 M_\odot$ within the half-light radius. This means that DM could be comprised of objects with mass as high as $\sim 10^4-10^5 M_\odot$, and still be consistent with observations of galactic dynamics!  Indeed, there are models of scalar field DM where the field condenses into dense and compact massive objects, known as a boson star or an axion star in the case of axions. For reviews on the subject, see Refs.~\cite{Schunck:2003kk,Braaten:2018nag}. Due to the conventions of the field, this is not regarded as an independent DM candidate, since not all of the DM energy density is necessarily stored in the boson stars. 

{\it Primordial black holes.}  ---  The high density limit of a DM clump is a black hole, and here the constraints are much more severe. Dark matter consisting of black holes is known as  primordial black hole (PBH) DM, since they must have been formed and present well before recombination.  To the extent that we would like to consider any possible candidate that makes up $O(10 \%)$ of the total DM density, then PBHs are viable up to masses of about $\approx 50\, M_\odot$. Massive PBHs  can accrete matter in the early universe, leading to emission of ionizing radiation which is strongly constrained by CMB observations~\cite{Ali-Haimoud:2016mbv}. At the low mass end, PBHs lighter than $\sim 10^{-18} M_\odot$ can evaporate in a time comparable to the age of the universe, while PBHs of mass $\sim 10^{-17} M_\odot$ evaporate enough that they are not sufficiently dark. These are therefore not good DM candidates. 
In between $10^{-17} M_\odot$ and $50\, M_\odot$, there are searches for lensing by PBH dark matter in the Milky Way, which bound the fraction of DM in PBHs at the 10\% level. However, there remain open windows for PBHs to be all of the dark matter for $10^{-16} M_\odot - 10^{-14} M_\odot$ and $10^{-13} M_\odot - 10^{-11} M_\odot$, see for example Ref.~\cite{Katz:2018zrn}. (Note that all statements above assume a PBH mass function which is strongly peaked at one mass.)  For recent reviews on PBHs, see Refs.~\cite{Carr:2016drx,Sasaki:2018dmp}.

{\it Superheavy candidates and composite objects.}  ---  In between the $O(10)$ TeV scale and $\sim 10^{16}$ g, the theoretical landscape is more sparse and has not been explored as thoroughly in the literature.  Up to the Planck mass $M_{\rm pl}$, it is still straightforward to consider fundamental particles as DM; such superheavy candidates are typically called {\emph{WIMPzillas}}. As we will discuss below, they could not have been in thermal equilibrium and must have been produced non-thermally or through gravitational particle production at the end of inflation~\cite{Chung:2001cb}. For masses larger than $M_{\rm pl}$, DM candidates include composite objects -- bound states or nuggets of lighter fundamental particles. The boson stars mentioned above can populate this mass range. A related possibility is that of {\emph{Q-balls}}, solitonic states carrying baryon number that appear in supersymmetric models~\cite{Kusenko:1997si}. {\emph{Nuggets}} of baryons or other fermions have also been considered, where recent work has explored formation in phase transitions or by fusion processes (see Ref.~\cite{Bai:2018dxf} and references therein, for example).

\subsection{WIMPs and thermal candidates}

Like human beings, dark matter candidates tend to gather densely populated regions of the landscape. We lastly turn to the mass range of keV up to $\sim 100$ TeV, which includes the most thoroughly and frequently explored models of DM -- thermal candidates and WIMPs. The rest of these lectures will be devoted to this type of DM, and here we only briefly highlight the motivations and the classes of models.

Thermal DM candidates refer to those where the DM was in thermal equilibrium with the SM thermal bath. Then the early universe density and relic density can be determined by only a few quantities. This is appealing in that other relic densities (of neutrinos, photons, nuclei) are also determined by early universe thermodynamics, with the important caveat that we don't know what sets the baryon-to-photon ratio. Furthermore, the assumption of thermal equilibrium implies some level of interaction between DM and the SM, leading to a variety of interesting and testable signatures. Thermal candidates can be as light as $\sim$~keV, as discussed in the context of the warm dark matter bound in Sec.~\ref{sec:WDMlimit}. Thermal DM more massive than $\sim 100$ TeV suffers from what is known as the unitarity bound or an overclosure problem~\cite{Griest:1989wd}. Within these limits, one can make a few further categorizations:
\begin{itemize}
	\item 10 GeV -- 10 TeV: roughly the mass range for weakly interacting massive particles (WIMPs), where we have taken the mass range of DM candidates in supersymmetric extensions of the SM at the TeV scale. WIMPs are also used to mean thermal DM candidates as a whole, although we will not use that terminology here. For comprehensive reviews of dark matter centered on WIMPs, see Refs.~\cite{Bertone:2004pz,Plehn:2017fdg}.
	\item keV -- 10 GeV: usually called ``light DM'', in contrast to the bosonic field DM discussed earlier (which is instead ``ultralight''). Thermal candidates here are typically examples of dark sector models, and will be the emphasis of these lectures.
\end{itemize}

Within this mass range, there are also interesting candidates where the DM was never in equilibrium with the SM thermal bath. These include sterile neutrino models, which could simultaneously explain neutrino mass and produce DM with the correct relic abundance. These candidates are typically at the keV scale. A recent comprehensive review can be found in Ref.~\cite{Adhikari:2016bei}. Freeze-in DM is another interesting class of dark sector models; we will return to this later.

In Lecture~\ref{sec:thermal}, we will review early universe cosmology and freezeout of thermal DM candidates. We apply these results to work through the implications for particle models of DM. For thermal DM candidates below the GeV-scale, generic ingredients that appear are new light mediators or other weakly coupled states. This often goes under the name dark sectors, an active and vibrant area of DM searches with many new experimental directions. Many models of dark sectors also predict signatures in the DM power spectrum or small scale structure, at the frontier of our understanding of DM clustering. Lecture~\ref{sec:sectors} is an overview of dark sector phenomenology, while Lectures~\ref{sec:dd}-\ref{sec:newdd} provide an introduction to direct detection. 

\clearpage

\section{Thermal dark matter candidates \label{sec:thermal} }

The relic abundance of cold DM obtained from the {\it Planck} 2018 analysis of CMB observations is~\cite{Aghanim:2018eyx} 
\begin{align}
	\Omega_{c} h^2 = 0.120 \pm 0.001
\end{align}
where $h=H_0$ in units of 100 km/s/Mpc and $H_0$ is the rate of expansion of the universe today. Using $h = 0.68$, then $\Omega_{c} \approx 0.259$ and DM is 26\% of the energy density today.  The average DM energy density today is $\rho_c = \Omega_{c} \rho_{\rm crit} = \Omega_{c} 3 H^2/(8 \pi G_N)$ and in physical and natural units, 
\begin{align}
	\boxed{  \rho_{c,0} = 1.26 \times 10^{-6}\ \GeV/\textrm{cm}^3 = 0.97 \times 10^{-11}\ \eV^4. } 
	\label{eq:DMdensity}
\end{align} 
These are helpful numbers to remember for doing quick estimates. Philosophically, this energy density need not be dominated by just one species/particle. However, as a starting point, it is certainly reasonable to study candidates that have the flexibility to be either all of the DM or at least an $O(1)$ fraction of it.

Where does this relic abundance come from? All of the ordinary matter (baryons, leptons, photons) was in thermal equilibrium in the very early universe, and today there are relic photons, neutrinos, electrons, protons, and light elements formed during Big Bang Nucleosynthesis.  Thermal DM candidates are also assumed to be in equilibrium with the SM thermal bath at early times. As the universe cooled, eventually the DM would have been too heavy to be produced in the thermal bath. The number density of the DM would have dropped, both due to the expansion of the universe and due to DM annihilation. The interactions of the DM then become slow compared to the expansion rate and ``freeze out'', leaving a relic abundance that is observed today.

\subsection{Cosmologist's Digest}

We provide here a very brief review of some key quantities and results in early universe thermodynamics (see Ref.~\cite{Kolb:1990vq} for more details). The goal is to be able to quickly perform estimates for DM relic abundances, and use these results to get a feel for the landscape of DM models: broadly speaking, what masses, couplings, or other model features would be sufficient? (Of course, a proper approach requires starting from the Boltzmann equation, and is especially needed in models with more complex dynamics at the time of freezeout. This approach is covered in many other works~\cite{Gondolo:1990dk,Kolb:1990vq,Bertone:2004pz}.)

 There are a few useful relations we'd like to derive:
\begin{itemize}
	\item Estimating when freezeout occurs and what the number density is at freezeout
	\item How to relate the number density at freezeout to the observed density today
\end{itemize}
which we'll then apply to a few examples. 

\begin{table}[t]
\begin{center}
\begin{tabular}{|p{0.99\linewidth}|}\hline 
\rule{0pt}{1ex}
We will be working with the standard FRW metric
\begin{align}
    	ds^2 = dt^2 - a(t)^2 d\bfx^2
\end{align}
where the Hubble rate of expansion is $H = (da/dt)/a$. $H^{-1}$ is a time scale, which is connected to the age of the universe; $c H^{-1}$ therefore corresponds to a horizon size. In the usual convention where $a({\rm today}) = 1$, the scale factor is related to redshift by $1+z = 1/a(t)$.
\ \\
\ \\
The phase space distributions for particles in thermodynamic equilibrium are given by
 \begin{align}
    	f({\bfp}) = \frac{g}{\exp \left( \frac{E(\bfp) -\mu}{T} \right)  \pm 1}. \label{eq:fp}
\end{align}
with the + (-) sign for Fermi-Dirac (Bose-Einstein) statistics.
    Integrating over the momentum gives
\begin{align} 
\label{eq:numberdensity}
{\rm Number \ density:}  \quad n = \int \frac{d^3 \bfp}{(2\pi)^3} f(\bfp) \quad \quad \to 
\quad n = \left[ \frac{3}{4} \right] \frac{\zeta(3)}{\pi^2} g T^3  \\
{\rm Energy \ density:}  \quad \rho = \int \frac{d^3 \bfp}{(2\pi)^3} E(\bfp) f(\bfp) \quad \to 
\quad \rho = \left[ \frac{7}{8} \right] \frac{\pi^2}{30} g T^4  \label{eq:energydensity}
 \end{align}  where the result in the last step is for $m, |\mu| \ll T$, the number in the bracket is the factor needed for fermions, and $\zeta$ is the Riemann zeta function. Similar results can be given for the pressure, which can be written as $p = w \rho$.  \ \\
  \ \\
The rate of expansion is related to the energy density, $H^2 = 8 \pi G_N \rho/3$. During the radiation-dominated era, the energy density is given by
\begin{align}
	\rho = \frac{\pi^2}{30} T^4  \left( \sum_i g_i \left[ \frac{7}{8} \right]  \frac{T_i^4}{T^4} \right) \equiv \frac{\pi^2}{30} g_*(T) \, T^4,
\end{align}
where $g_*(T)$ is the effective number of relativistic degrees of freedom.
It will be useful to remember the temperature scaling of $H$ during the radiation-dominated era:
\begin{align}
    	\quad H \approx 1.66 \sqrt{g_*(T)} \frac{T^2}{\Mpl}, \quad {\rm with} \quad \Mpl = 1.22 \times 10^{19} \, \GeV.
	\label{eq:H_rad}
\end{align}
 An important quantity is the entropy density $s = (p + \rho)/T =  (1+w) \rho/T$. Using $w = 1/3$ for a relativistic species, we obtain
  \begin{align}
 	s = \frac{2\pi^2}{45} T^3  \left( \sum_i g_i \left[ \frac{7}{8} \right]  \frac{T_i^3}{T^3} \right) \equiv \frac{2\pi^2}{45} g_{*,S}(T) \,  T^3.
	\label{eq:entropy}
 \end{align}
 The comoving entropy $s a(t)^3$ is conserved for particles in equilibrium. The number of SM particles that are in equilibrium is a function of $T$, so $g_{*}(T)$ and $g_{*,S}(T)$ are temperature dependent. A redshifting temperature $T \propto 1/a(t)$ is only true when $g_{*,S}(T)$ is constant: across thresholds where $g_{*,S}(T)$ changes, $T$ decreases less slowly. The evolution of both $g_*(T)$ and $g_{*,S}(T)$ are shown in Fig.~\ref{fig:gstar}.
 \\\hline
\end{tabular}
\end{center}
\caption{Summary of early universe thermodynamics. \label{tab:thermo} }
\end{table}
\FloatBarrier

\begin{figure}[t]
\includegraphics[width=0.7\textwidth]{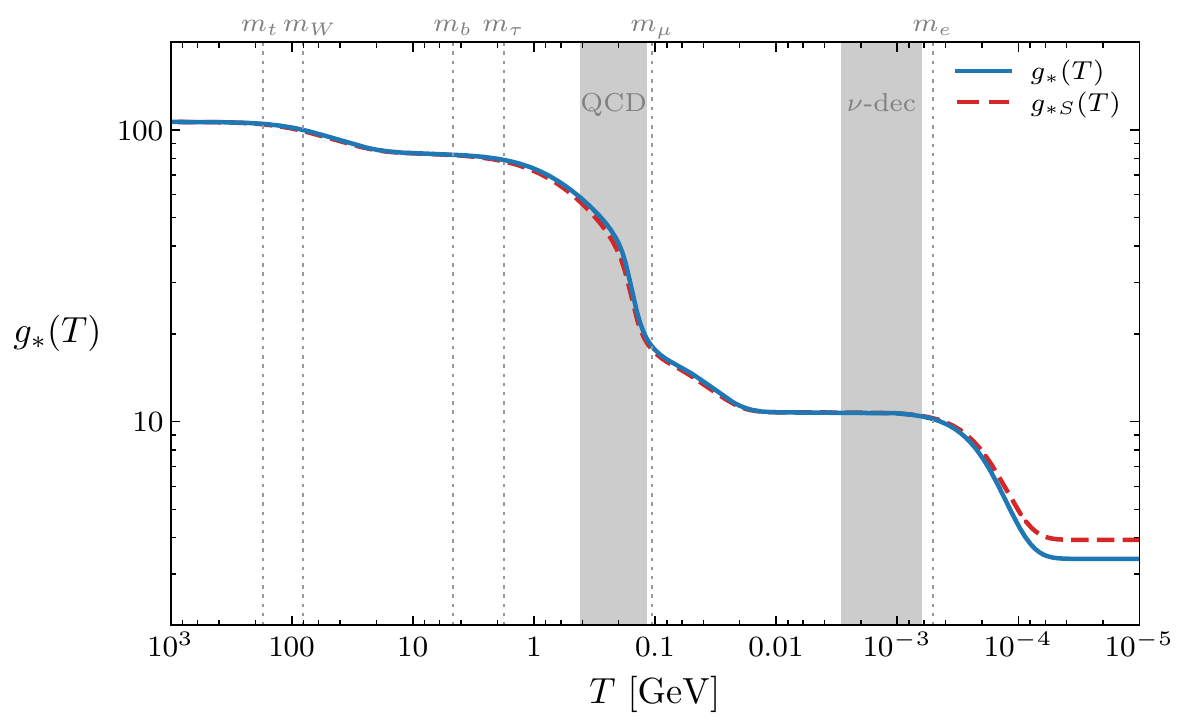}
 \caption{ Reproduced from Ref.~\cite{Wallisch:2018rzj}, the evolution of the effective number of degrees of freedom in the energy density, $g_*(T)$, and in the entropy density, $g_{*,S}(T)$, assuming only the particle content of the Standard Model. Both $g_*(T)$ and $g_{*,S}(T)$ drop when heavy particles annihilate away; some of these masses scales are indicated by the dashed lines. At $ T = 10^3$ GeV when all SM particles are in equilibrium,  $g_*(T) = g_{*,S}(T) = 106.75$. The shaded bands indicate the QCD phase transition and neutrino decoupling.  \label{fig:gstar}}
\end{figure}

The most important definitions and conventions are given in the boxed text in Table~\ref{tab:thermo}.  The time scale $H^{-1}$ is the relevant quantity to estimate when processes are important, and it depends on the effective number of relativistic degrees of freedom $g_*(T)$, shown in Fig.~\ref{fig:gstar}. Meanwhile, the entropy density is useful because $s \, a(t)^3$, the comoving entropy or entropy in a comoving volume, is conserved for a system in equilibrium. This allows us to track $a(t)$ in terms of $g_{*,S}(T)$ and $T$, even at times when the usual redshifting of the photon temperature $T\sim 1/a(t)$  does not hold. This is important during epochs when the number of degrees of freedom in equilibrium with the photon thermal bath is changing, such as when electrons and positrons annihilate away. The temperature evolution of $g_{*,S}(T)$ is shown in Fig.~\ref{fig:gstar}. In order to make use of the conservation of comoving entropy, we need one more quantity:
\begin{align}
	 \boxed{  s_{\gamma,0} = 1.1 \times 10^{-11}\ \eV^3 }
\end{align} 
which is the entropy density contained in the CMB photons, obtained using Eq.~\ref{eq:entropy}.

Define the quantity $Y \equiv n/s$, where $n$ is the number density of a particle species; then $Y$ is proportional to the comoving number density $ n \, a(t)^3$. We will refer to $Y$ as the abundance. If the rate of number-changing processes is sufficiently slow compared to the expansion of the universe, then $Y$ is approximately constant in time. In the spirit of our approach, we will define a specific time or temperature $T_{\rm fo}$ when freeze-out occured. Given $\Yfo = Y(T_{\rm fo})$, we obtain the relic energy density by multiplying by the total entropy density in photons today and the DM mass $m_\chi$:
\begin{align}
	\rho_{\rm DM} \approx m_\chi \Yfo \, s_{\gamma, 0}
\end{align}
where as mentioned above, the relation is only approximate depending on the specific thermal history.

Requiring that this is equal to the total observed DM density, Eq.~\ref{eq:DMdensity}, we find the remarkably simple requirement
\begin{align}
	\boxed{  \Yfo \simeq \frac{\eV}{m_\chi} \, .}
	\label{eq:Yfo}
\end{align}
Since we will consider thermal DM candidates keV or heavier, we need to find a way to have $\Yfo \ll 1$. As we discuss in the following section, a particle which was relativistic during freezeout typically has $\Yfo \sim O(1)$.

\FloatBarrier

\subsection{Neutrino freezeout \label{sec:neutrinofreezeout} }

The story of neutrino freezeout is a testament to the success of the standard cosmology, and we are now in the precision cosmology era where detailed physics of neutrino mass and interactions can be measured\footnote{For example, see {\emph{Neutrinos in Cosmology}} in the PDG review~\cite{PhysRevD.98.030001}.}. We begin by reviewing this story, and apply it to any species that was relativistic when it decoupled from the SM thermal bath (and relativistic today). 

Neutrinos are kept in equilibrium by electroweak processes such as $e^+ e^- \leftrightarrow \nu \bar \nu$ through an off-shell $Z$:
\begin{center}
\begin{tikzpicture}[line width=1.5 pt, scale=1.2]
	\draw[fermionbar] (-140:1)--(0,0);
	\draw[fermion] (140:1)--(0,0);
	\draw[vector] (0:1)--(0,0);
	\node at (-140:1.22) {$\bar  \nu$};
	\node at (140:1.22) {$\nu$};
	\node at (.5,.3) {$Z$};	
\begin{scope}[shift={(1,0)}]
	\draw[fermion] (-40:1)--(0,0);
	\draw[fermionbar] (40:1)--(0,0);
	\node at (-37:1.22) {$e^+$};
	\node at (37:1.27) {$e^-$};	
\end{scope}
\end{tikzpicture}
\end{center}
While neutrinos are in equilibrium, the thermally-averaged cross section times relative velocity for such processes can be estimated as $\sigv \approx G_F^2 T^2$ by dimensional analysis, where $G_F$ is the Fermi constant.  (Note that the proper boost-invariant definition of $\sigv$ is given in the classic paper of Ref.~\cite{Gondolo:1990dk}.) Then the rate per neutrino is $\Gamma \approx n_\nu \sigv \approx G_F^2 T^5$.

The quantity we use in estimating freezeout is $\Gamma/H$. There are a few ways to think about this. One is that the decrease in $n_\nu$ due to expansion is given by $dn_\nu/dt = - 3 H n_\nu$. When $\Gamma/H$ is small, the fractional change in $n_\nu$ due to collisions or interactions is negligible compared to the dilution of the number density by expansion. Another reason is that $1/H$ is a time scale for the age of the universe (in the standard FRW cosmology) so that $\Gamma/H$ gives the possible fractional change in $n_\nu$.  

In this case, using Eq.~\ref{eq:H_rad}, we find a quantity that drops with temperature:
\begin{align}
	\Gamma/H \approx G_F^2 T^3 \Mpl  \, .
\end{align}
When $T$ drops below $(G_F^2 \Mpl)^{-1/3} \approx 1\, \MeV$, processes that change the neutrino number are slow and the neutrinos drop out of equilibrium.\footnote{We have focused the discussion on processes that change the neutrino number, which establish chemical equilibrium. Scattering processes that change neutrino temperature are required for kinetic equilibrium and have a similar scaling in the rate, so they also become ineffective around this time. Thus we can say the neutrinos are no longer in equilibrium with the photons.} In reality, the process is not instantaneous and different neutrino flavors decouple at different times, though all at around $T \sim$1--10~MeV.

Having identified the freezeout temperature $T_{\rm fo} \approx \MeV$, we can obtain the comoving abundance for a relativistic fermion with $g$ degrees of freedom. The number density at the time of freezeout is 
$n = \frac{3 \zeta(3) g}{4\pi^2} T^3$, with a corresponding abundance:
\begin{align}
	 \Yfo = \frac{135\, \zeta(3) g}{8\pi^4 g_{*,S} }.
\end{align}
For active neutrinos, $g=2$ for a single species, and  $g_{*,S}(T_{\rm fo}) = (2 + 4 \times 7/8) = 11/2$ for electrons and photons.  The above quantity is conserved until today.  Now suppose that neutrinos are nonrelativistic today -- can they be a valid DM candidate? We know that Eq.~\ref{eq:Yfo} gives the correct DM relic density. So dividing the result above by $\eV/m_\chi$, we obtain the relic abundance of nonrelativistic neutrinos
\begin{align}
	\label{eq:neutrino_Omega}	
	\Omega_\nu h^2 \simeq 0.12 \times \frac{ g}{ g_{*,S}} \left(\frac{ \sum m_\nu}{4.1\ \eV} \right).
\end{align}
Plugging in the numbers for a single neutrino species, we get the familiar result that $\Omega_\nu h^2 \approx 1$ when $m_\nu = 94$ eV. Conversely, this implies that for $m_\nu \approx 0.1$ eV (the current bound for the neutrino mass scale), a single massive neutrino species is at most 0.2\% of the matter density today.

Now consider instead a new particle $\chi$ which was relativistic at the time of decoupling. Suppose $\chi$ freezeout occurs at a temperature $T_{\rm fo}$ before neutrino decoupling. We have to be a little careful about how we track the entropy. After neutrino freezeout, the entropy of the neutrino sector and that of the photon and electron sector are separately conserved. This can be accounted for in an instantaneous decoupling approximation, where all neutrinos freezeout at exactly $T = \MeV$. Tracking the abundance up until $T = \MeV$ gives
\begin{align}
	n_\chi(T = \MeV) = \frac{135\, \zeta(3) g}{8\pi^4 g_{*,S}(T_{\rm fo})} s(T = \MeV^+)
\end{align}
where the $^+$ superscript indicates we evaluate the entropy of the entire SM thermal bath at temperatures just before neutrino freezeout, so $s(T = \MeV^+) = (2 + 4 \times 7/8 + 6 \times 7/8) \tfrac{2 \pi^2}{45} T^3 = \tfrac{43}{4} \tfrac{2 \pi^2}{45} T^3$.  To track the density from just after $T = \MeV$ to today, we should determine the abundance using just the entropy in the photon bath, since this scales as $a(t)^{-3}$:
\begin{align}
	Y_\chi =  \frac{n_\chi(T = \MeV)}{s_\gamma(T = \MeV^-)} = \frac{135\, \zeta(3) g}{8\pi^4 g_{*,S}(T_{\rm fo})} \frac{43}{22}
\end{align}
where  the entropy in the photon plus positron/electron bath is $s_\gamma(T = \MeV^-) = \tfrac{11}{2} \tfrac{2 \pi^2}{45} T^3$.  This result has an additional factor of $43/22$ compared to the one we obtained for neutrinos; however, this compensated by the fact that $g_{*,S}(T_{\rm fo})$ is larger when $T_{\rm fo} > \MeV$, so that the largest possible value of $Y_\chi$ is that of the abundance for neutrinos. The relic abundance is then given by
\begin{align}
	\Omega_\chi h^2 \simeq 0.12 \times \frac{ g}{ g_{*,S}(T_{\rm fo})} \left(\frac{m_\chi}{2\, \eV} \right).
\end{align}

Earlier, we determined that a viable thermal dark matter candidate should have mass $m_\chi \gtrsim 1-10$ keV.  However, if freezeout occurs when the DM is relativistic, then we obtain $\Omega_\chi h^2 = 0.12$ only when $m_\chi \simeq 1-10\, \eV$. Here it is assumed that $g_{*,S}(T_{\rm fo}) \sim O(10)$ and $g\sim 2$.   Larger $m_\chi$ would lead to an excess of matter density and would result in $\sum_i \Omega_i > 1$, known as overclosure. Alternatively, one can increase $g_{*,S}(T_{\rm fo})$. From the result above, we see that for $m_\chi > \keV$, we would need $g_{*,S}(T_{\rm fo}) \gtrsim 1000$ -- many more degrees of freedom than is present in the Standard Model!\footnote{In fact, this is the assumption in most searches for warm dark matter, meaning the resulting bounds are quite conservative.}  The lesson from this exercise is that freezeout of a relativistic species can give a cold dark matter candidate -- but only in nonstandard cosmologies. Otherwise, the relic number density is simply too high.

\subsection{Thermal freezeout and the WIMP miracle \label{sec:wimpfreezeout}}

\begin{figure}[t]
\includegraphics[width=0.6\textwidth]{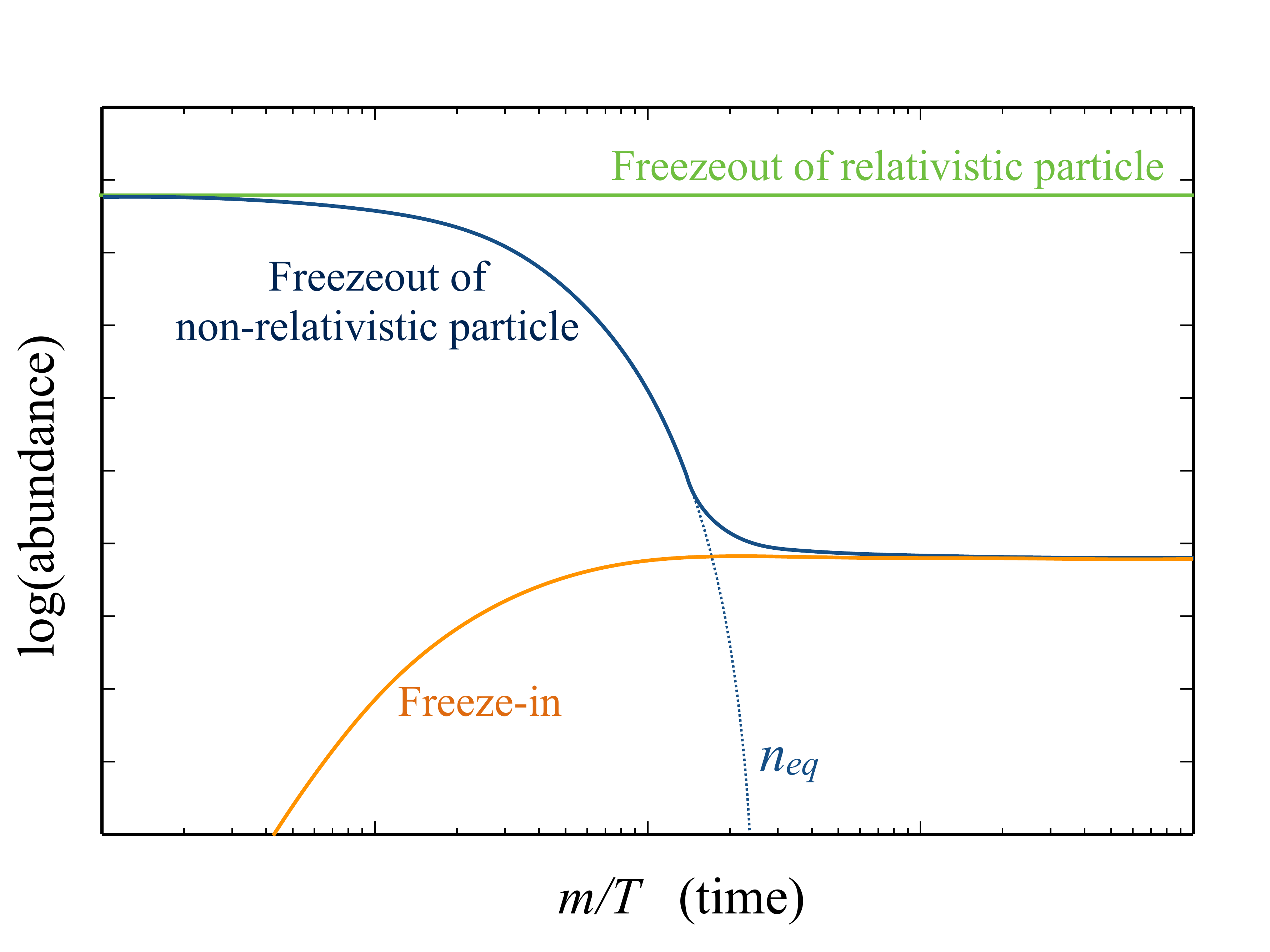}
 \caption{ An illustration of the resulting abundance from freezeout of relativistic particle (Section~\ref{sec:neutrinofreezeout}), freezeout of a nonrelativistic particle (Section~\ref{sec:wimpfreezeout}), and freeze-in (Section~\ref{sec:freezein}). The line labelled $n_{eq}$ assumes the number density for a particle in thermal equilibrium and with zero chemical potential.
 \label{fig:relic_mechanisms}}
\end{figure}

The first and minimal modification we can make to the above arguments is considering a species that is \emph{non-relativistic} at the time of freezeout. Using Eqs.~\ref{eq:fp}-\ref{eq:numberdensity}, the behavior of the equilibrium number density when $T \ll m_\chi$ is given by
\begin{align}
	n_\chi^{eq} \approx g \left( \tfrac{ m_\chi T}{2 \pi} \right)^{3/2} e^{-m_\chi/T}.
\end{align}
The exponential suppression allows us to obtain $Y_{\rm fo} \ll 1$, whereas we saw above that $\Yfo \sim O(1)$ is much too large.

Following the approach above, we can do a quick and dirty calculation to obtain the ``miraculous'' thermal relic WIMP annihilation cross section.  The details of solving these equations are reviewed quite extensively elsewhere in this school or in reviews and books, and we will feel good about getting at the same answer (to an order of magnitude) with not much work.

We first evaluate the condition for freezeout, by comparing the rate of annihilation per DM particle with the Hubble expansion:
\begin{align}
	\Gamma = n_\chi^{eq} \sigv = H
\end{align}
where again $\sigv$ is the thermally averaged cross section times velocity.
With this, we can write the comoving abundance at freezeout
\begin{align}
	  \Yfo = \frac{n_\chi^{eq}}{s} = \frac{H}{s \sigv} \simeq \frac{\sqrt{g_*}}{g_{*,S}} \frac{1}{\sigv T_{\rm fo} \Mpl} \quad .
\end{align} 
$\Yfo \ll 1$ is possible for freezeout of a nonrelativistic species,  as long as the the annihilation cross section $\sigv$ is sufficiently large. A larger $\sigv$ means that interactions of the DM persist for a somewhat longer time, which leads to a further decrease in $n_\chi^{eq}$.

Since the number density (and hence annihilation rate) drops exponentially below $T \approx m_\chi$, we know that $T_{\rm fo}$ should be somewhat below $m_\chi$, but not too far below. For the estimate here, let's take $T_{\rm fo} \simeq m_\chi/10$. Again using Eq.~\ref{eq:Yfo}, we find that the full DM relic abundance is obtained when the annihilation cross section is given by:
\begin{align}
	\boxed{ \sigv \simeq \frac{\sqrt{g_*}}{g_{*,S}} \frac{10}{\eV \times \Mpl } \simeq \frac{1}{10^{9}\ \GeV^2}, }
	 \label{eq:sigv_thermal}
\end{align}
where we estimated the result with typical values for $g_*, g_{*,S}$. This is a \emph{minimum} annihilation cross section needed for a thermal DM candidate, in order to avoid an overabundance. It also presents an interesting target for indirect searches for dark matter, where the often-used benchmark is~\cite{Steigman:2012nb}
\begin{align}
	\sigv \approx 2 \times 10^{-26} {\rm cm}^3/{\rm s},
	\label{eq:sigv_number} 
\end{align}
and we have written the result in the relevant units for those searches.

We have now established that freezeout of a non-relativistic species is a viable way to get the observed relic abundance. As long as $m_\chi \gtrsim 1-10$ keV, we also expect that it is possible to satisfy the warm dark matter bounds discussed in Section~\ref{sec:WDMlimit}. 

\exercise{Suppose that the annihilation rate had a temperature dependence $\sigv = \sigv_0 \, T/m_\chi$ ($p$-wave cross section). Estimate the value of $\sigv_0$ required to the saturate the observed DM relic density. What is the predicted annihilation rate for indirect detection searches in the Milky Way, compared to $s$-wave annihilation cross section above?} 

\subsubsection{Implications for DM models}

Next, let us ponder on the implications of Eq.~\ref{eq:sigv_thermal} for model-building  and the resulting restrictions on DM mass range. We will take as an illustrative example annihilation that occurs through an $s$-channel mediator with mass $m_V$ (remaining for the moment agnostic as to the identity of $V$):
\begin{center}
\begin{tikzpicture}[line width=1.5 pt, scale=1.2]
	\draw[fermionbar] (-140:1)--(0,0);
	\draw[fermion] (140:1)--(0,0);
	\draw[vector] (0:1)--(0,0);
	\node at (-140:1.22) {$\bar \chi$};
	\node at (140:1.22) {$ \chi$};
	\node at (.5,.3) {$V$};	
\begin{scope}[shift={(1,0)}]
	\draw[fermion] (-40:1)--(0,0);
	\draw[fermionbar] (40:1)--(0,0);
	\node at (-37:1.22) {$\bar  f$};
	\node at (37:1.27) {$f$};	
\end{scope}
\end{tikzpicture}
\end{center}
where the vector $V$ has coupling $g_\chi$ with the DM and coupling $g_f$ with the final state fermions.
We will neglect the mass of the final state fermions, as before. In the non-relativistic limit, the cross section for this process is given by
\begin{align}
	\sigma = \int d\Omega_{\rm cm} \, \frac{|\bfp_f|}{16 \pi^2 E_{\rm cm}^3 |\bfv_1 - \bfv_2|} |{\cal M}|^2 = \int d\Omega_{\rm cm} \, \frac{1}{|\bfv_1 - \bfv_2|} \frac{|{\cal M}|^2}{32 \pi^2 s} 
	\label{eq:sigmaCM}
\end{align}
where $\Omega_{\rm cm}$ are center of mass scattering angles, the center of mass energy is $s = E_{\rm cm}^2= 4 m_\chi^2 + O(m_\chi T) + ...$, and we used that $|\bfp_f| \approx E_{\rm cm}/2$ in the limit of massless fermions $f$. Using this result, we can approximate the thermally averaged $\sigv$ for annihilation by
\begin{align}
	\sigv \simeq \frac{|{\cal M}|^2}{32 \pi m_\chi^2} .
\end{align}
Assuming Dirac fermion DM, a single flavor/color of the fermion, and a vector mediator, the spin-averaged matrix element squared of the process is given by
\begin{align}
	|{\cal M}|^2 \approx g_\chi^2 g_f^2 \frac{32 m_\chi^4 }{(s - m_V^2)^2}
\end{align}
in the nonrelativistic limit.

\begin{itemize} 
	\item {\bf $m_V > m_\chi$}: In this case, the heavy $V$ state generates a four-fermion interaction with amplitude $g_\chi g_f/ m_V^2$. The annihilation cross section can be estimated as
	\begin{align}
		\sigv \simeq  \frac{16\pi \alpha_\chi \alpha_f m_\chi^2}{m_V^4} 
		\label{eq:sigv_LeeWeinberg}
	\end{align}
	with $\alpha_\chi \equiv g_\chi^2/(4\pi)$ and $\alpha_f \equiv g_f^2/(4\pi)$.
	\item {\bf $m_V < m_\chi$}:  We find that the annihilation $\chi \bar \chi \to f \bar f$ is:
	\begin{align}
		\sigv \simeq \frac{\pi \alpha_\chi \alpha_f}{m_\chi^2} .
		\label{eq:sigv_lightmed}
	\end{align}
	However, a new process is then kinematically allowed, $\chi \bar \chi \to V V$. If $m_V \ll m_\chi$ then the only mass scale in the problem is $m_\chi$ and we obtain:
	\begin{align}
		\sigv_{\chi \bar \chi \to V V} \simeq \frac{\pi \alpha_\chi^2}{m_\chi^2} . \label{eq:secluded}
	\end{align}
	If $\alpha_\chi \gg \alpha_f$, then the relic abundance may be primarily determined by this latter process. The ``secluded" scenario is where $\alpha_\chi \gg \alpha_f$ and where $V$ is a new mediator not already present in the Standard Model~\cite{Pospelov:2007mp}. Then we may regard the $\chi$ and $V$ states as comprising a ``dark sector'', and couplings to the SM thermal bath are not important for thermal freezeout. 
\end{itemize}
We find, generally, that the thermally-averaged cross section is bounded from above by 
\begin{align}
		\sigv \lesssim \frac{\pi \ \textrm{max}(\alpha_\chi \alpha_f, \alpha_\chi^2)}{m_\chi^2} 
		\label{eq:upper_sigv}
\end{align}
since for the case $m_V > m_\chi$ there is an additional suppression in the cross section by $(m_\chi/m_V)^4$. (An exception is if there is a resonance in the $s$-channel, $m_V \approx 2 m_\chi$.) 

If the desired cross section is that of Eq.~\ref{eq:sigv_thermal}, then there are a few important lessons to draw from this. First, one can set an upper bound on the DM mass, if we assume perturbative couplings  -- this is also known as a perturbative unitarity bound~\cite{Griest:1989wd}. Taking $\alpha_{\chi,f} \to 1$ in Eq.~\ref{eq:upper_sigv} then gives
\begin{align}
		m_\chi \lesssim 50-100\ \textrm{TeV}.
\end{align}
More specific bounds can be obtained in particular models~\cite{Hedri:2014mua}. The basic idea is that a heavier DM candidate that was in thermal equilibrium could not annihilate away sufficiently, resulting in an overabundance of DM. We already alluded to the fact that there are possible exceptions to this analysis, and we will return to this in Sec.~\ref{sec:miracles}.

The second lesson is that a TeV-scale DM candidate is a viable possibility, where $\alpha_{\chi,f}$ can be on the order $10^{-2}-10^{-1}$. Indeed, the WIMP miracle is the observation that $\sigv$ above can be rewritten in the form:
\begin{align}
	\sigv \approx \frac{\alpha_w^2}{1 \, {\rm TeV}^2}
	\label{eq:WIMP_miracle}
\end{align}
with $\alpha_w \approx 0.03$ for  $SU(2)_L$ weak interactions.  This is an inviting cross section for particle theorists accustomed to the sizes of the SM gauge couplings and to thinking about extensions of the SM at the TeV scale. Indeed, thermal candidates that are electroweak multiplets (such as a pure Higgsino in supersymmetric extensions of the SM) remain compelling DM candidates~\cite{Cirelli:2005uq,2013JHEP...05..100C,Bramante:2015una,Beneke:2016jpw,Krall:2017xij}. In the most minimal scenario, no other new particles or gauge interactions are required for TeV thermal relics. 

\subsubsection{Mediators for light DM}

In turning to lighter DM below the weak scale, the minimalist approach becomes insufficient at a certain point. Let us take $V$ to be a weak scale mediator, which could be the Higgs or an EW gauge boson. We have a cross section given by Eq.~\ref{eq:sigv_LeeWeinberg}, which can be rewritten as
\begin{align}
	\sigv \approx \frac{m_\chi^2}{\GeV^2} \frac{\alpha_\chi \alpha_f}{\alpha_w^2} \frac{1}{10^{9}\, \GeV^2}
\end{align}
where we have taken $m_V \approx 100\ \GeV$. If the couplings are the weak gauge couplings, $\alpha_\chi \alpha_f = \alpha_w^2$, then the cross section drops below the desired thermal relic cross section when $m_\chi < \GeV$. Weak interactions would thus lead to an overabundance of sub-GeV DM, a conclusion commonly known as the Lee-Weinberg bound~\cite{Lee:1977ua}.

The implication is that for sub-GeV DM, {\emph{new mediators below the weak scale are required.}} But wait -- there is one more candidate for $V$ in the Standard Model, the photon. Since $m_V \ll m_\chi$, the Lee-Weinberg bound does not apply. Although this would appear to go against the notion of dark matter as ``dark'', we must work through this possibility in a more quantitative way.

Consider a DM candidate which has a small, fractional electric charge $Q \ll 1$, otherwise known as a millicharged or minicharged DM candidate.\footnote{For a discussion of specific models realizing this, see for example Refs.~\cite{Dubovsky:2003yn,Vogel:2013raa,Dvorkin:2019zdi}.} The annihilation cross section $\bar \chi \chi \to f \bar f$ for a single charged species $f$ is given by $\sigv \approx \pi \alpha_{\rm em}^2 Q^2/m_\chi^2$; this should be summed over all final states with appropriate final state charges and color factors. Neglecting such factors for the purpose of estimation, one finds that 
\begin{align}
	Q \simeq 10^{-3} \left( \frac{m_\chi}{\GeV} \right)
	\label{eq:millicharge_thermalrelic}
\end{align}
in order to obtain the desired thermal relic cross section. This time $\bar \chi \chi \to \gamma \gamma$ can be neglected since the cross section has an additional $Q^2$ suppression. Furthermore, we ought to require $m_\chi > m_e$ for there to be a final state to annihilate into.

\begin{wrapfigure}{r}{0.15\textwidth}
\begin{tikzpicture}[line width=1.5 pt, scale=1.2]
\begin{scope}[shift={(4,-0.5)}]
	\begin{scope}[rotate=90]
			\draw[fermionbar] (-140:1)--(0,0);
			\draw[fermion] (140:1)--(0,0);
			\draw[vector] (0:1)--(0,0);
			\node at (-140:1.2) {$f$};
			\node at (140:1.2) {$f$};
			\node at (.5,.3) {$\gamma$};	
		\begin{scope}[shift={(1,0)}]
			\draw[fermionbar] (-40:1)--(0,0);
			\draw[fermion] (40:1)--(0,0);
			\node at (-40:1.2) {$\chi$};
			\node at (40:1.2) {$\chi$};	
		\end{scope}
	\end{scope}
\end{scope}
\end{tikzpicture}
\end{wrapfigure}

Such charges can potentially be constrained in accelerator experiments. However, the strongest test of charged DM arises from its behavior in the early universe. At redshifts $z \gtrsim 1000$, the universe was mostly ionized in the form of free protons and electrons. 
A millicharged DM candidate can scatter off the protons and electrons with a Rutherford-type cross section, leading to both a suppression of the growth of DM structure~\cite{McDermott:2010pa} as well as a DM-baryon drag force which leaves an imprint on the CMB anisotropies~\cite{Dvorkin:2013cea,Xu:2018efh,Slatyer:2018aqg,Boddy:2018wzy}. We can obtain an estimate of the bound arising from the first effect, although the second effect leads to an even stronger constraint over the entire mass range~\cite{Xu:2018efh}.

DM-baryon scattering has a Rutherford-type cross section given by $d\sigma/d\Omega = Q^2 \alpha_{\rm em}^2/\left(4\mu_{\chi f} v^2 \sin^2 (\theta/2) \right)^2$, where $\mu_{\chi f}$ is the DM-baryon reduced mass and $v$ is the relative velocity. This scattering rate diverges in the $\theta \to 0$ limit and is regulated by a Debye screening angle $\theta_D \equiv m_D/(\mu_{\chi f} v)$, with the Debye mass given by $m_D = \sqrt{4 \pi \alpha n_e/T_\gamma} \approx 4 \times 10^{-6}\,  T_\gamma$. Furthermore, in the limit of forward scattering the momentum transfer between the DM and baryon goes to zero. As a result, the physically relevant quantity is the transfer cross section $\sigma_T$, defined as:
\begin{align}
	\sigma_T \equiv \int d\Omega \, \frac{d\sigma}{d\Omega} (1 - \cos \theta) = \frac{ 4 \pi Q^2 \alpha^2_{\rm em}}{ \mu_{\chi f}^2 v^4}  \log \left( \frac{2}{\theta_D} \right) \, .
	\label{eq:sigma_transf}
\end{align}

\exercise{Consider a gas of DM particles (with temperature $T$) incident with average velocity ${\bf V}$ on a gas of protons with the same temperature. Assuming nonrelativistic particles, what is the momentum transfer in a single collision? Average over the DM and proton velocity distributions to obtain the force $m_\chi d{\bf V}/dt$ experienced by the DM gas, and show that it is proportional to the transfer cross section. }

We now make a number of approximations in order to obtain a quick estimate. Let us focus on DM-proton scattering, since the transfer cross section scales as $1/v^4$ and the protons are much slower moving than the electrons. Next, for sub-GeV DM we can take $\mu_{\chi p} \approx m_\chi$. For the velocities we can make a few choices: we can assume that the relative velocities $v$ are dominated by the baryon thermal velocities, $v \sim \sqrt{T_\gamma/m_p}$, while the DM is much colder. Alternatively, we can assume that the DM is at a similar temperature, in which case $v \sim \sqrt{T_\gamma/m_\chi}$ since the DM is much lighter. Since the cross section is inversely proportional to $v^4$, we will make the conservative assumption that $v \sim \sqrt{T_\gamma/m_\chi}$ because this suppresses the scattering rate and will give the weakest constraint (this was the case considered in Ref.~\cite{McDermott:2010pa}). Finally, because the Debye angle $\theta_D$ is usually tiny, we approximate the forward scattering log as $\log \tfrac{2}{\theta_D} \sim O(10)$.  Then the rate per DM particle is
\begin{align}
	\Gamma = n_p \sigv \approx  6\times 10^{-10} \, n_\gamma \frac{40 \pi Q^2 \alpha^2_{\rm em}}{ \sqrt{m_\chi T_\gamma^3} } \approx 10^{-8} \, Q^2 \alpha^2_{\rm em} \sqrt{ \frac{ T_\gamma^3}{ m_\chi }}
\end{align}
Here we used that the baryon-to-photon asymmetry is $\eta \approx 6 \times 10^{-10}$ and $n_\gamma \approx  T_\gamma^3 /4$.

We will compare the rate against the Hubble expansion at $T_\gamma \approx$ eV: at these temperatures, we are well within the epoch where we require DM structure formation to proceed according to $\Lambda$CDM. At this time, the universe is just barely still radiation dominated and using $H \approx T_\gamma^2/\Mpl$, we find that $\Gamma/H \ll 1$ implies
\begin{align}
	Q \ll 2 \times 10^{-6} \left( \frac{m_\chi}{\GeV} \right)^{1/4}
	\label{eq:Qbound}
\end{align}
This bound excludes the charges needed for a thermal relic, given in Eq.~\ref{eq:millicharge_thermalrelic} for $m_\chi > m_e$.

Concluding our survey of millicharged DM, we return to the main point: a model for sub-GeV DM which obtains its relic abundance by thermal freezeout generically requires additional new sub-GeV states for sufficient DM annihilation. We have shown that the SM electroweak gauge bosons and photon are not viable possibilities. (We have not discussed the strong force, as there are even more experimental constraints!) This will motivate the study of dark sectors for sub-GeV DM, that contain both the DM and other light states. The excess of energy and entropy density in a dark sector may be excluded by other cosmological considerations, requiring that the excess be deposited back into the SM thermal bath. As a result, we will often consider new light {\emph{mediators}} to the SM. We will explore these arguments in more detail in Lecture 3. 

\subsection{Many miracles \label{sec:miracles} }

The discussion up to this point has motivated thermal DM with relic abundance set by annihilations,  viable in the mass range of keV up to 100 TeV.    Within this mass range, the ``magic number'' of Eq.~\ref{eq:sigv_thermal} and the WIMP miracle seems like a wonderful coincidence in scales, and suggestive of a physics connection. But it is a coincidence between two model paradigms, not two demonstrated phenomena. Let's see what happens when we examine more closely some of the assumptions leading to Eq.~\ref{eq:sigv_thermal}:
\begin{itemize}
	\item  {\emph{No chemical potential.}}  It was assumed that the annihilation rate is proportional to the equilibrium number density of $\chi$ particles, but there could be an asymmetry in the number density of $\chi$ and $\bar \chi$, similar to the baryons. The relic density then depends on both the annihilation rate and the asymmetry~\cite{Graesser:2011wi,Lin:2011gj}. An additional appeal of asymmetric DM models~\cite{Zurek:2013wia} that they can explain why $\Omega_c$ is of the same order as the baryon density $\Omega_b$.
	\item  {\emph{No resonances or threshold behavior.}} It is assumed that the cross section is approximately constant throughout freezeout. The presence of threshold behavior or resonances leads to a strong $T$-dependence in the cross section, affecting the desired couplings (see for example Ref.~\cite{Griest:1990kh} on ``three exceptions'' to the standard freezeout story).
	\item  {\emph{Annihilations of DM DM $\to$ SM SM interactions dominate.}} The annihilation rate is proportional to one power of the DM density with this assumption. Exceptions include co-annihilation, where DM annihilates against another (heavier) state~\cite{Griest:1990kh};  co-scattering~\cite{DAgnolo:2017dbv}, where inelastic DM scattering processes determine the relic abundance; and SIMP~\cite{Hochberg:2014dra} models where $ 3\to 2$ annihilations are most important in setting the relic density. If the annihilation is DM DM DM $\to$ DM DM, this can be considered an example cannibal DM~\cite{Carlson:1992fn,Farina:2016llk}, where the dark sector consumes itself to stay warm.
	\item  {\emph{Annihilation during the radiation-dominated era in the standard cosmology. }} A long-lived massive particle present in the early universe could result in a matter dominated era. If freezeout occurs during this era, then the annihilation cross section could in principle be much smaller, since the DM density is diluted after the decay of the massive particle. Examples include Refs.~\cite{Kane:2015qea,Drees:2017iod}. Note that this could also be applied to get around the need for new light mediators for sub-GeV thermal candidates. 
\end{itemize}
The list goes on.  
All of these have been applied in dark sector models, showing the great breadth of freezeout phenomenology.

So far we have outlined ways to decrease $\Yfo$, motivated by the overabundance of any massive particle that was relativistic as freezeout. Of course, another way to suppress $\Yfo$ is if the DM had never been in equilibrium in the early universe. The difficulty with this scenario is that the relic abundance can be quite sensitive to initial conditions (UV-dominated) and other assumptions at early times. However, it is possible to have a non-thermal DM candidate that is not UV-dominated, {\emph{and}} potentially observable direct detection experiment. In the remainder of this section, we describe this candidate.

\subsubsection{Freeze-in \label{sec:freezein}}

There is one more scenario, which strictly speaking is not a thermal candidate.  Freeze-in~\cite{Hall:2009bx} is a mechanism whereby rare interactions within the SM thermal bath slowly build up an abundance of DM. (In the usual freeze-in story, it is thus assumed that dark sector particles are not produced at an appreciable level through decay of the inflaton during reheating.) As a specific example, let's look at freeze-in by $s$-channel annihilation of SM particles, such as $e^+ e^-$, into DM particles. The coupling of the DM particles is assumed to be sufficiently feeble, that the reaction is never in equilibrium. 

In the ``UV-dominated'' scenario, the production cross section depends on a high scale $\Lambda$; for example, for scalar DM the interaction is a dimension-5 operator which we can parameterize as $\tfrac{g_\chi g_e v_H}{\Lambda^2} \chi^2 \bar e e$.  We have included the factor of the Higgs vacuum expectation value, $v_H = 246$ GeV, to account for the fact that the operator is not $SU(2)_L$ invariant. Then the cross section goes as
\begin{align}
	\sigv \simeq \frac{\alpha_\chi \alpha_e v_H^2}{\Lambda^4}.
\end{align}
The rate of producing a DM particle per electron is $\Gamma_{e^+ e^- \to \chi \chi} = n_e \sigv \sim \alpha_\chi \alpha_e T^3 v_H^2/\Lambda^4 \,$ for $T \gg m_e, m_\chi$. Thus, the number of DM particles  created per electron in a Hubble time is $\Gamma H^{-1}$.  The abundance of total newly-created DM at any given time is 
\begin{align}
	Y_\chi = \frac{n_\chi}{s} \simeq \frac{ n_e \Gamma H^{-1}}{s} \simeq  \frac{\Gamma}{g_{*,S} \, H} \simeq  \frac{\alpha_\chi \alpha_e \, v_H^2 \, \Mpl T}{\sqrt{g_*} g_{*,S} \, \Lambda^4},
\end{align}
with the greatest abundance produced at the highest $T$ (as long as $T < \Lambda$). The relic density is sensitive to the reheating of the universe and the maximum available temperatures. Assuming that this process gives all of the dark matter and requiring that the highest $T > \MeV$ and $m_\chi > \keV$ gives a lower bound on  $\Lambda/(\alpha_\chi \alpha_e)^{1/4} \gtrsim 10^6 \, \GeV$. With such high scales or small couplings, the prospects for laboratory detection in the near future are quite limited. 

\exercise{What is the maximum DM-electron scattering cross section for this case? In order to obtain an estimate, assume an incident DM with velocity $v \sim 10^{-3}$ and take the electron to be at rest. Then the electron scattering cross section is roughly $\sigma_e \sim \mu_{\chi e}^2 \alpha_\chi \alpha_e v_H^2/(m_\chi^2 \Lambda^4)$, where $\mu_{\chi e}$ is the DM-electron reduced mass. What if we instead consider a dimension-6 operator (the heavy $V$ limit of the vector interaction considered earlier) for freeze-in?}

On the other hand, if the mediator is lighter than the DM, then we can have ``IR-dominated'' freeze-in. Going back to the example introduced above Eq.~\ref{eq:sigmaCM} where DM is coupled to a light vector mediator, the production cross section for $e^+e^- \to \chi \bar \chi$ has the form
\begin{align}
	\sigv \simeq \frac{\alpha_\chi \alpha_e}{T^2}
\end{align}
in the limit of $m_V \ll m_\chi$.
Similar to above,  the  comoving abundance of total newly-created DM is given by
\begin{align}
	Y_\chi = \frac{n_\chi}{s}  \approx  \frac{\Gamma}{g_{*,S} \, H} \simeq  \frac{\alpha_\chi \alpha_e \, \Mpl}{\sqrt{g_*} g_{*,S} \, T}
\end{align}
so that most of the DM is produced at around the lowest $T$ where the process is kinematically accessible. Either the DM becomes too heavy ($T< m_\chi$) which suppresses the production rate, or  the electrons become too dilute ($T< m_e$). While the total abundance should be obtained by integrating the production at all times, we can estimate the relic abudance by taking $\Yfo = Y(T)$ at the lowest $T$.
There are two possibilities depending on the DM mass, which results in the following condition on the couplings
\begin{align}
		\alpha_\chi \alpha_e  \simeq \begin{cases}
		 \sqrt{g_*} g_{*,S} |_{T=m_\chi} \times  \frac{ \eV}{\Mpl} \approx 3\times 10^{-27} - 10^{-26} , &   m_\chi > m_e \\
		\sqrt{g_*} g_{*,S} |_{T=m_e}  \times \frac{ \eV}{\Mpl} \frac{m_e}{m_\chi}  \approx 3 \times 10^{-27} \times \frac{m_e}{m_\chi} , &   m_\chi < m_e 
		 \end{cases}.
\end{align}
If we take the mediator to be the SM photon, this is then an example of a sub-GeV DM candidate that does not require any new mediators beyond the SM! In particular, the couplings required satisfy the bound given in Eq.~\ref{eq:Qbound}. (Note that for $m_\chi < m_e$ and the SM photon as the mediator, there is an additional large production mechanism whereby the in-medium plasma oscillations can decay to $\chi \bar \chi$; this modifies the coupling constants above by about an order of magnitude, depending on the mass~\cite{Dvorkin:2019zdi}.) Despite the tiny couplings, this scenario is potentially detectable with direct detection or indirect searches when the mediator mass is much smaller than the DM mass. We will explore this further in the following lectures.

In addition to freeze-in via annihilation-type processes, another possibility is freeze-in of DM through the decay of heavy particles. Since the heavy particle can have relatively larger coupling with the SM, they could be produced in colliders and decay with a long lifetime, giving an interesting observational handle of these types of freeze-in models. For studies of the long-lived particle collider signatures from freeze-in, see for example Refs.~\cite{Co:2015pka,Calibbi:2018fqf,Belanger:2018sti}. A recent review of freeze-in models and their associated signatures can be found in Ref.~\cite{Bernal:2017kxu}.

\clearpage

\section{Dark sectors and light mediators \label{sec:sectors} }  

What we have learned from the previous lecture is that models of sub-GeV thermal relic DM generically require some new mediator states below the weak scale. This is so that there is a sufficiently large annihilation cross section, and the DM is not too abundant. The new state (and the DM itself, of course) must have a sufficiently small coupling to SM particles, otherwise we could have seen it in numerous laboratory experiments. For instance, the new state could mediate SM-SM interactions as well, and at a minimum the effects should be smaller than weak interactions. This motivates the {\bf dark sector, or hidden sector} framework, where we view the DM and mediator as part of a separate ``Dark Standard Model'', and where there is a portal link to the SM. If the dark sector is thermally populated during the early universe, the portal link also allows for any excess energy/entropy to be deposited back to the SM thermal bath. The ingredients of the dark sector can be fairly minimal, consisting of a feebly coupled dark force and a dark fermion charged under that force. At the other end of the spectrum of possibilities, there are composite dark sectors or a multitude of new states (such as a large number of copies of the SM). 

The reasons for studying dark sectors are the following:
\begin{itemize}
	\item They represent a theoretical generalization of thermal DM candidates beyond WIMPs or supersymmetric DM 
	\item They provide a collection of benchmarks and useful targets for experimental tests of the thermal relic scenario
	\item They represent new directions in the signatures and phenomenology of DM that is driven more by the desire to expand experimental searches and data, rather than by top-down considerations such as naturalness, the strong CP problem, etc.
\end{itemize}
While we do not aim to provide a complete review of the literature, we refer the reader to the some of the well-known early works proposing dark sectors~\cite{Boehm:2003hm,Pospelov:2007mp,Feng:2008mu,ArkaniHamed:2008qn} and to some recent white papers~\cite{Alexander:2016aln,Battaglieri:2017aum}. 

Here we explore realizations of dark sectors, the implications for cosmology, and the myriad ways to search for the portal link. We begin by asking the question -- is any (detectable) link to the SM {\it{required}}?

\subsection{Cosmology of secluded sectors} 

In the case of a secluded sector (see Eq.~\ref{eq:secluded} and below), the DM candidate $\chi$ annihilates to a mediator state $V$ with $m_\chi > m_V$. Ostensibly, interactions with the SM are not required for thermal freezeout. However, there are interesting and strong constraints even in this scenario.

\begin{figure}[t]
\begin{center}
\begin{tikzpicture}[line width=1.5 pt, scale=1.2]
	\draw[fermionbar] (-0.8,-0.6)--(0,0);
	\draw[fermionbar] (0,0)--(0,1);
	\draw[fermion] (-0.8,1.6)--(0,1);
	\draw[vector] (1.0,-0.2)--(0,0);
	\draw[vector] (1.0,1.2)--(0,1.0);
	\node at (-1.1,-0.6) {$\bar \chi$};
	\node at (-1.1,1.6) {$ \chi$};
	\node at (.5,.3) {$V$};	
	\node at (0,-1.2) {$m_\chi> m_V$, $m_V \ll$ eV};
\end{tikzpicture} \hspace{1cm}
\begin{tikzpicture}[line width=1.5 pt, scale=1.2]
	\draw[fermionbar] (-0.8,-0.6)--(0,0);
	\draw[fermionbar] (0,0)--(0,1);
	\draw[fermion] (-0.8,1.6)--(0,1);
	\draw[vector] (1.0,-0.2)--(0,0);
	\draw[vector] (1.0,1.2)--(0,1.0);
	\node at (-1.1,-0.6) {$\bar \chi$};
	\node at (-1.1,1.6) {$ \chi$};
	\node at (.5,.3) {$V$};	
	\node at (2.3,-1.2) {$m_\chi> m_V > 2 m_{\nu_D}$, $m_{\nu_D} \ll$ eV};
\begin{scope}[shift={(1,-0.2)}]
	\draw[fermion] (-25:0.7)--(0,0);
	\draw[fermionbar] (25:0.7)--(0,0);
	\node at (0.9,-0.3) {$\bar  \nu_D$};
	\node at (0.9,0.3) {$\nu_D$};	
\end{scope}
\begin{scope}[shift={(1,1.2)}]
	\draw[fermion] (-25:0.7)--(0,0);
	\draw[fermionbar] (25:0.7)--(0,0);
	\node at (0.9,-0.3) {$\bar  \nu_D$};
	\node at (0.9,0.3) {$\nu_D$};	
\end{scope}
\begin{scope}[shift={(4.5,0.5)}]
	\draw[fermion] (0,0)--(0.5,0.5);
	\draw[fermionbar] (0,0)--(0.5,-0.5);
	\draw[vector] (-1.0,0)--(0,0);
	\draw[fermion] (-1.5,0.5)--(-1,0);
	\draw[fermionbar] (-1.5,-0.5)--(-1.0,-0);
	\node at (-1.7,-0.55) {$\bar \chi$};
	\node at (-1.7,0.55) {$\chi$};	
	\node at (0.75,-0.55) {$\bar  \nu_D$};
	\node at (0.75,0.55) {$\nu_D$};	
	\node at (-0.5,-0.35) {$V$};	
\end{scope}
\end{tikzpicture}
\end{center}
\caption{({\bf left}) A secluded sector with a very low mass $m_V \ll $ eV, where the relic abundance is determined by the annihilation process $\bar \chi \chi \leftrightarrow V V$. ({\bf right}) A secluded sector where the vector $V$ decays to $\bar \nu_D \nu_D$, with $m_{\nu_D} \ll$ eV. The relic abundance depends on both diagrams, where the dominant process depends on the couplings. \label{fig:secluded} } 
\end{figure}
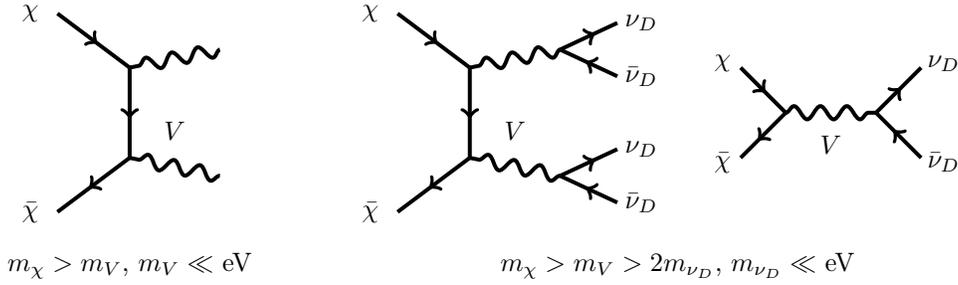

As a starting point, suppose the dark sector was populated during reheating, leaving behind a separate thermal plasma of secluded sector states that don't interact with the SM thermal bath during freezeout. Let us call the temperature ratio between the two sectors $\xi$:
\begin{align}
	\xi \equiv \frac{T_{d}}{T_\gamma},
\end{align}
where this quantity is time-dependent. In order for the freezeout story to be more or less unmodified, the DM must interact with a relativistic thermal bath. If $m_V \ll m_\chi$, then the thermal population of $V$ itself could play this role. An alternative simple extension is to introduce an additional light species $\nu_D$ in the dark sector, where $m_\chi > m_V \gg m_{\nu_D}$. The annihilation processes for these two possibilities is shown in Fig.~\ref{fig:secluded}.

\exercise{How are the estimates of relic abundance modified for this secluded sector? As part of this exercise, determine how the relic abundance calculation changes when $\xi \neq 1$.}

The excess entropy and energy density in the dark sector must go somewhere, however. Suppose $V$ is stable, as in the first scenario in Fig.~\ref{fig:secluded}.  Then $V$ is a component of the total dark matter relic density, with a number density proportional to $\xi^3 T^3$. Assuming that $\xi \sim O(1)$, we can use Using Eq.~\ref{eq:neutrino_Omega} as a guide for the relic abundance in $V$. Then if $m_V \gg$ eV, there would be an overabundance of DM, but for $m_V$ on the order of eV there would be a non-negligible quantity of hot DM. As a result, it must be that $m_V \ll$ eV. (For more quantitative bounds on specific hot DM scenarios, see for example Ref.~\cite{DiValentino:2015wba}.) It is then the case that $V$ behaves as an additional component of radiation during the early universe. 

The presence of additional radiation during BBN or recombination leaves an imprint on the light element abundances and the CMB power spectrum. Any free-streaming radiation not contained in the photon bath is usually encapsulated in the quantity $\Neff$, the effective number of relativistic degrees of freedom. At leading order, the relevant quantity is the energy density in this relativistic species, so let us write down the total energy density in radiation, including  neutrinos while they are still relativistic:
\begin{align}
	\rho_{\rm tot} &= \frac{2 \pi^2}{30} \, T_\gamma^4 + \frac{7 \Neff^{\rm SM} }{4}  \frac{\pi^2}{30} \,T_\nu^4 + \rho_{d} 
\end{align}
where $\rho_{d}$ is the dark sector contribution. The second term is from neutrinos, using Eq.~\ref{eq:energydensity} with $g=2$. Recall that in the instantaneous decoupling approximation that $T_\nu = (4/11)^{1/3} T_\gamma$ is the neutrino temperature after electron decoupling in the standard cosmology.  In the standard model, $\Neff^{\rm SM} \approx 3.046$, accounting for deviations from instantaneous decoupling.

Additional sources of radiation are parametrized by  $\Delta \Neff$, the deviation in $\Neff$ from the standard model value. Assuming that the neutrino temperature is the same as in the SM, then
\begin{align}
	\rho_{\rm tot}  \equiv \rho_\gamma + \left(\Neff^{\rm SM} + \Delta \Neff \right) \frac{7}{4} \frac{\pi^2}{30} \,T_\nu^4 \ .
	\label{eq:rhotot_Neff}
\end{align}
For the dark sector, this implies
\begin{equation}
	\rho_{d} = \sum_i g_i \frac{\pi^2}{30}  T_i^4  \quad \rightarrow \quad \Delta \Neff=\frac{4}{7} \sum_i g_i \left(\frac{T_i}{T_\nu}\right)^4,
\end{equation} 
where  $g_i$ and $T_i$ are the effective degrees of freedom and the temperature of the various relativistic species in the dark sector.

Clearly, $\Delta \Neff$ is a time-dependent quantity, while the constraints are at specific epochs: $\Delta \Neff^{\rm BBN}$ is the energy density in radiation for $ T \sim 10\, \keV - 1\, \MeV$ and $\Delta \Neff^{\rm CMB}$ is the energy density $T < \, \eV$ for $\Delta \Neff^{\rm CMB}$. At the time of writing, BBN abundance measurements~\cite{Cyburt:2015mya} and {\em Planck} observations of the CMB (combined with baryon acoustic oscillations) indicate~\cite{Aghanim:2018eyx}, respectively,
\begin{align}
	 \Neff^{\rm BBN} &= 2.89 \pm 0.28 \\
	 \Neff^{\rm CMB} &= 2.99 \pm 0.17 \label{eq:NeffCMB}
\end{align}
It is important to note that the central values depend on the particular dataset and assumed parameters, while the error bars are also quite model-dependent. For instance, accounting for a possible mass scale in the neutrino/dark sector, or strong self-interactions can lead to error bars larger by an $O(1)$ factor. Accounting for possible model independence, this roughly indicates that currently  $|\Delta \Neff^{\rm BBN, CMB}| \lesssim 0.5-0.7$  at the 2$\sigma$ level.

It is exciting to note that measurements of $\Neff$ will improve significantly with future CMB observations. The target CMB-S4 sensitivity is $\sigma(\Delta \Neff) \approx 0.03$~\cite{Abazajian:2016yjj}. This is remarkable because it is a probe of any relativistic degree of freedom that was in equilibrium with the SM thermal bath, up to the QCD phase transition (or even up to the EW phase transition, for fermions or vectors).

What does this mean specifically for dark sector models? Let us take as an example a fairly minimal secluded scenario, where all the degrees of freedom are scalars (to have the smallest $g$). Suppose equilibrium was established at some early time  by some higher dimension operator, but the two thermal baths later decoupled at a temperature $T_{dec}$. We assume that the decoupling is well before BBN, and occurs when the DM is still relativistic. The {\emph{minimum}} energy density in the dark sector at this time was 
\begin{align}
	\rho_d^{min}(T_{dec}) = \frac{2\pi^2}{30} T_{dec}^4 
\end{align}
where we have taken 2 as the minimum number of degrees of freedom, for one scalar DM and one scalar mediator state.  At decoupling, we may assume $T = T_\gamma = T_\nu$ but afterward the two thermal baths evolve separately.  

\begin{figure}[t]
\includegraphics[width=0.6\textwidth]{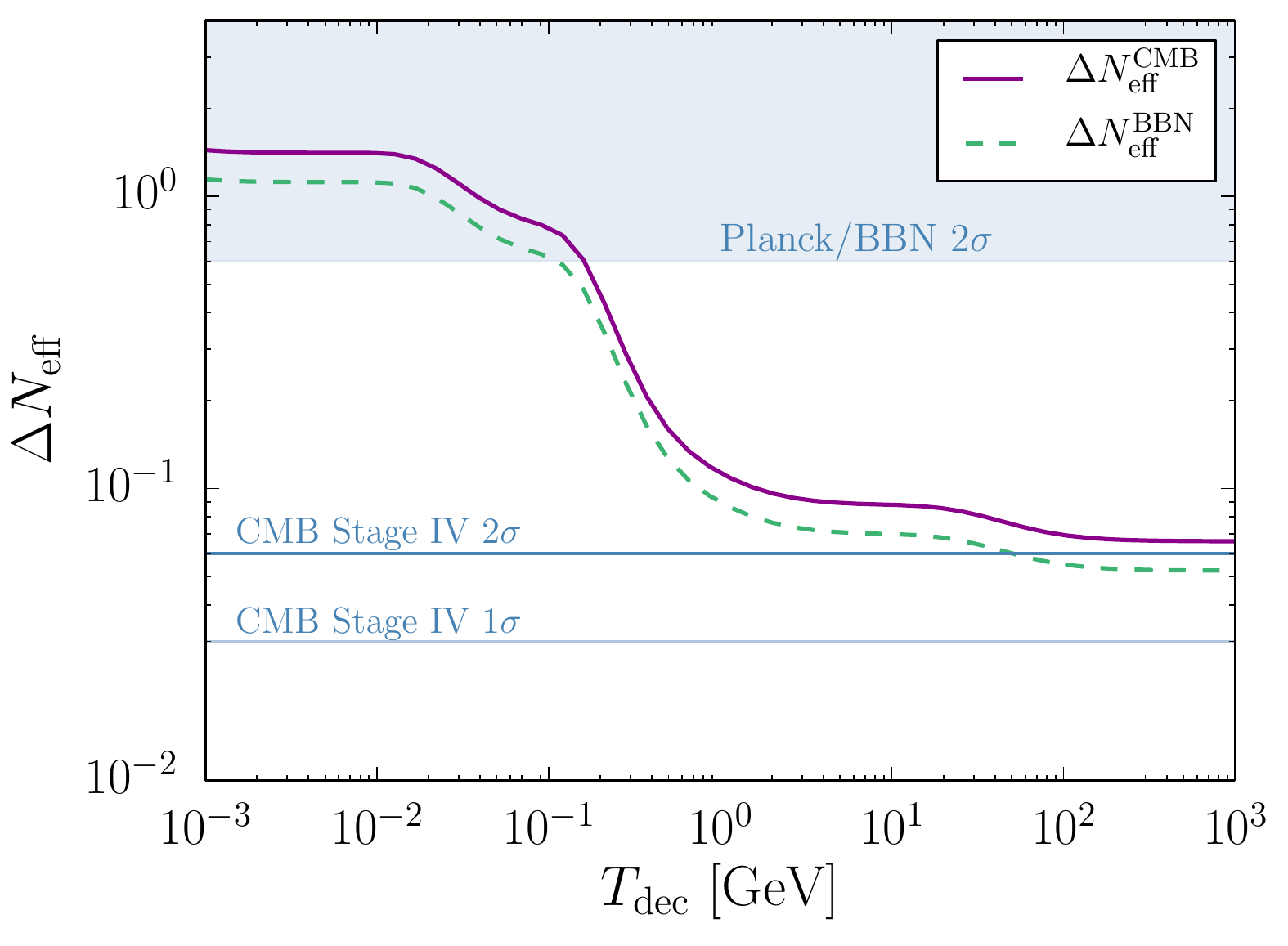}
 \caption{ A secluded sector where the relic abundance is set entirely by annihilations to dark radiation still contributes to $\Delta \Neff$. Here it is assumed there is a minimal dark sector with two degrees of freedom, with DM mass below the MeV scale. The difference in $\Delta \Neff^{\rm CMB}$ and $\Delta \Neff^{\rm BBN}$ comes from the fact that the DM is still in equilibrium during BBN and undergoes dark sector freezeout sometime before recombination. The shaded region corresponding to the 2$\sigma$ bounds from Planck/BBN is approximate, see the discussion around Eq.~\ref{eq:NeffCMB}.
 \label{fig:Neff_Tdec}}
\end{figure}

For the sake of being concrete, suppose that the mass of both the DM and mediator are {\emph{below}} the MeV scale, so that we may assume they are relativistic at $T=\MeV$. As the universe cools, massive particles are annihilating away in the SM thermal bath, so that there is a relative increase in the photon+neutrino temperature by $\left[g_{*}^{\rm SM}(T_{dec})/g_{*}^{\rm SM}(\MeV) \right]^{1/3}$ from $T_{dec}$ until neutrino decoupling. Here $g_{*}^{\rm SM}(T)$ refers to the effective number of relativistic SM degrees of freedom, shown in Fig.~\ref{fig:gstar}. To find $\Delta \Neff$, we track the dark sector temperature relative to the neutrino temperature:
\begin{align}
	\rho_d^{min}(\MeV) = \frac{2 \pi^2}{30} T_\nu^4 \left[ \frac{g_{*}^{\rm SM}(\MeV)}{g_{*}^{\rm SM}(T_{dec})} \right]^{4/3}  .
\end{align}
Comparing with Eq.~\ref{eq:rhotot_Neff}, we obtain a value of $\Delta  \Neff^{\rm BBN} = 0.053$ if $T_{dec} > 200$ GeV. This is a minimum value of $\Delta \Neff^{\rm BBN}$, assuming there are no other degrees of freedom that are in equilibrium with the SM. Fig.~\ref{fig:Neff_Tdec} shows $\Delta \Neff^{\rm BBN}$ as a function of $T_{dec}$.

Below $T \sim \MeV$, the neutrino temperature redshifts in the standard cosmology. However, life still goes on in the dark sector. At some point, the DM freezes out by annihilation to the mediator states, with the number of degrees of freedom changing from 2 to 1. Entropy is conserved in this process, leading to a relative increase in the dark sector temperature by  $2^{1/3}$. Therefore the energy density at the time of recombination is
\begin{align}
	\rho_d^{min}(T_{\rm rec}) = \frac{\pi^2}{30} T_\nu^4 2^{4/3} \left[ \frac{g_{*}^{\rm SM}(\MeV)}{g_{*}^{\rm SM}(T_{dec})} \right]^{4/3} 
\end{align}
giving $\Delta  \Neff^{\rm CMB} = 0.067$ if $T_{dec} > 200$ GeV! Again, this Fig.~\ref{fig:Neff_Tdec} shows $\Delta \Neff$ as a function of $T_{dec}$, comparing the values at CMB and BBN. Even this minimal scenario could lead to an observable $\Delta  \Neff^{\rm CMB}$ with CMB-S4!

The lesson to be drawn from this is that future cosmological datasets will provide a strong test of the secluded dark sector through gravitational effects. However, this is no guarantee: a null result could be consistent with a high $T_{dec}$, the presence of additional degrees of freedom in equilibrium with the SM thermal bath (e.g. in a supersymmetric model), or a modified cosmology. For example, Ref.~\cite{Adshead:2016xxj} explores the limitations on reheating scenarios that can populate a secluded sector consistent with bounds on $\Neff$. A positive detection of $\Delta \Neff$, conversely, is not necessarily a signature of DM. But it would be strongly suggestive of a light relic that decoupled from the SM thermal bath before BBN, thus giving a probe into a new regime of the early universe.

We leave it as an exercise to determine $\Delta  \Neff^{\rm CMB}$ for a secluded sector consisting of Dirac fermion DM and vector mediator. Additional dark sector scenarios for $\Delta \Neff$ can be found
Refs.~\cite{Ackerman:mha,Baumann:2016wac,Green:2017ybv,Knapen:2017xzo,Berlin:2017ftj,Cui:2018imi}. There are also phenomenological possibilities beyond just $\Delta  \Neff^{\rm CMB}$ and $\Delta  \Neff^{\rm BBN}$. For a study of how time-dependent $\Delta \Neff$ during BBN affects light element abundances, see Ref.~\cite{Berlin:2019pbq}. In addition, an interacting relativistic dark sector has a different imprint on the CMB compared to free-streaming dark sector radiation~\cite{Brust:2017nmv}.

\subsubsection{DM self-interactions \label{sec:SIDM}}

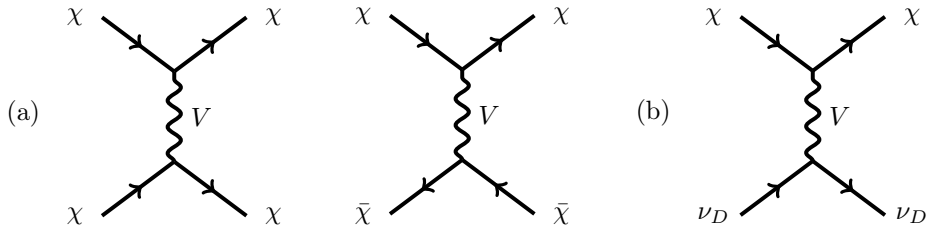
\begin{figure}[t]
\centering
\begin{tabular}{m{0.5\linewidth}m{0.3\linewidth}}
(a)
\raisebox{-1.5cm}{
    \begin{tikzpicture}[line width=1.5 pt, scale=1.2]
    	\draw[fermion] (-0.8,-0.6)--(0,0);
    	\draw[vector] (0,0)--(0,1);
    	\draw[fermion] (-0.8,1.6)--(0,1);
    	\draw[fermionbar] (0.8,-0.6)--(0,0);
    	\draw[fermionbar] (0.8,1.6)--(0,1.0);
    	\node at (-1.1,-0.6) {$\chi$};
    	\node at (-1.1,1.6) {$\chi$};
    	\node at (.3,.5) {$V$};
    	\node at (1.1,-0.6) {$\chi$};
    	\node at (1.1,1.6) {$\chi$};
    \end{tikzpicture}
    \hspace{0.5cm}
    \begin{tikzpicture}[line width=1.5 pt, scale=1.2]
    	\draw[fermionbar] (-0.8,-0.6)--(0,0);
    	\draw[vector] (0,0)--(0,1);
    	\draw[fermion] (-0.8,1.6)--(0,1);
    	\draw[fermion] (0.8,-0.6)--(0,0);
    	\draw[fermionbar] (0.8,1.6)--(0,1.0);
    	\node at (-1.1,-0.6) {$ \bar \chi$};
    	\node at (-1.1,1.6) {$ \chi$};
    	\node at (.3,.5) {$V$};
    	\node at (1.1,-0.6) {$ \bar \chi$};
    	\node at (1.1,1.6) {$ \chi$};
    \end{tikzpicture}
}
\hspace{1.0cm}
&
(b)
\raisebox{-1.5cm}{
    \begin{tikzpicture}[line width=1.5 pt, scale=1.2]
    	\draw[fermion] (-0.8,-0.6)--(0,0);
    	\draw[vector] (0,0)--(0,1);
    	\draw[fermion] (-0.8,1.6)--(0,1);
    	\draw[fermionbar] (0.8,-0.6)--(0,0);
    	\draw[fermionbar] (0.8,1.6)--(0,1.0);
    	\node at (-1.1,-0.6) {$ \nu_D$};
    	\node at (-1.1,1.6) {$\chi$};
    	\node at (.3,.5) {$V$};
    	\node at (1.1,-0.6) {$ \nu_D$};
    	\node at (1.1,1.6) {$\chi$};
    \end{tikzpicture}
}
\end{tabular}
\caption{ Example diagrams contributing to (a) DM self-scattering, important for the structure of DM halos and (b) DM scattering on dark radiation, which can imprint dark acoustic oscillations on the matter power spectrum and CMB. \label{fig:DMselfint} }  
\end{figure}

Interactions within the dark sector can also be imprinted in the clustering of DM. Self-scattering of DM, shown in Fig.~\ref{fig:DMselfint}a,  can lead to thermalization of DM in halos, suppression of small scale structure, and a reduction of central densities within galaxies and galaxy clusters. Indeed, self-interacting DM (SIDM) was originally proposed~\cite{Spergel:1999mh} in order to explain the ``cusp vs. core'' problem, where the central profiles of dwarf galaxies were observed to be much shallower than predicted in DM-only simulations, and the ``missing satellites'' problem, where the observed number of satellites of the Milky Way appeared to be far too low. Recent work has also pointed out the related ``too big to fail''~\cite{2011MNRAS.415L..40B} and ``diversity of rotation curves''~\cite{Oman:2015xda,Kamada:2016euw} problems. Improvements in simulations that include baryons suggest that baryonic physics may be sufficient to explain the halo profiles of dwarfs, and more satellites are now being discovered. Still, DM self-interactions remain an intriguing way to explain other properties of galaxies, and the predictions for SIDM in the presence of baryons are still being understood; for a recent review, see Ref.~\cite{Tulin:2017ara}.

A figure of merit for when self-interactions are important is when the transfer cross section  $\sigma$ (cross section weighted by momentum transfer, see definition in Eq.~\ref{eq:sigma_transf}) over DM mass has a typical value of
\begin{align}
	\sigma/m_\chi \sim 1 \frac{{\rm cm}^2}{ \rm g }= 2\times 10^{-24} {\rm cm}^2 \, \left( \frac{\rm GeV}{m_\chi} \right) \, .
\end{align}
 Assuming a typical DM density $\rho_{\rm DM} \approx$ GeV/cm$^3$ and DM velocity $v \sim 10^{-3}$, this gives a scattering rate per DM particle
\begin{align}
	\Gamma \approx \frac{\rho_{\rm DM}}{m_\chi} \sigma v \approx \frac{1}{10^9 \, {\rm year} }.
\end{align}
The dynamical time scale for a typical MW-like galaxy or its massive subhalos is around $10^8-10^9$~years, so this is around the cross section needed to affect the halo dynamics. In the inner regions of halos, the interaction rate can be significantly higher. 

DM scattering via light mediators can easily give rise to such large cross sections. As a result, self-interactions are an important constraint on low mass dark sectors and especially the secluded scenario where, by definition, the mediator is lighter than the DM, $m_V < m_\chi$.
Specifically, consider the secluded scenario with an ultralight mediator with mass $m_V \ll \eV$, shown in the left panel of Fig.~\ref{fig:secluded}. In this limit, self-interactions are especially restrictive. Since  $m_V \ll m_\chi v$ for $m_\chi \gtrsim \keV$, there is a forward scattering enhancement and the Born limit for the transfer cross section is given by 
\begin{align}
	\sigma \approx  \frac{8 \pi \alpha_\chi^2}{m_\chi^2 v^4} \ln \frac{ (m_\chi v)^2}{m_V^2} .
\end{align}
$\alpha_\chi = g_\chi^2/(4\pi)$ is the coupling of the DM with the mediator and  the specific numerical prefactor depends on the spin of the DM and mediator. Taking $\sigma < 1 {\rm cm}^2/{\rm g}$, this translates to an SIDM bound of
\begin{align}
	\alpha_\chi \lesssim {\rm few} \times 10^{-10} \left( \frac{ v}{10^{-3}} \right)^2 \left( \frac{m_\chi}{\MeV} \right)^{3/2} \left( \frac{10}{\ln ( m_\chi^2 v^2/m_V^2)} \right)^{1/2}.
	\label{eq:SIDM_ultralightmed}
\end{align}
Now, if we compare with the couplings needed to set the relic abundance, the SIDM constraint {\emph{excludes}} thermal freezeout if $m_\chi$ is below the GeV scale.
\exercise{Using the annihilation cross section given in Eq.~\ref{eq:secluded}, verify the above statement that thermal freezeout via annihilation into an ultralight mediator (left panel of Fig.~\ref{fig:secluded}) is strongly constrained by self-interactions in galaxies. }
Of course, there are ways around these conclusions. If there is a large splitting for the Dirac fermion, then self-interactions are kinematically suppressed. For a discussion of self-scattering for inelastic DM, see for example Refs.~\cite{Schutz:2014nka,Zhang:2016dck,Vogelsberger:2018bok}.   Alternatively, additional hidden sector states could be introduced for thermal freeze-out  of sub-GeV secluded sectors, for example with processes such as that shown in the right panel of Fig.~\ref{fig:secluded}. Then it is not necessary to take the ultralight $m_V \ll \eV$ limit. If the mediator mass $m_V \gg m_\chi v$, then the self-interaction cross section can be estimated as $\sigma \approx \pi \alpha_\chi^2 m_\chi^2/m_V^4$. Then SIDM bounds are still  an important consideration for dark sector models, but much less restrictive. in particular, it does not exclude the possibility of thermal freezeout.

\begin{figure}[t]
\includegraphics[width=0.6\textwidth]{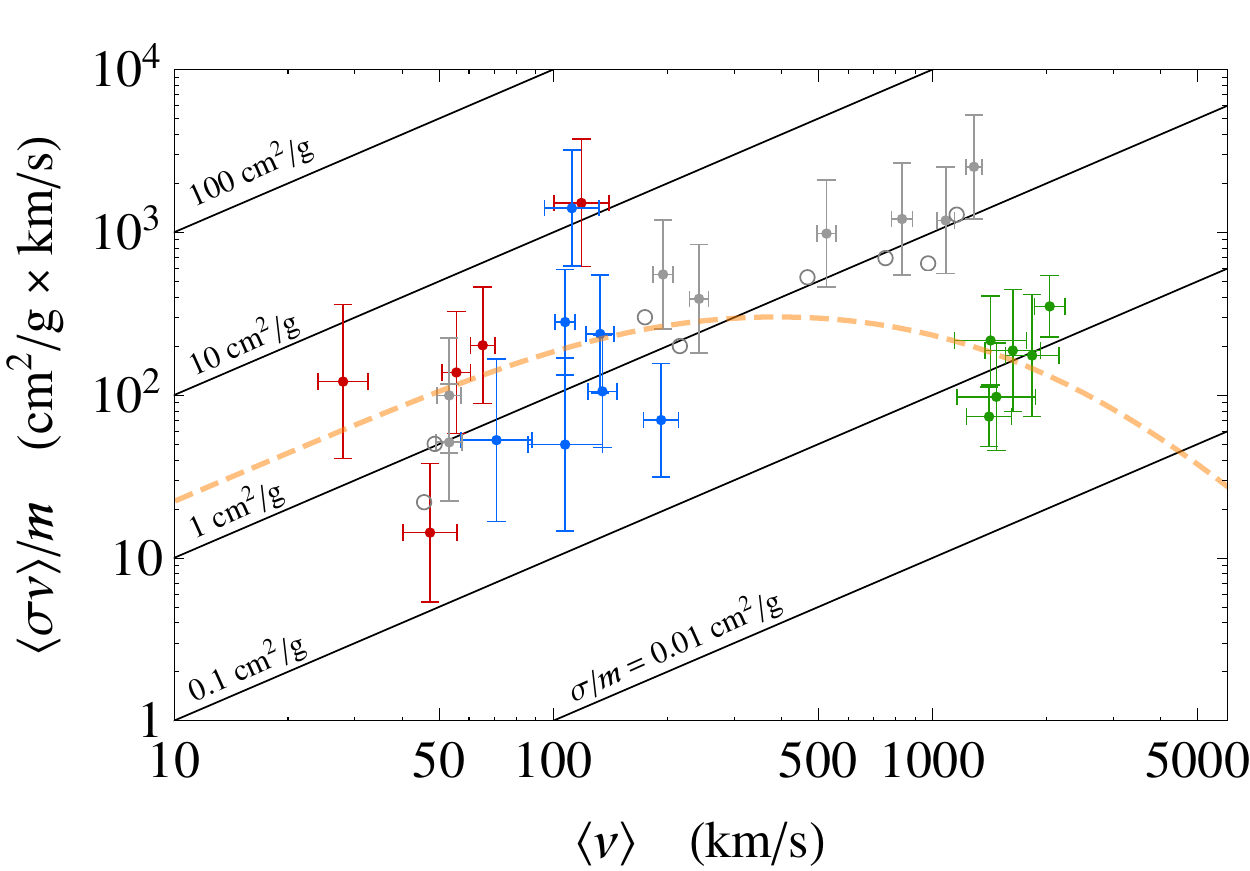}
\caption{Reproduced from Ref.~\cite{Kaplinghat:2015aga}, the DM self-interaction cross section obtained from fits to stellar kinematics data from dwarf galaxies (red), low surface brightness galaxies (blue), and galaxy clusters (green). The fits are done with an analytic model for halo profiles with SIDM, and the results are shown for velocity-weighted cross section as a function of mean collision velocity. The gray data points are from fitting to simulated SIDM halos with $\sigma = 1$~cm$^2$/g (with no baryons in the simulation). The dashed orange line comes from a model of dark photon mediated scattering with $m_\chi = 15\, \GeV$ and dark photon mass of $17 \, \MeV$. \label{fig:sidm} }
\end{figure}

In the above discussion, we have considered the scattering cross section only in a few limits where the Born approximation is valid. In the general parameter space of $\alpha_\chi, m_V, $and $m_\chi$, there are additional resonant enhancement effects and one must solve the general quantum mechanical scattering problem. A number of works have considered the parameter space of $\alpha_\chi, m_V, m_\chi$, subject to the requirement of explaining the DM relic density~\cite{Tulin:2013teo,Kaplinghat:2013yxa}.  Interestingly, a combined fit to the halo properties across a range of velocity dispersions (from dwarf galaxies to galaxy clusters) has found preferred parameters of DM with mass $\sim 10$ GeV and scattering through mediator with mass $\sim 10$ MeV~\cite{Kaplinghat:2015aga}. This is shown in Fig.~\ref{fig:sidm}. (See Ref.~\cite{Huo:2017vef} for the best fits in the asymmetric DM case.) The results were obtained in analytic models of SIDM halos, and there is much to be done in interpreting observations of galaxies. At the order of magnitude level, Fig.~\ref{fig:sidm} also gives a useful summary of the typical cross sections allowed in various astrophysical systems, since we expect the cross section cannot be significantly larger than their best fit.

Finally, let us briefly discuss DM sector interactions in the early universe. If there are dark sector states that were relativistic during the early universe, then DM scattering can lead to dark acoustic oscillations. An example is shown in Fig.~\ref{fig:DMselfint}b, where the $\nu_D$ is a sub-eV relativistic particle. The presence of sizeable self-interactions may then be tied to CMB or large scale structure signatures. For example, Ref.~\cite{Huo:2017vef} has explored the predictions for dark acoustic oscillations within the SIDM parameter space, focusing on a minimal secluded dark sector with dark radiation. Another example is atomic dark matter~\cite{CyrRacine:2012fz}, where the dark sector consists of a dark proton, a dark electron, and a dark photon; an example matter power spectrum was shown Fig.~\ref{fig:powerspectrum}.

We conclude this subsection by noting that we have made a fairly conservative assumption of no coupling with the SM, in order to illustrate a limiting case. In many models of secluded sectors small couplings with the SM may still be present, for example if the mediator $V$ decays back to SM particles. This can change the $\Neff$ prediction, however the SIDM considerations still apply. Indeed, SIDM bounds are also important in the direct coupling scenarios discussed below, as well as in many other dark sector models that don't fall in the simplistic classification given here.

\subsection{Cosmology of direct coupling scenarios} 

At the other end of dark sector scenarios discussed here, the DM is in thermal equilibrium with the SM plasma throughout freezeout. The DM annihilates into SM particles (or annihilates into particles which decay to SM particles) and the energy density of the dark sector is thus deposited back into the SM thermal bath. This would appear to get around the $\Neff$ bounds from BBN and the CMB discussed above, if the DM is sufficiently heavy. However, there are additional effects on BBN abundances and on the CMB which lead to strong constraints.

{\emph{BBN}} ---- If the thermal freezeout occurs close to the time of BBN (i.e. DM mass is close to the MeV scale), this can modify the predicted abundance via both the baryon-to-photon ratio and $\Neff$.  For instance, suppose DM is in equilibrium with the $e^\pm$ and $\gamma$ thermal bath after neutrino decoupling. When the electrons and positrons annihilate away, some of the energy density goes towards heating the DM. This would temporarily {\emph{reduce}} the energy density in the photons, and increase the baryon-to-photon ratio $\eta$.  An increase in $\eta$ allows more efficient conversion of deuterium into helium, since the fusion rate competes with dissociation from the high energy photons. As a result, the effect of the DM is reduced D and increased $^4$He. (Meanwhile, an additional source of radiation in $\Neff$ primarily affects BBN by changing the expansion rate and thus increasing the relic $n/p$ ratio; this leads to more D and $^4$He.)
Refs.~\cite{Boehm:2013jpa,Nollett:2013pwa,Nollett:2014lwa} used measurements of abundances and the baryon density to set lower bounds on the DM mass $m_\chi \gtrsim 1-10$ MeV, with the specific number depending on the number of degrees of freedom of the candidate. (It is also interesting to note that these bounds apply generally to any MeV-scale particle that was in equilibrium with the SM around the time of BBN, and does not make any assumption about the particle's relic abundance.)

\begin{figure}[t]
\includegraphics[width=0.95\textwidth]{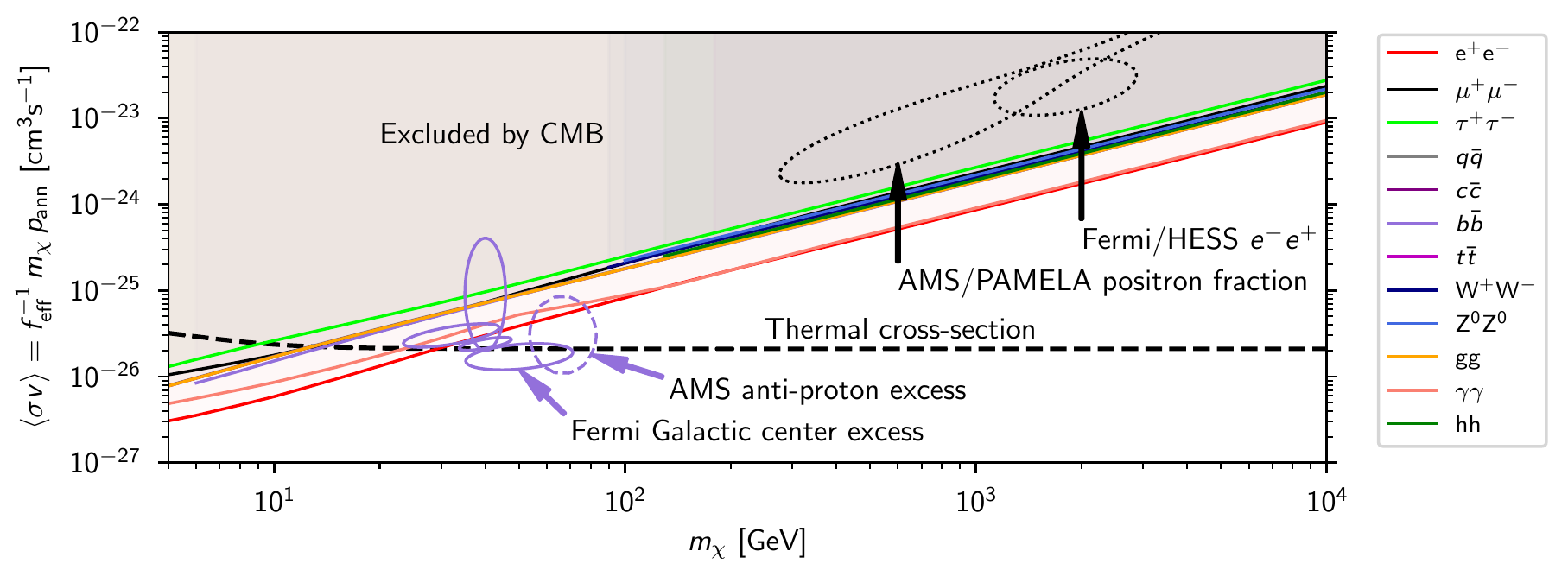}
\caption{Reproduced from Ref.~\cite{Aghanim:2018eyx}, the  {\emph{Planck 2018}} constraints on DM annihilation cross section as a function of DM mass. The different shaded regions/lines correspond to constraints on different final states in the DM annihilation. Also shown are preferred regions obtained from fits of cosmic ray and gamma ray data; for discussion of these excesses, see Refs.~\cite{Slatyer:2017sev,Hooper:2018kfv}. \label{fig:planck_dmann} }
\end{figure}

{\emph{CMB}} ---- After freezeout, DM annihilation continues all the way through the epoch of recombination, with a rate given by $\rho_{\rm DM}^2 \langle \sigma v\rangle /(2 m_\chi^2)$ for Majorana particles. This is a meager rate that barely changes the DM density, but it is a source of energy injection which does affect the CMB. Accounting for the energy released in the annihilation $(2 m_\chi)$, the rate of energy deposited per volume and time  is parametrized as
\begin{align}
	\frac{dE}{dV \, dt}(z)  = \rho_c^2 \Omega_{\rm cdm}^2 (1+z)^6   \frac{\langle \sigma v\rangle}{m_\chi}  f(z)
\end{align} 
where $f(z)$ is typically an $O(1)$ number characterizing the efficiency of energy deposition at redshift $z$~\cite{Slatyer:2009yq,Slatyer:2015jla}. For instance, annihilation into photons in a ``transparency window'' (energies and redshifts where the photon cooling time is relatively slower compared to the expansion rate) or annihilation into neutrinos would reduce the efficiency. The physical effect of the late energy injection is primarily to modify the ionization history: there is an increase in residual ionized hydrogen. CMB photons scatter off the free electrons, leading to damping at small scales in the CMB power spectra ($\ell \gtrsim 200$) and an increase in the power for CMB polarization (EE) at large scales ($\ell \lesssim 200$)~\cite{Padmanabhan:2005es}.

CMB bounds on DM annihilation are quoted in terms of an effective efficiency $f_{\rm eff}$, which is the leading coefficient when $f(z)$ is decomposed into a set of basis functions characterized by their effect on the CMB~\cite{Slatyer:2015jla}; this approach has been shown to capture most of the effects of DM annihilation~\cite{Finkbeiner:2011dx}.  The most recent result from {\emph{Planck 2018} is~\cite{Aghanim:2018eyx} 
\begin{align}
	f_{\rm eff}   \frac{\langle \sigma v\rangle}{m_\chi} <  \frac{2.2 \times 10^{-26} \, {\rm cm}^3/{\rm s}}{69 \, {\rm GeV} } \, 
\end{align}
for $s$-wave annihilation. This can be compared with the thermal relic cross section given in Eq.~\ref{eq:sigv_number}. For DM annihilating into SM fermions with the thermal relic cross section, this excludes masses below $\sim 10-30$ GeV, depending on the final state! Fig.~\ref{fig:planck_dmann} shows the bounds for a range final states in the annihilation. This is thus an important consideration models of light DM. However, these bounds have limited applicability to some classes of models, including asymmetric DM, models with an inelastic splitting so that annihilation is kinematically forbidden at recombination, and those where the DM annihilation is primarily $p$-wave (scales as $v^2$). The latter two examples utilize the fact that DM is generally extremely cold by the time of recombination.

An one example, consider complex {\emph{scalar}} DM $\phi$ which annihilates through a vector $V$ into SM fermions.  The coupling of the  scalar with the vector is given by $g_{\phi} V^\mu \left( i \phi^* \partial_\mu \phi  + {\rm h.c.} \right)$, and we assume an interaction with SM fermions given by $g_f V^\mu \bar f \gamma_\mu f$.
The amplitude is proportional to the difference in 4-momenta of the initial states, and in limit of non-relativistic DM, this gives a factor of $m_\phi v_{\rm rel}$ where $v_{\rm rel}$ is the relative velocity of the initial states. The annihilation cross section for $v_{\rm rel} \ll c$ and $m_f \ll m_\phi$ is given by
\begin{align}
	\sigv = \sigma v_{\rm rel} = \frac{1}{6\pi} \frac{ g_\phi^2 g_f^2 m_\phi^2 v_{\rm rel}^2}{(4 m_\phi^2 - m_V^2)^2 + m_V^2 \Gamma_V^2}
\end{align}
where $\Gamma_V$ is the width of the $V$. For further discussion of this benchmark as well the one with an inelastic splitting, see for example Ref.~\cite{Izaguirre:2015yja}.

In addition to the CMB, there are numerous other ways to search for DM annihilation, which are beyond the scope of this discussion. BBN bounds on energy injection from DM annihilation have also been considered~\cite{Depta:2019lbe}, and are generally weaker for $s$-wave annihilation. Annihilation in the Milky Way is a source of gamma rays and cosmic rays; efforts to detect such signals are described within this lecture series~\cite{Hooper:2018kfv}. 

From the perspective of DM models, the BBN and CMB bounds have the strongest implications. Summarizing Lecture 3 so far, we find:
\begin{itemize}
	\item DM candidates below the GeV scale can fall in a few classes, including: secluded, such that there is little or no annihilation to SM particles; asymmetric; and models with $p$-wave or kinematic suppression.
	\item DM candidates below the MeV scale must be thermally decoupled from the SM thermal bath at the time of BBN and freezeout. (An exception is DM which comes into equilibrium with neutrinos after neutrino decoupling~\cite{Berlin:2017ftj}.) Sub-MeV models are therefore often secluded or not thermal (see freeze-in discussion in Section~\ref{sec:freezein}), and $\Neff$ bounds are important. 
\end{itemize}

\subsection{Portals with the SM}

Having outlined the general features for the cosmology of light DM, we next turn to specific mediators to the SM. In the direct coupling scenario, a mediator interaction with SM states is essential for setting the relic abundance and there is a minimum coupling size needed, which can be translated into a minimal predicted signal in experiments. In the secluded scenario, a mediator-SM interaction can assist in depleting excess energy/entropy and provides a way to search for DM in laboratory experiments, but the signal strength may vary much more.

Commonly studied mediator portals with the SM include:
\begin{itemize}
	\item {\emph{Kinetic mixing portal}} $-$ a vector $V$ has an interaction $\frac{\kappa}{2 \cos \theta_w} V_{\mu \nu} B^{\mu \nu}$ where $B^{\mu\nu}$ is the hypercharge field strength. Since this is a renormalizable operator consistent with SM gauge invariance, we can regard it as an operator that is generically present in the effective field theory. The kinetic mixing is also generated from loops of heavy fermions. After electroweak symmetry breaking, the operator results in a kinetic mixing with the photon $\frac{\kappa}{2} V_{\mu \nu} F^{\mu \nu}$ and $V$ is typically called a dark photon. (Note the kinetic mixing of a vector with the SM photon is variously called $\varepsilon$, $\kappa$, or $\chi$ in the literature. We will stick with the notation $\kappa$ here.)
	\item  {\emph{Higgs portal}} $-$ suppose a scalar mediator $\phi$ has interactions $\phi^2 H^\dagger H$ and $\phi H^\dagger H$  with the Higgs doublet $H$. These interactions can generate a mixing of $\phi$ with the Higgs boson $h$ after electroweak symmetry breaking, which in turn leads to couplings of the scalar with SM fermions $\propto y_f \phi \bar f f$.   There are also extensions of this with a two-Higgs doublet model (2HDM).
	\item  {\emph{Axion portal}} $-$ a pseudoscalar (or axion-like particle) $a$ can have couplings to SM fermions, $\frac{\partial_\mu a}{f_a} \bar f \gamma^\mu \gamma^5 f$, or to gauge bosons $\frac{a}{f_a} G \tilde G$, $\frac{a}{f_a} F \tilde F$, etc.
	\item {\emph{Neutrino portal}} $-$ this refers to the interaction $\bar L H \nu_s$ where $L$ is a SM lepton doublet and $\nu_s$ is a sterile neutrino. While sterile neutrinos are considered as DM candidates themselves, they could also serve as a mediator to a dark sector (consisting minimally of an additional dark fermion and scalar that can couple to $\nu_s$).
\end{itemize}
These portals are also the lowest-dimension operators that connect new particles (that are neutral with respect to the SM) to the SM. We next give an overview of some of the constraints and searches for light mediators. 

\subsubsection{Vector portal \label{sec:vector}}

Perhaps the most often studied mediator in recent times is kinetically-mixed dark photon, owing to the appeal of a simple $U(1)$ extension with a rich phenomenology and the absence of any flavor problems. Focusing on the interactions with just the photon (most relevant for the low-energy phenomenology), the vacuum interactions for this portal are:
\begin{align}
	{\cal L} \supset -\frac{1}{4} F_{\mu \nu} F^{\mu \nu} -\frac{1}{4} V_{\mu \nu} V^{\mu \nu} +\frac{\kappa}{2} F_{\mu \nu} V^{\mu \nu}  + \frac{1}{2} m_V^2 V_\mu V^\mu + e A_\mu J^\mu_{\rm EM} + g_\chi V_\mu J^\mu_{D}
	\label{eq:vacuumL_darkphoton}
\end{align}
where $J^\mu_{\rm EM}$ is the electromagnetic current and $J^\mu_{D}$ is a dark current with gauge coupling $g_\chi$. The kinetic mixing parameter $\kappa$ can be positive or negative, though constraints are typically shown on the absolute value $|\kappa|$. In addition, there are the kinetic or mass terms for any dark charged particles, which aren't written explicitly in order to be general. The vector mass could arise from a dark Higgs mechanism, where the dark Higgs boson is extremely massive and has been integrated out of the theory (i.e, a Stueckelberg mass term). However, in some dark sector models the dark Higgs is also a light degree of freedom and important to the phenomenology.

First consider the case that $m_V = 0$. Then we can make a field redefinition $\tilde V_\mu = V_\mu - \kappa A_\mu$. This eliminates the kinetic mixing term, and we are left with the following interactions:
\begin{align}
	{\cal L} \supset -\frac{1}{4} (1 -\kappa^2) F_{\mu \nu} F^{\mu \nu} -\frac{1}{4} \tilde V_{\mu \nu} \tilde V^{\mu \nu}  + e A_\mu J^\mu_{\rm EM} + g_\chi \left(\tilde V_\mu + \kappa A_\mu \right) J^\mu_{D}
	\label{eq:masslessV_basis}
\end{align}
where the factor of $(1 - \kappa^2)$ in the photon kinetic term can be eliminated by a field (or electric charge) redefinition.
In the absence of a dark current $J^\mu_D$, we would have a completely decoupled vector $\tilde V$ with no observable effects. Hence in the massless vector limit, we expect that the only limits would come from effects that involve the DM. Now suppose there is a DM particle, for example $J^\mu_{D} = \bar \chi \gamma^\mu \chi$. In this basis, it is clear that the DM couples to the photon with an effective charge $\kappa g_\chi$ or millicharge $Q = \kappa g_\chi/e$. This model gives an explicit realization of millicharged DM, discussed earlier in these lectures around Eq.~\ref{eq:millicharge_thermalrelic}.

Now we examine what happens when $m_V \neq 0$, starting with the vacuum Lagrangian. Another often-used basis comes from making the field redefinition $\tilde A_\mu = A_\mu - \kappa V_\mu$, which eliminates the kinetic mixing term:
\begin{align}
	{\cal L} \supset -\frac{1}{4} \tilde F_{\mu \nu} \tilde F^{\mu \nu} -\frac{1}{4} (1 - \kappa^2)V_{\mu \nu} V^{\mu \nu} + \frac{1}{2} m_V^2 V_\mu V^\mu + e (\tilde A_\mu + \kappa V_\mu) J^\mu_{\rm EM} + g_\chi V_\mu J^\mu_{D}
	\label{eq:massiveV_basis}
\end{align}
where we see that dark photon mass eigenstate $V$ couples to SM charged particles.  Of course, the physics is independent of any field redefinitions or change of basis for the $A, V$ fields\footnote{One could also use this basis for the massless $m_V$ limit. Then in the absence of a dark current, there is no phenomenological difference from regular QED.  One can then check that Coulomb scattering, bremsstrahlung etc, are all the same as in QED up to an overall redefinition of electric charge squared as $e^2 (1 + \kappa^2)$ in the limit $\kappa \ll 1$. This redefinition of electric charge is identical to the charge renormalization in the basis of Eq.~\ref{eq:masslessV_basis}. }.
This basis is most often used in collider studies of dark photon phenomenology, where the vacuum assumption is valid. However, one must be more careful when considering dark photons in a dense medium, for example in the early universe, in stars, or in a solid state material.

The in-medium Lagrangian for the vector portal must account for the photon dispersion and mass, which leads to modified couplings. In Appendix~\ref{sec:inmedium}, we derive the effective Lagrangian in detail, accounting for the in-medium photon properties. To illustrate why this is relevant for the constraints on a dark photon, consider emission of dark photons in stars. This emission would the stellar luminosity and lifetime, and forms the basis for an important class of constraints on dark photons. The photon dispersion in typical stars has an effective mass given by the plasma frequency $\omega_p \simeq 0.1-10$ keV. Accounting for in-medium effects, the total emission rate is proportional to $(\kappa m_V/\omega_p)^2$ in the limit of $m_V \ll \omega_p$~\cite{An:2013yua}; this would not be the case if we just took the vacuum couplings.  (For a derivation of this scaling with $m_V$, see Appendix~\ref{sec:inmedium}.)

The stellar constraints on kinetic mixing $\kappa$ and dark photon mass $m_V$ are shown in Fig.~\ref{fig:darkphoton}, where we see the $1/m_V$ scaling in the bounds labeled `solar lifetime', HB (for horizontal branch stars) and RG (for red giants). The $1/m_V$ dependence can also be seen in the bounds from emission of dark photons in SN1987a~\cite{Chang:2016ntp}. All of these results assume {\emph{no coupling to DM, and no other hidden sector states}}. If there is a large population of DM in the star or if the dark photon decays quickly to DM, this might affect whether the dark photons escape. If there is a sufficiently low mass dark Higgs in the theory, the constraints also dramatically change and the bounds on $\kappa$ are independent of $m_V$ in the low mass limit~\cite{An:2013yua}.

\begin{figure}[t!]
\includegraphics[width=0.9\textwidth]{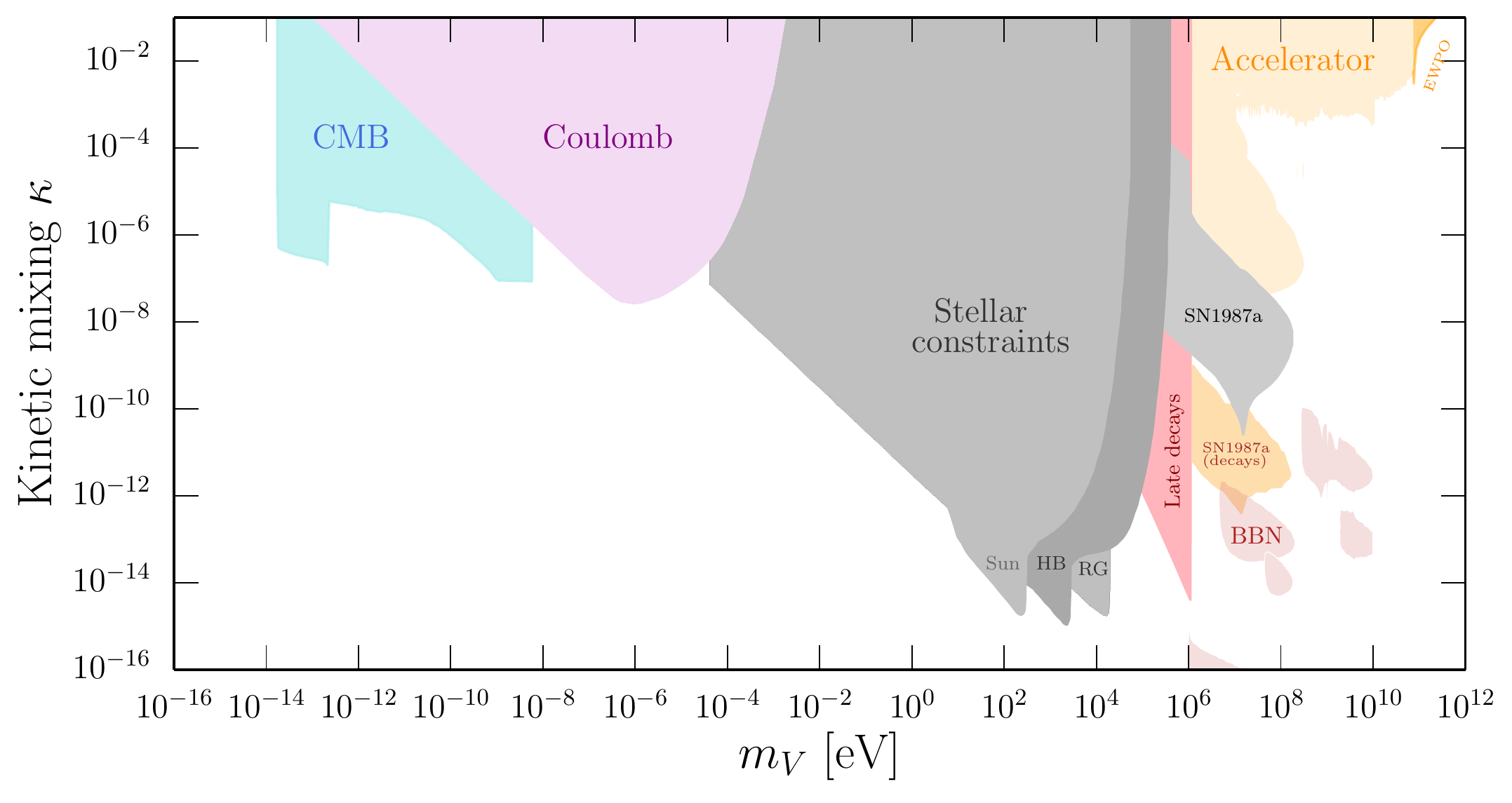}
 \caption{Collected constraints on a kinetically mixed vector with mass $m_V$ and kinetic mixing parameter $\kappa$. There are stellar emission bounds from  the Sun, HB stars, RG stars (adapted from Ref.~\cite{Redondo:2013lna}, and first derived in Refs.~\cite{An:2013yfc,An:2013yua}) and SN1987a~\cite{Chang:2016ntp}. Other bounds at low masses are from resonant conversion of CMB photons into dark photons~\cite{Mirizzi:2009iz}, and from fifth force searches (Coulomb)~\cite{PhysRevLett.61.2285}. 
 For $m_V \lesssim 10^5\, \eV$, the bounds shown assume the dark photon is either long-lived or decays to invisible hidden sector states. For $m_V \gtrsim 10^5\, \eV$, the bounds shown assume that the dark photon does {\emph{not}} decay to hidden sector states. Then there are constraints from late decays of dark photons into $3\gamma$ for $m_V < 2 m_e$~\cite{Redondo:2008ec} or from energy injection in the BBN or CMB for $m_V > 2 m_e$~\cite{Fradette:2014sza}. The region labeled `SN1987a (decays)' comes from dark photons which were produced in SN1987a and decays outside to produce a cosmic or gamma ray signal~\cite{DeRocco:2019njg}.
 For dark photon masses above MeV, we show constraints from electroweak precision observables (EWPO)~\cite{Hook:2010tw} and accelerator bounds on a visibly-decaying dark photon~\cite{Andreas:2012mt,Lees:2014xha,Anastasi:2015qla,Batley:2015lha,Aaij:2017rft}. The accelerator bounds also change if the dark photon can also decay to DM; see for example the plots of Ref.~\cite{Berlin:2018bsc} for collected constraints and projections on both visibly-decaying and invisibly-decaying dark photons.
 \label{fig:darkphoton}
 }
\end{figure}

For $m_V \gtrsim 10^5$ eV, there are a variety of bounds shown for a visibly-decaying dark photon (i.e. assuming the dark photon does not decay to other dark sector states). In particular, there are accelerator bounds for $m_V > 2 m_e$, where the constraints at $\kappa < 10^{-3}$ come from beam dump experiments. There is a vibrant experimental program to search for visibly-decaying and invisibly-decaying dark photons, which aims to cover a significant fraction of open parameter space for MeV $< m_V < $ GeV and $\kappa > 10^{-5}$. We refer the reader to works such as Refs.~\cite{Battaglieri:2017aum,Berlin:2018bsc,Ilten:2018crw} which more clearly describe the state of the art in accelerator bounds and prospects.

For $m_V < 10^{-6}$~eV the most important bounds are from resonant conversion of CMB photons into dark photons~\cite{Mirizzi:2009iz} and from fifth force searches~\cite{PhysRevLett.61.2285}. There are also superradiance bounds on gravitational production of dark photons in the small coupling regime and in the mass ranges $6\times 10^{-20} \, \eV < m_V < 2\times 10^{-17} \, \eV$ and $5\times 10^{-14}\,  \eV < m_V < 2\times 10^{-11}\,  \eV$~\cite{Baryakhtar:2017ngi,Cardoso:2017kgn}; they are not shown on the plot as the small coupling regime still needs to precisely quantified. Aside from the superradiance bounds, there are very few limits in the ultralight regime where $m_V \lesssim 10^{-14} $ eV. This is an important feature specific to dark photon phenomenology. As the vector mass goes to zero, emission of dark photons goes to zero due to the in-medium effects discussed above. Meanwhile, searches for fifth forces are also more challenging: traditional searches for fifth forces rely on the large coherent charge of macroscopic objects, which doesn't apply for the dark photon since most objects are neutral. In the low mass limit, the Compton wavelength of the dark photon is also so large that laboratory searches for modifications to Coulomb's law are ineffective. Thus ultralight vectors with fairly large kinetic mixing $\kappa$ is possible. 

Finally, note that we have shown here the most competitive bounds on a general dark photon. A few other results are not shown to keep the plot simpler; some of the other results can be found in the recent summary plot in Ref.~\cite{Jaeckel:2013ija}, for example. If the dark photon itself forms a large fraction of the dark matter, there are additional constraints.

The situation is more difficult for other vector mediators. A gauged $B-L$ is strongly constrained for $m_V \lesssim$ eV by fifth force searches, since in this case a macroscopic neutral material can have a large $B-L$ charge. (Summary plots of the constraints can be found in Refs.~\cite{Heeck:2014zfa,Knapen:2017xzo}, for example.) Other anomaly-free combinations of lepton number could also be considered, such as $L_\mu - L_\tau$~\cite{Bauer:2018onh,Escudero:2019gzq}. Alternatively, it is possible to gauge an anomalous global symmetry such as $U(1)_B$; while this may still be viable in the high mediator mass regime,  there are strong constraints for sub-GeV masses (for example, see Refs.~\cite{Dobrescu:2014fca,Dror:2017ehi}).
Going beyond pure vector couplings, one could consider a massive vector with axial couplings. However, a massive axial vector is not gauge invariant by itself, and additional ingredients must be supplied in the model; once these are included, there are also severe constraints~\cite{Kahn:2016vjr}. 

As a result, the dark photon remains the most compelling and viable vector mediator, especially for sub-GeV dark sectors where low mass mediators are needed to set the DM relic abundance. When $m_{V} > m_\chi$, then direct annihilation of DM DM $\to e^+ e^-$ is sufficient to set the relic abundances. In this direct coupling scenario, DM with mass $m_\chi \gtrsim $ few MeV is viable given the BBN considerations discussed earlier. If $m_{V} < m_\chi$, DM annihilation to mediators is possible; then the DM mass may be as low as the 1-10 keV scale, which is consistent with warm DM and BBN limits as long as the dark sector is decoupled from the SM plasma and sufficiently cold. And in the limit of ultralight mediator $m_V \ll \eV$, the DM behaves effectively like a millicharged particle and freeze-in of DM through an ultralight vector is an interesting benchmark. In addition to being instrumental for setting the DM relic abundance, the vector portal is a useful ingredient in models where some heat exchange with the SM sector is needed to dump excess energy/entropy.  In the last subsection of this lecture, we will summarize with two specific benchmarks for DM coupled to a dark photon mediator, where the correct relic abundance is obtained.

\subsubsection{Scalar mediator portals}

A scalar mediator with coupling to SM fermions could arise from the Higgs portal mentioned above, or from other extensions of the SM. Let's parameterize the interactions as $y_n \phi \bar n n$ (where we assume equal coupling to neutrons and protons) and $y_e \phi \bar e e$, where $y_n$ and $y_e$ need not be directly related.  The DM also has an interaction $y_\chi \phi \bar \chi \chi$.

Fig.~\ref{fig:scalar} shows the allowed parameter space for the interactions, taken from Ref.~\cite{Knapen:2017xzo}. The stellar emission constraints discussed for the vector portal are also important for the scalar portals, with the difference that the constraints are independent of the mediator mass for masses below $\sim$10 keV. (The precise cutoff of the stellar constraints is quite sensitive to the assumed temperatures and plasma frequencies in the stars.) For masses well below 1 eV, fifth force constraints on the new Yukawa interaction start to become the most important bound for nucleon interactions.

\begin{figure}[t!]
\includegraphics[width=0.49\textwidth]{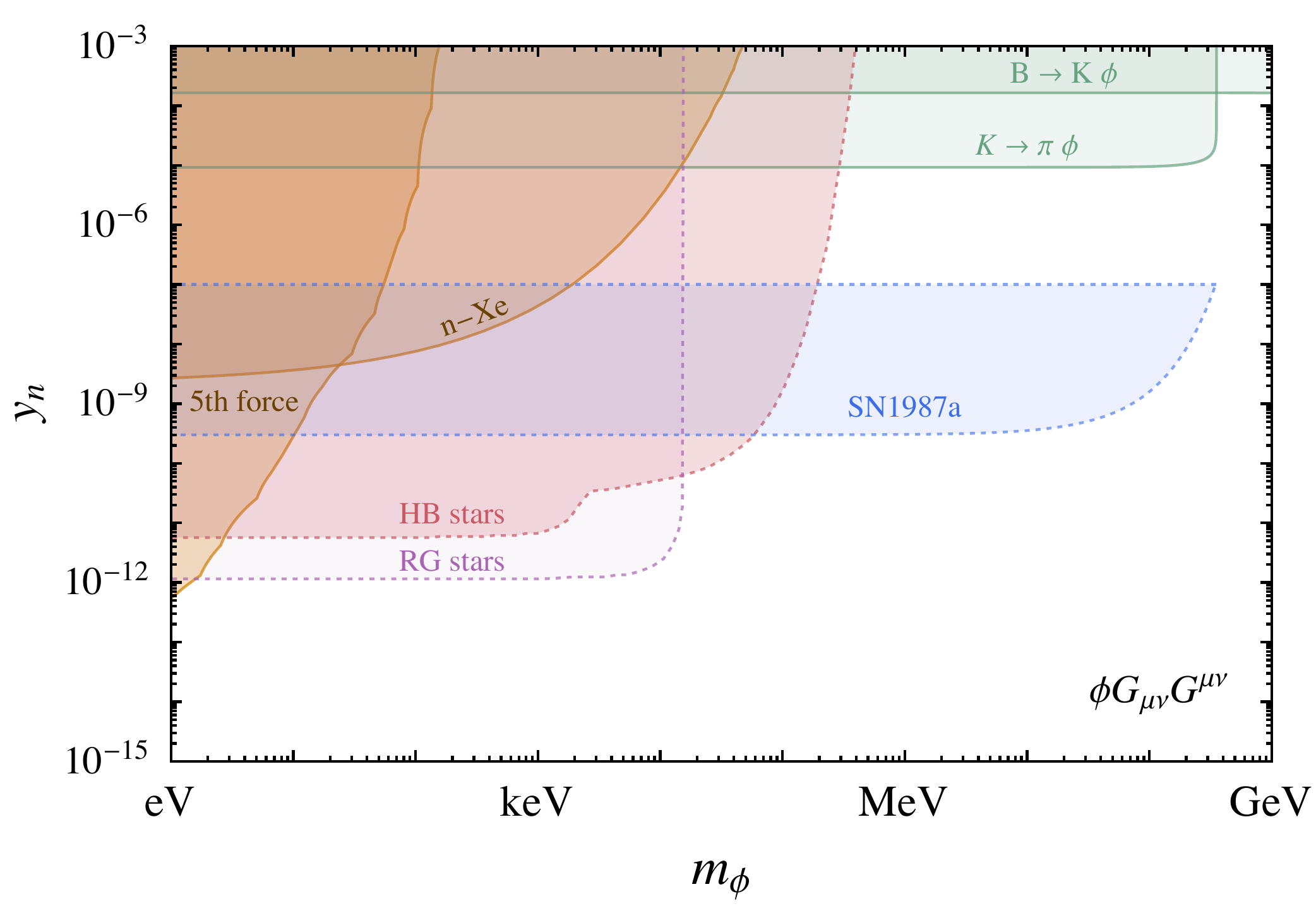}
\includegraphics[width=0.49\textwidth]{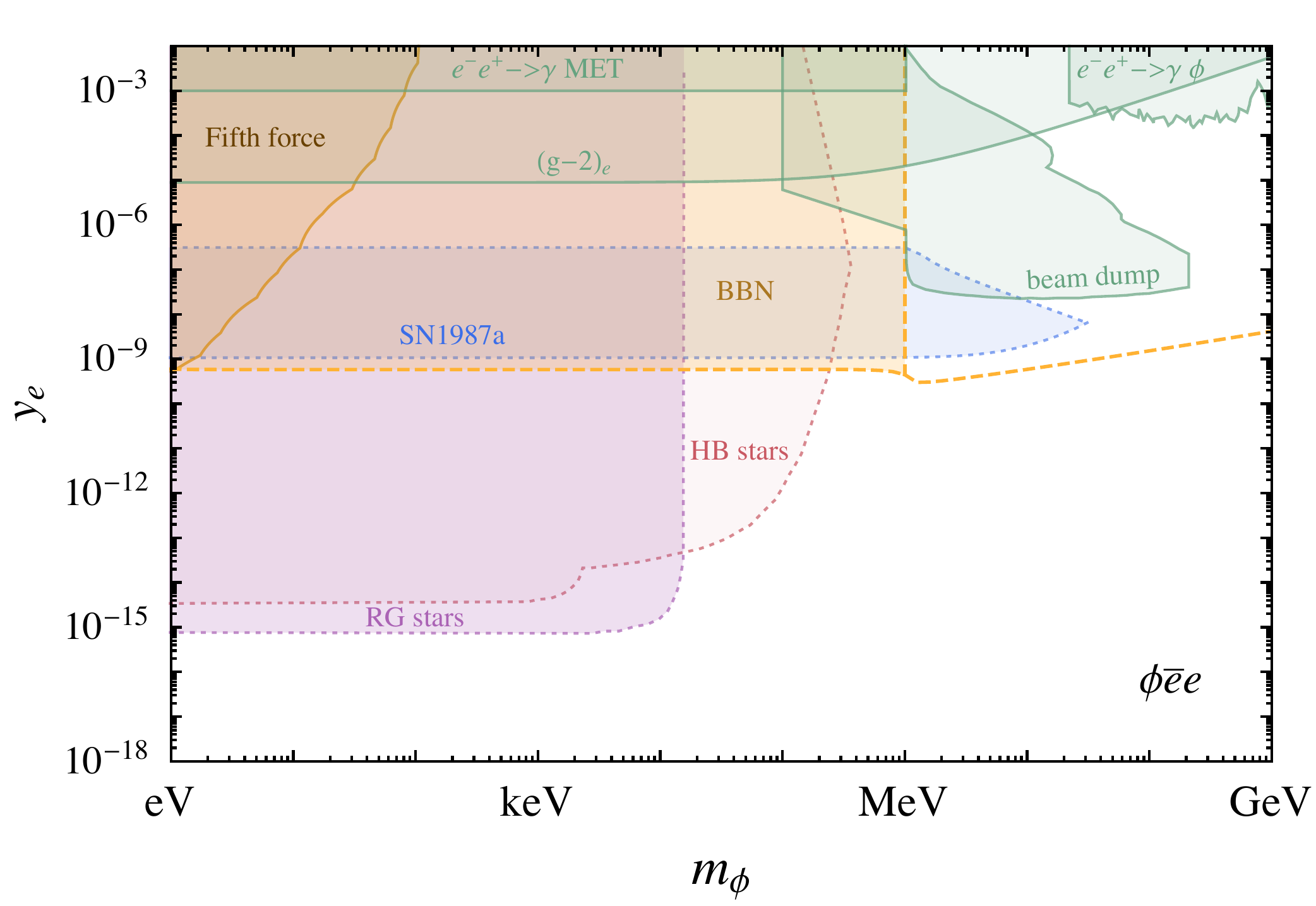}
 \caption{From Ref.~\cite{Knapen:2017xzo}, constraints on sub-GeV scalar mediators assuming $\phi$ does not decay to hidden sector states.  ({\bf left}) The scalar has a coupling only to nucleons, $y_n \phi \bar n n$, which is generated by the operator $\phi G G$. The collider bounds on the scalar coupling to nucleons can differ if $\phi$ couples to SM quarks. We show bounds from stellar cooling, rare meson decays, neutron scattering~\cite{Leeb:1992qf}, and fifth force searches.
  ({\bf right}) The scalar has a coupling only to electrons, $y_e \phi \bar e e$. We show limits from fifth force searches, accelerators~\cite{Essig:2013vha,Lees:2014xha,Liu:2016qwd} and $(g-2)_e$~\cite{Liu:2016qwd,Liu:2016mqv}.  For $y_e \gtrsim 10^{-9}$ (dashed yellow line), $\phi$ can thermalize with the SM plasma before electron decoupling, while the shaded region is excluded by BBN constraints. In both figures, the RG and HB star constraints are from Ref.~\cite{Hardy:2016kme} and the fifth force bounds are adapted from Ref.~\cite{Murata:2014nra}. The SN1987a regions are only approximate bounds and estimated in Ref.~\cite{Knapen:2017xzo}.
\label{fig:scalar}}
\end{figure}

The bounds on scalars for $m_\phi \gtrsim$ MeV are from accelerator searches. For the $\phi \bar e e$ operator, the results shown assume $m_\phi < 2 m_\chi$ so that the mediator decays to visible states. For the $y_n \phi \bar n n$ operator, the bounds from rare meson decays $B \to K \phi$ and $K \to \pi \phi$ assume the mediator only has a coupling to the gluons; these can differ if the scalar couples to any SM quarks. For example, see Refs.~\cite{Knapen:2017xzo,Batell:2018fqo} for a scalar which couples to top quarks only or up quarks only. Ref.~\cite{Fradette:2018hhl} considered freeze-in (and subsequent decay) of the scalar Higgs portal mediator  to set additional bounds, but we don't show them here as they are specific to the Higgs portal and depend on the reheating scale.

Due to the strong stellar and fifth force constraints, thermal freezeout of DM in the direct coupling scenario is only possible in a limited region of the parameter space shown. Of course, if there are only $\phi \bar n n $ interactions, then sub-GeV DM has no kinematically accessible annihilation channels. If there are $\phi \bar e e$ interactions, then thermal freezeout from $\chi \bar \chi \to e^+ e^-$ (though an off-shell $\phi$) is only viable for $m_\chi, m_\phi \gtrsim 10\, \MeV$. Comparing with the right panel of  Fig.~\ref{fig:scalar}, the coupling $y_e$ can be sufficiently large if we consider the open parameter space above the beam dump and SN constraints. This is also consistent with the BBN bounds discussed earlier. 

If $m_\phi < m_\chi$, then thermal freezeout is possible for a larger range of DM masses, as low as the keV scale. Additional hidden sector states are typically needed for thermal freezeout of sub-MeV DM through a scalar mediator~\cite{Green:2017ybv,Knapen:2017xzo}. As discussed in Section~\ref{sec:SIDM}, the self-interacting DM bounds are a strong constraint on the coupling of $\phi$ with DM, and exclude the couplings $y_\chi$ where DM annihilation to an ultralight mediator can set the relic abundance. These issues can be resolved if $\phi$ decays to light hidden sector states or if the relic abundance is set by DM annihilation to other states which don't mediate DM self-interactions or DM-SM interactions. However, the stellar and fifth force bounds make the prospect of {\emph{detecting}} DM which interacts via a sub-MeV scalar mediator quite challenging.

For the scalar Higgs portal mediator specifically, dark sector models are only viable in the mass range $m_\chi \gtrsim$ MeV and for $m_\phi < m_\chi$~\cite{Krnjaic:2015mbs} (assuming no additional hidden sector states beyond the mediator and DM). Freeze-in of light DM through a Higgs portal mediator is also challenging due to the stellar constraints, and here it was shown that the entire DM relic abundance can be obtained only for DM masses of $O(\GeV)$~\cite{Krnjaic:2017tio}. 

\subsection{Summary and benchmarks}

In this section, we have presented the general constraints and considerations for sub-GeV dark sector model building, summarizing the possibilities for just a few portals. Here we identify some specific models that will motivate our final lectures on direct detection searches.  To be concrete, consider the following two benchmarks with the kinetically-mixed vector portal:
\begin{itemize}
	\item $m_\chi = 10$ MeV, $m_V$ = 30 MeV, $g_\chi = 0.3$ and $\kappa = 10^{-4}$ (thermal relic, direct coupling)
	\item $m_\chi = 1$ MeV, $m_V = 10^{-12}$ eV, $g_\chi = 3 \times 10^{-6}$, and $\kappa = 10^{-6}$ (freeze-in)
\end{itemize}
where $g_\chi$ is the coupling of the dark photon with Dirac fermion DM.
\exercise{Check that the above parameters can satisfy the existing constraints on dark photons. Check that the first benchmark is a good candidate for a thermal relic, and that the entire relic abundance can be produced by freeze-in for the second benchmark.}

The spin-averaged matrix element squared and cross section for DM-electron scattering can be written as
\begin{align}
	 |{\cal M}|^2  & = \frac{16 g_\chi^2 \kappa^2 e^2 m_\chi^2 m_e^2}{((q_\mu^2) - m_V^2)^2} \approx  \frac{16 g_\chi^2 \kappa^2 e^2 m_\chi^2 m_e^2}{(\qmod^2 + m_V^2)^2}.  \\
	\bar  \sigma_e  & \equiv \frac{ 16 \pi \mu_{\chi e}^2 \kappa^2 \alpha_\chi \alpha}{( (\alpha m_e)^2 + m_V^2)^2}
	\label{eq:electron_scattering}
\end{align}
where $\alpha$ is the fine structure constant, $m_e$ is the electron mass, $\mu_{\chi e}$ is the DM-electron reduced mass, and $\alpha_\chi = g_\chi^2/(4\pi)$. $|\bfq|$ is the momentum transfer, and a typical value for electron scattering is $|\bfq| \simeq \alpha m_e$. We will motivate this definition in more detail in Lecture 5. Interestingly, the two benchmarks give comparable values for $\bar \sigma_e \approx 10^{-37} {\rm cm}^2$. Can light DM with this cross section leave an observable signature in direct detection experiments? The last two lectures deal with this question.

This is not to say that direct detection is the only way to search for these candidates. Furthermore, these are certainly not the only detectable candidates! There are also exciting and promising ideas to use accelerator searches, astrophysical probes, cosmological probes, and so on for a variety of dark sector models. We encourage the reader to dive into these topics, after reading the rest of these lecture notes.

\clearpage

\section{Introduction to direct detection \label{sec:dd} }

Direct detection is one of cornerstones of DM searches, as it is a laboratory probe for particle interactions of the DM in the Milky Way. The goal is to detect and record the rare occasions that particle DM scatters off a target material, ideally in a very quiet environment. As the solar system moves through the Milky Way, we encounter a DM ``wind'' with a typical velocity of $v \sim 10^{-3}$ and a velocity dispersion on the same order. The properties of the DM wind are determined by gravitational interactions, as long as we can treat DM as collisionless particles in the Milky Way. As a result, direct detection probes a very different energy scale for DM interactions compared to the annihilation processes that could set the relic abundance. Indirect detection and accelerator searches can better target those energy scales: indirect detection aims to observe the visible products of DM annihilation, while in accelerators the inverse process of DM pair production is possible. However, they also have drawbacks. There are large astrophysical backgrounds for indirect detection, and accelerator searches for DM can only ascertain that new particles are long-lived on detector timescales, rather than on cosmological timescales. 

\begin{figure}[t]
	\begin{tikzpicture}[line width=1.5 pt, scale=2.5]
		\draw[fermion](145:1) -- (145:.3cm);
			\node at (145:1.2) {$\chi(p)$};
		\draw[fermion](215:1) -- (215:.3cm);
			\node at (215:1.2) {$N(k = 0)$};
		\draw[fermionbar](35:1) -- (35:.3cm);
			\node at (35:1.22) {$\chi(p')$};
		\draw[fermionbar](-35:1) -- (-35:.3cm);
			\node at (-35:1.2) {$N(q = p - p')$};
		\draw[fill=black] (0,0) circle (.3cm);
		\draw[fill=white] (0,0) circle (.29cm);
		\begin{scope}
	    	\clip (0,0) circle (.3cm);
	    	\foreach \x in {-.9,-.8,...,.3}
				\draw[line width=1 pt] (\x,-.3) -- (\x+.6,.3);
	  	\end{scope}
	 \end{tikzpicture}	
\caption{Scattering of DM candidate $\chi$ on nucleus $N$. We will build up to this rate, starting from DM interactions with quarks or gluons. \label{fig:NR} }
\end{figure}
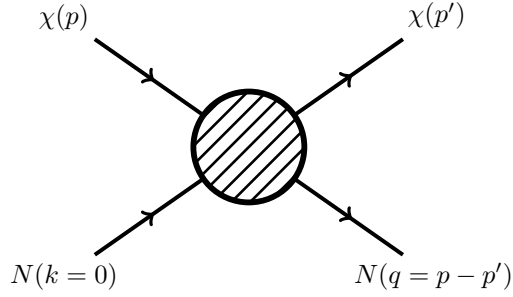

Direct detection of DM deals with low-energy (sub-MeV) scattering events, where the recoiling target particle is detected. There are numerous experimental challenges in distinguishing a DM signal from radioactive backgrounds, and we refer the reader to Refs.~\cite{Undagoitia:2015gya,Schumann:2019eaa} for recent in-depth reviews of experimental approaches. In this lecture, we introduce the theory fundamentals needed to translate a DM model into a signal prediction. The goal will be to develop intuition for determining whether a DM candidate has spin-independent vs. spin-dependent scattering, the qualitative behavior of the recoil spectrum, and how the DM velocity distribution impacts expected signals.  We will begin by focusing on traditional direct detection, where the emphasis is on DM mass above $\sim$ few GeV and WIMPs. Here the signal of DM scattering is nuclear recoils. Originally proposed in Ref.~\cite{Goodman:1984dc}, direct detection of WIMPs is an immense field and additional classic reviews that emphasize the theoretical ingredients can be found in Refs.~\cite{Lewin:1995rx,Bertone:2004pz,Jungman:1995df}.  Lecture 5 will then introduce some of the new tools and techniques needed for direct detection of sub-GeV DM.

\subsection{Kinematics of nuclear recoil}

The relevant physics can be identified if we examine the kinematics of DM-nucleus scattering more closely. The typical kinetic energy of a nucleus in an experiment is $O(kT)$, where $T$ is usually well below room temperature, so we can treat the nucleus as a particle at rest initially. Using the momentum labels in Fig.~\ref{fig:NR}, the initial total energy and momentum are
\begin{align}
	E_i = \frac{\bfp^2}{2 m_\chi}, \quad \bfp = m_\chi {\bf v}
\end{align}
while the final energy is
\begin{align}
	E_f = \frac{ (\bfp - \bfq)^2 }{2m_\chi} + \frac{\bfq^2}{2m_N}  \, .
\end{align}
The deposited energy (recoil energy of the nucleus) is $E_R = \bfq^2/(2m_N)$ and $\bfq$ is the momentum transfer. Defining $\cos \theta = \hat \bfp \cdot \hat \bfq$, energy conservation gives the requirement that
\begin{align}
	\frac{ | \bfp| |\bfq| \cos \theta}{m_\chi} =  \frac{\bfq^2}{2 \mu_{\chi N} } 
\end{align}
where $\mu_{\chi N} = m_\chi m_N/(m_\chi + m_N)$ is the reduced mass for the DM-nucleus system. There is a maximum momentum transfer $|\bfq|_{\rm max} = 2 \mu_{\chi N} |\bfp|/m_\chi = 2  \mu_{\chi N} v$ where $v \sim 10^{-3}$. For a WIMP scattering off a typical nucleus, then $\mu_{\chi N} \simeq 10-100$ GeV and $|\bfq|_{\rm max} \simeq 20-200\, \MeV$. The corresponding maximum recoil energy is
\begin{align}
	E_R^{\rm max} = \frac{|\bfq|_{\rm max}^2}{2 m_N} = \frac{2 \mu_{\chi N}^2 v^2}{m_N} \simeq 20-200 \, \keV .
\end{align}
In this estimate, we have used only the typical velocity of the DM, $v \sim 10^{-3}$, and one could imagine larger recoils from DM with greater speeds. For a given recoil energy, the minimum incoming speed for the DM would then be $v_{\rm min} = \sqrt{ m_N E_R / 2 \mu_{\chi N}^2 }$. However, the likelihood of an incident DM with larger velocities eventually becomes exponentially suppressed and it is typically assumed there is essentially no DM above the local escape velocity, which translates to $v \lesssim 3 \times 10^{-3}$ in the lab frame. A commonly-used DM velocity distribution is given Section~\ref{sec:dm_fv} below.

The energy and momentum scales elucidate the relevant nuclear and atomic physics. Translating the typical momentum transfer to a length scale, we have $1/|\bfq| \sim 1-10 \, {\rm fm}$. This is on the order of the radius of a typical nucleus, which implies that we need a form factor for the DM-nucleus interactions. In the limit of $1/|\bfq| \gg 10 \, {\rm fm}$, then the DM interacts with the whole nucleus coherently. If the DM-nucleon interaction is constructive over the whole nucleus, then the rate scales as $A^2$ in the small $|\bfq|$ limit, where $A$ is the atomic number. However, there is a form factor suppression at larger $|\bfq|$ depending on the target nucleus. 

Another assumption implicit in the kinematic considerations above is that we can neglect the effect of exciting bound atomic or nuclear states. The splitting for nuclear excited states is around $\sim 30-1000$ keV. Rates for inelastic scattering via nuclear excitations are generally much smaller than for elastic recoils (see for example~\cite{Ellis:1988nb,Baudis:2013bba,McCabe:2015eia,Suzuki:2018xek}) because the splittings are large compared to the typical recoil energy, and because there isn't an $A^2$ coherent enhancement.   Meanwhile, the binding energy for atomic states is around 10 eV -- 10 keV. This case is not as clear. Traditionally, it has been assumed that the bound electrons can immediately follow the recoiling nucleus. However, the actual process is more complicated. Recent treatments of atomic excitations/ionizations in DM scattering can be found in Refs.~\cite{Kouvaris:2016afs,Ibe:2017yqa,Dolan:2017xbu}, where it was shown to have a large impact on the detectability of sub-GeV DM.  The effects are modest for DM with mass above $\sim 10$ GeV, so we will not pursue this further.  We have also only considered the primary DM-nucleus interaction here: following this interaction, the recoiling atom then collides with other atoms in the target, giving rise to the detected signals in scintillation light, charge yield, or phonons~\cite{Undagoitia:2015gya}.

DM-nucleus scattering can thus be treated in several steps: given some microscopic interactions, we must determine the correct theory at the QCD scale. Operators describing DM interactions with quarks and gluons must then be matched onto operators with nucleons. The nucleons inside a nucleus can be treated in the nonrelativistic limit, since the momentum scale is set by the radius of the nucleus $R_N$, where $1/R_N \simeq 1- 10 \, \MeV \ll m_{n,p}$. We must then determine the nuclear matrix elements for nonrelativistic nucleon operators -- this is encoded inside nuclear form factors or response functions. Finally, the differential scattering cross section is integrated against the DM velocity distribution in the lab frame in order to determine the differential rate of nuclear recoils.

\subsection{From microscopic operators to nucleon couplings}

For DM annihilation and other processes in the early universe, the interactions are typically at an energy scale of $\sim m_\chi$ or greater. For direct detection processes, we must instead consider the non-relativistic limit and match onto interactions with the nuclei present in targets. The first step in this process is to go from quark and gluon interactions to determine the interactions with protons and neutrons. (The entire procedure differs for DM interactions with electrons, where the wavefunctions of bound electrons in materials must be accounted for in the calculation. This will be discussed in the last lecture.)

We begin by considering interactions of fermionic DM mediated by weak-scale mediators. Below the weak scale $\sim m_Z$, these are present as dimension-6 operators:
\begin{align}
	\frac{C_V}{m_Z^2} \bar \chi \gamma^\mu \chi \, \bar q \gamma_\mu q + \frac{C_A}{m_Z^2} \bar \chi \gamma^\mu \gamma^5 \chi \, \bar q \gamma_\mu \gamma^5 q + \frac{C_S m_q}{m_h^2 \, v} \bar \chi  \chi \, \bar q q + ...
\end{align}
where $C_i$ are arbitrary coupling-dependent Wilson coefficients and $v$ is the Higgs vacuum expectation value. The dimensionful scales in the operator coefficients have been normalized to the SM boson masses and Yukawa couplings (proportional to quark masses) for convenience and to be suggestive of WIMP interactions. For WIMPs, one should construct the effective field theory at the appropriate high scale and run down to the QCD scale. Note that the operators can mix, so that even if some operators are zero at tree level, they will be generated at loop level. It is important to check the loop-induced interactions, since the constraints on spin-independent DM-nucleon couplings are much stronger than on spin-dependent couplings. We refer the reader to some of the literature on matching and running down to the QCD scale~\cite{DEramo:2014nmf,Hill:2014yka,Hill:2014yxa,Bishara:2018vix,Mohan:2019zrk}, and as our starting point assume that $C_i$ are Wilson coefficients at the QCD scale.

The next step is that quark level operators must get matched onto nucleon level operators with a similar tensor structure. Therefore we expect that $\bar q \gamma_\mu q \sim \bar n \gamma_\mu n$ and so on for other operators. The matching can be written more precisely by taking the matrix elements of the partonic operators with nucleon states~\cite{DelNobile:2013sia,Hill:2014yxa,Bishara:2017pfq,Bishara:2017nnn}. Considering the vector current for one of the light quark flavors only ($u,d,$ or $s$), 
\begin{align}
	\langle n(k') |\bar q \gamma^\mu q | n(k) \rangle =  \bar u_n(k') \left[ F_1^{q,n}(q^2) \gamma^\mu +   \frac{i}{2 m_n} F_2^{q,n}(q^2) \  \sigma^{\mu \nu} q_\nu \right] u_n(k)
	\label{eq:vectorcurrent_spinor}
\end{align}
where the momentum transfer $q = k'-k$ and we use $n$ for nucleon (either proton $p$ or a neutron $n$, we will not distinguish the couplings here although they can differ). This is analogous to the form factor for the vector current in quantum electrodynamics. Since direct detection takes place in the non-relativistic regime, we can take the limit $q^2 \to 0$. Then the  form factor $ F_1^{q,n}(0)$ captures the quark content
\begin{align}
	F_1^{u,p}(0) = 2, \ \ F_1^{d,p}(0) = 1, \ \ F_1^{s,p}(0) = 0 . 
\end{align} 
The second form factor $F_2^{q,n}(0)$ describes the contribution of that quark flavor to the nucleon anomalous magnetic moment, and can be determined by a combination of lattice QCD, isospin symmetry, and the magnetic moment of the proton and neutron. Since the typical $q^\mu/m_n \ll 1$, this second piece is subdominant and we will just focus on the first term. 

We next take the nonrelativistic limit of the nucleon operator. To do this, we can expand the operators according to the nonrelativistic symmetry of Galilean invariance~\cite{Fitzpatrick:2012ix,Fan:2010gt}. The possible quantities appearing in the non-relativistic limit are: 
\begin{align}
	1, \quad \bfq \equiv \bfp - \bfp\, ' = \bfk' - \bfk, \quad \bfv_{\rm rel} \equiv \bfv_{\chi,i} - \bfv_{n,i}, \quad {\bf S}_\chi, \quad {\bf S}_n
\end{align}
where $\bfv_{\chi,i}$ and $\bfv_{\chi,i}$ are the initial velocities of the DM and nucleon. For an orthogonal set of quantities, we can take $\bfv_{\rm rel} \to \bfv^\perp \equiv \bfv_{\rm rel} - \bfq/(2 \mu_{\chi n})$, where $\bfv^\perp \cdot \bfq = 0$ by conservation of energy.  The first operator above is just the scalar, the next two quantities are three-vectors, and the spin operators ${\bf S}_{\chi,n}$ are pseudo-vectors or parity-odd three-vectors. Note that since the theory is actually Lorentz invariant, there are also operators that are not Galilean-invariant which appear at higher order~\cite{Bishara:2017pfq}.

For the matrix element in Eq.~\ref{eq:vectorcurrent_spinor}, we split $\bar u_n(k') \gamma^\mu u_n(k)$ into a scalar with respect to the symmetry in the nonrelativistic limit, $\bar u_n(k') \gamma^0 u_n(k)$, and a three-vector, $\bar u_n(k') \gamma^i u_n(k)$. For a nonrelativistic Dirac fermion, at leading order we can take $u_n(k) \to \sqrt{m_n} \left( \xi_s \ \xi_s \right)^T$ with $\xi_s$ a spinor.  Then we see explicitly that the scalar part is given by $ \bar u_n(k') \gamma^0 u_n(k) = 2 m_n \xi_{s'}^\dagger \xi_s$ where $s$ and $s'$ are the initial and final spin states. For the three-vector $\bar u_n(k') \gamma^i u_n(k)$, a similar expansion can be done in terms of the spinors. One might guess that the three-vector reduces to $\bfq \xi_{s'}^\dagger \xi_s$, since the operator depends only on the nucleon momenta $\bfk'$ and $\bfk$ and because $\bfq$ is Galilean invariant. The actual  nonrelativistic expansion gives an operator proportional to $(\bfk'+ \bfk) \xi_{s'}^\dagger \xi_s$, which illustrates breaking of Galilean invariance when we expand the operator $\bar u_n(k') \gamma^\mu u_n(k)$ beyond leading order. However, since $|\bfk|/m_n, |\bfk'|/m_n \ll 1$, we can approximate the operator $\bar u_n(k') \gamma^\mu u_n(k)$ as just the scalar operator.

Applying the same argument to the DM part of the operator $\bar \chi \gamma^\mu \chi$ and labeling the spinors as $\zeta_t$, we conclude that the leading behavior of the matrix element for vector exchange is given by:
\begin{align}
	{\cal M} = \frac{12 C_V}{m_Z^2} m_n m_\chi  \, \zeta_{t'}^\dagger \zeta_t \  \xi_{s'}^\dagger \xi_s  \,  .
	\label{eq:M_vector}
\end{align}
where we summed over all light quarks.
This type of scattering is known as {\emph{spin-independent}} scattering because it does not depend on the spin of the target nucleus (i.e., it does not involve a spin flip of the nucleon). It should be clear that the $\bar \chi \chi \, \bar q q$ operator leads to a similar form, although with different matching coefficients. The coupling is the same with all nucleons in a nucleus, so this leads to a coherent enhancement in scattering rates by $A^2$ in the long-wavelength limit, where $A$ is the mass number.

The last operator above, $\bar \chi \gamma^\mu \gamma^5 \chi \, \bar q \gamma_\mu \gamma^5 q$, is an example of {\emph{spin-dependent}} scattering. Again considering only the light quark flavors, we start with the matching to nucleon matrix elements:
\begin{align}
	\langle n(k') |\bar q \gamma^\mu  \gamma^5 q | n(k) \rangle =  \bar u_n(k') \left[ F_A^{q,n}(q^2) \gamma^\mu \gamma^5 +   \frac{1}{2 m_n} F_{P'}^{q,n}(q^2) \ \gamma^5 q^\mu \right] u_n(k) \, ,
\end{align}
The $q^2=0$ limit of these form factors are determined by a combination of semileptonic decays of hadrons, $\nu p$ scattering, and/or lattice data. We give values for $F_A^{q,n}(0)  = \Delta_{q}^n$ in Table~\ref{tab:DDmatrixelements}, since this is most important piece for direct detection.

As before, we consider only the leading term,  $\Delta_{q}^n \, \bar u_n(k') \gamma^\mu \gamma^5 u_n(k)$.  The corresponding nonrelativistic operator can be diagnosed in a similar way as above. This is a parity-odd coupling,  so we can split it into a pseudoscalar part $\bar u_n(k')  \gamma^0 \gamma^5  u_n(k)$ and a pseudo three-vector part $\bar u_n(k') \gamma^i \gamma^5 u_n(k)$. The only available pseudoscalar that we can form out of the nucleon spin and momenta is $\bfq \cdot {\bf S}_n$. For the parity odd three-vector,  $m_n {\bf S}_n$ is good candidate which has the correct dimensions. The momentum-dependent part suffers an additional suppression, so we only need to consider the pseudo three-vector. Combining this with a similar argument about the DM part of the operator, the leading behavior of the matrix element must go as 
\begin{align}
	{\cal M} & =  \sum_q \frac{16 C_A \, m_\chi m_n }{m_Z^2} \, \Delta_{q}^n \, \zeta_{t'}^\dagger {\bf S_\chi} \zeta_t \,  \cdot \, \xi_{s'}^\dagger  {\bf S}_n \xi_s  \\
	&=  \sum_q  \frac{4 C_A \, m_\chi m_n }{ m_Z^2} \,  \Delta_{q}^n \, \zeta^\dagger \sigma^i \zeta \, \xi^\dagger \sigma^i \xi \quad \textrm{ for spin 1/2 fermions}
\end{align}
and the spatial indices of ${\bf S_\chi}$ and ${\bf S}_n$ are contracted. In the second equation, we have replaced the spin operators with the corresponding Pauli matrices for spin 1/2 fermions. This is called spin-dependent scattering because it is proportional to the nucleon spin and it involves spin flips. Note that the spins of the nucleons do not add coherently inside a nucleus and typically there are only a few unpaired nucleons. For example, some of the isotopes with sensitivity to spin-dependent interactions are $^{129}$Xe (ground state with nuclear spin 1/2), $^{131}$Xe (spin 3/2), and $^{133}$Cs (spin 7/2). One would think that the scattering rate is then proportional to the square of total nuclear spin, $\propto J_N (J_N+1)$. However, this is not really the case because  the total nuclear spin is a sum of the nucleon spin and angular momentum. The scattering rate is then typically normalized as $\propto (J_N+1)/J_N$. The direct detection limits are thus much weaker compared to spin-independent scattering.

\exercise{Which type of scattering is mediated by heavy pseudoscalar exchange, where the operator is $\bar \chi \gamma^5 \chi \, \bar q \gamma^5 q$? Find the nonrelativistic operator, up to $O(1)$ factors. What about $\bar \chi \gamma_\mu \gamma^5 \chi \, \bar q \gamma^\mu q$ and $\bar \chi \gamma_\mu \chi \, \bar q \gamma^\mu \gamma^5  q$?}

\def\arraystretch{1.2}\setlength{\tabcolsep}{10pt}
\begin{table}[t]
\begin{tabular}{|c|c|c|c|c||c|c|c|} 
\hline
  & $f_{T,g}^n$ & $f_{T,u}^n$ & $f_{T,d}^n$ & $f_{T,s}^n$  & $\Delta_u^n$ & $\Delta_d^n$ &  $\Delta_s^n$ \\ \hline
Neutrons & 0.910(20) & 0.013(3) & 0.040(10) & 0.037(17) & -0.46(4) & 0.80(3) & -0.12(8) \\
Protons & 0.917(19) & 0.018(5) & 0.027(7) & 0.037(17) & 0.80(3)  & -0.46(4) & -0.12(8) \\
\hline
\end{tabular}
\caption{Matrix elements for scalar and axial light quark operators in nucleon states. The numbers for the scalar operators come from Ref.~\cite{Ellis:2018dmb}, where the number in parentheses is the one-sigma uncertainty. For the axial-vector operator we use the shorthand $\Delta_{q}^n = F_A^{q,n}(0)$ and take values from Ref.~\cite{Hill:2014yxa}. Additional recent compilations of nucleon matrix elements can be found in Refs.~\cite{Hill:2014yxa,DelNobile:2013sia,Bishara:2017pfq}. \label{tab:DDmatrixelements}}
\end{table}

\subsubsection{Heavy scalar exchange}

So far, we have dealt with light quark matrix elements in nucleons. What about operators such as $G^{\mu \nu} G_{\mu \nu}$ or those involving heavy quarks? We need to consider these matrix elements in order to properly treat the scalar operator $\frac{C_S m_q}{m_h^2 \, v} \bar \chi \chi  \bar q q$, with a sum over all quarks for Higgs exchange. Heavy quarks are integrated out at the scale given by the mass of the quark, where integrating out a single flavor generates the loop-induced operator 
\begin{align}
	- \frac{C_S}{m_h^2 \, v} \, \frac{\alpha_s}{12 \pi} \bar \chi \chi G^{\mu \nu} G_{\mu \nu}
\end{align} 
in the effective field theory. The DM coupling is then determined in terms of matrix elements for the light quarks and gluon field strength, given by
\begin{align}
	\langle n(k') | m_q \bar q q | n(k) \rangle  & =  m_n f_{T,q}^n  \, \bar u_n(k') u_n(k) \\
	\langle n(k') | \alpha_s G^{\mu \nu} G_{\mu \nu} | n(k) \rangle  & =  - \frac{8 \pi}{9} m_n f_{T,g}^n \, \bar u_n(k') u_n(k)
\end{align}
where the coefficients $f_{T,g}^n,  f_{T,q}^n$ can be determined through a combination of experimental data and lattice calculations. Recent discussions of the values and their uncertainties can be found in Ref.~\cite{Hoferichter:2017olk,Ellis:2018dmb}, and we reproduce the results in Table~\ref{tab:DDmatrixelements}.

For the scalar operator, the final matrix element for a given nucleon $n$ is then given by
\begin{align}
	{\cal M} &= \frac{C_S}{m_h^2 \, v}   m_n  \left( f_{T,u}^n + f_{T,d}^n  + f_{T,s}^n + \frac{2}{9} f_{T,g}^n \right) \bar u_\chi(p') u_\chi(p) \bar u_n(k') u_n(k) \\
	& \approx 4 m_\chi m_n \times \frac{ C_S \, m_n}{m_h^2 \, v} \times 0.29 \times \chi_{t'}^\dagger \chi_t \  \xi_{s'}^\dagger \xi_s  
\end{align}
where we have included the DM part as well and used the numerical values of the matrix elements. We will use this result below in estimating the typical cross sections for Higgs exchange.

\subsection{Nuclear recoils from DM scattering}

The next step in computing the scattering rate is to turn the nucleon-level matrix elements into matrix elements for DM-nucleus scattering in specific targets. The matrix element of the nucleon level operators with the nucleus state is encapsulated in nuclear form factors. For example,
\begin{align}
	|\langle N | {\bf S}_n( q) | N \rangle|^2 &= F^2_{\rm spin}(q) \\
	|\langle N | \bar n n | N \rangle|^2 &= F^2_{\rm mass}(q)
\end{align}
where again we just take the leading behavior of the matrix element. Computing these form factors requires a model of the bound nucleons inside a nucleus, and therefore involves complicated many-body physics\footnote{Similarly, when approaching direct detection of light DM, one needs to account for many-body effects that give rise to the electron and phonon excitations in a material. For instance, we will take a similar approach is defining phonon form factors in L5, except that we sum over all the ``bound state'' nuclei in a crystal lattice.}. A commonly used model is the shell model of nuclei. Let us consider just the scalar $\bar n n$ operator: in the non-relativistic limit, this is the number density operator for nucleons. The mass form factor can then be written as
\begin{align}
	F(q) = \int d^3 \bfx \, e^{- i \bfq \cdot \bfx} \ \frac{\rho_n(\bfx)}{m_n} \\
		\to A \quad \text{when} \quad q \to 0
\end{align}
where $\rho_n(\bfx)$ is the mass density of nucleons in the nucleus. A commonly used form factor is the Helm form factor~\cite{PhysRev.104.1466} since it has a simple analytic expression, $F(q) = \tfrac{3 j_1 (q r_n)}{q r_n} e^{ - (qs)^2/2}$, where $j_1$ is the first spherical Bessel function and an approximate fit for heavy nuclei can be obtained with $r_n \approx 1.14 A^{1/3}$ fm and $s \approx 0.9$ fm. Fig.~\ref{fig:dd_formfactor} shows example form factors for various target materials used in experiments. A standard reference with a clear explanation of the spin form factors can be found in Ref.~\cite{Ressell:1997kx}.

\begin{figure}[t]
\includegraphics[width=0.6\textwidth]{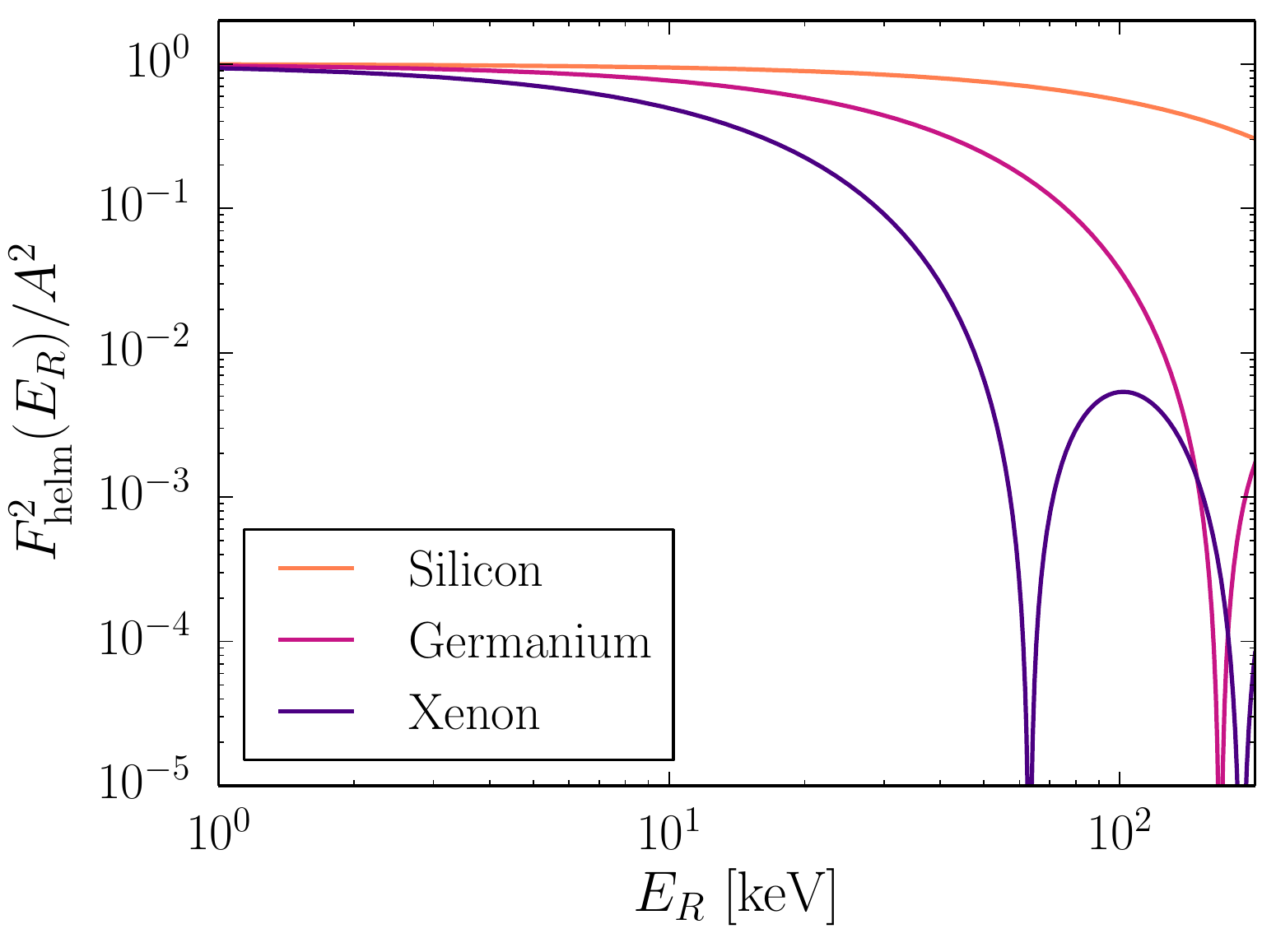}
\caption{ Nuclear recoil form factor for example target nuclei of silicon ($A=23$), germanium ($A=73$), and xenon ($A=131$). Plotted here is the Helm form factor, a simple analytic approximation. \label{fig:dd_formfactor} }
\end{figure}

We now embark on a calculation of the spin-independent scattering rate. For the vector operator, we start with the matrix element in Eq.~\ref{eq:M_vector}. Squaring, taking the trace, and averaging over final state spins gives
\begin{align}
	|{\cal M}|^2 = \left(\frac{3 C_V}{m_Z^2} \right)^2  \times (4 m_\chi m_n)^2 \equiv b_n^2 \, (4 m_\chi m_n)^2 
\end{align}
where we have assumed spin 1/2 fermions as above. The second equality defines $b_n$, the DM-nucleon vector coupling. We first determine the cross section for DM-nucleon scattering, where there is no nuclear form factor.  Taking the nonrelativistic limit and assuming the initial nucleon is at rest, the differential cross section is given by
\begin{align}
	d \sigma_n &= \frac{ |{\cal M}|^2}{4 m_\chi m_n v } \frac{ d^3 {\bf p}'}{(2\pi)^3 2 m_\chi} \frac{ d^3 {\bf q}}{(2\pi)^3 2 m_n} (2\pi)^4 \delta^{(4)}\left( p + k - p' - k' \right) \\
		& = \frac{ b_n^2 }{4 \pi \, v^2} d|\bfq|^2 \, d \cos \theta \, \delta\left(\cos \theta - \frac{|\bfq| }{2 \mu_{\chi n} v } \right)
\end{align}
where $v$ is the initial DM velocity, $\cos \theta = \hat \bfq \cdot \hat \bfp$, and $\mu_{\chi n}$ is the reduced mass of the DM and nucleon. Integrating over all kinematically allowed $|\bfq| \le 2 \mu_{\chi n} v$, the total cross section is
\begin{align}
	\sigma_n = \frac{\mu_{\chi n}^2 \, b_n^2 }{\pi} \, .   \label{eq:sigma_vector}
\end{align}

The differential cross section above can now be generalized to DM-nucleus scattering, where we use $N$ for nuclear quantities. Using the fact that $d|\bfq|^2 = 2m_N dE_R$ with $E_R$ the nuclear recoil energy and including the nuclear form factor, we obtain:
\begin{align}
	\frac{d\sigma_N}{dE_R} &= \frac{|{\cal M}|^2}{\pi} \frac{ m_N }{2 v^2}  \, F^2(q)\, d \cos \theta \, \delta\left(\cos \theta - \frac{|\bfq| }{2 \mu_{\chi N} v } \right)  \\
	&= \frac{\sigma_n}{\mu_{\chi n}^2} \frac{ m_N }{2 v^2} \, F^2(q)\,  \Theta\left(v - \sqrt{\frac{m_N E_R }{2 \mu_{\chi N}^2} } \right) \label{eq:dsigmaN} 
	\end{align}
In the second line, we have replaced the matrix element squared with the single nucleon cross section. The step function enforces the minimum incoming DM velocity for a recoil of energy $E_R$.

The differential rate per unit target mass and per unit time is defined as
\begin{align}
	\frac{dR}{dE_R} = N_T n_\chi \int \frac{d\sigma}{dE_R} v f({\bf v}) \, d^3 \bfv
\end{align}
with $N_T$ is the number of nuclei per unit target mass, $n_\chi = \rho_\chi/m_\chi$ is the DM number density, and $f(\bf v)$ is the DM velocity distribution in the lab frame and normalized to unity. Extracting just the dependence on the DM velocity, we can define an integral
\begin{align}
	g(v_{\rm min} ) \equiv \int   d^3 \bfv  \, \frac{f({\bf v})}{v} \Theta\left(v - \sqrt{\frac{m_N E_R }{2 \mu_{\chi N}^2} } \right)  \label{eq:gmin}
\end{align}
with $v_{\rm min} = \sqrt{m_N E_R /2 \mu_{\chi N}^2 }$. Finally, we account for the fact that the DM may have different couplings to protons or neutrons and rescale the total rate to obtain:
\begin{align}
	\boxed{  \frac{dR}{dE_R} = N_T \frac{\rho_\chi}{m_\chi} \frac{\sigma_n m_N}{2 \mu_{\chi n}^2}  \left( \frac{ b_p Z + b_n (A - Z) }{b_n A} \right)^2 F^2(q) \, g(v_{\rm min}) } \,
	\label{eq:dRdER}
\end{align} 
where in this formula $b_n$ is the neutron coupling and $b_p$ is the proton coupling.

Similar results can be obtained for axial-vector coupling and heavy scalar exchange. Summing over spins and averaging over initial spin, the cross section for spin-dependent scattering for spin-$1/2$ Dirac fermion DM  is given by 
\begin{align}
	\sigma_n = \frac{3 \mu_{\chi n}^2}{\pi} \left( \sum_q \Delta^n_q \frac{C_A}{m_Z^2} \right)^2 \equiv \frac{3 \mu_{\chi n}^2}{\pi} \left( \sum_q \Delta^n_q d_q \right)^2 .  \label{eq:sigma_axial}
\end{align}
For a detailed discussion of cross sections and rates for spin-dependent interactions, see Ref.~\cite{Agrawal:2010fh}. For heavy scalar exchange, the calculation proceeds in a similar way with 
\begin{align}
	\sigma_n = \frac{\mu_{\chi n}^2}{\pi} \left( \frac{0.29 \times C_S m_n}{m_h^2 v} \right)^2 \equiv \frac{\mu_{\chi n}^2 f_n^2 }{\pi} . \label{eq:sigma_higgs}
\end{align}
We have given the mapping to commonly used notation in these equations.

We've almost gotten most of the relevant formulae down. Before looking at the total scattering rate, we first discuss the DM velocity distribution and $g(v_{\rm min})$.

\begin{figure}[t]
\includegraphics[width=0.5\textwidth]{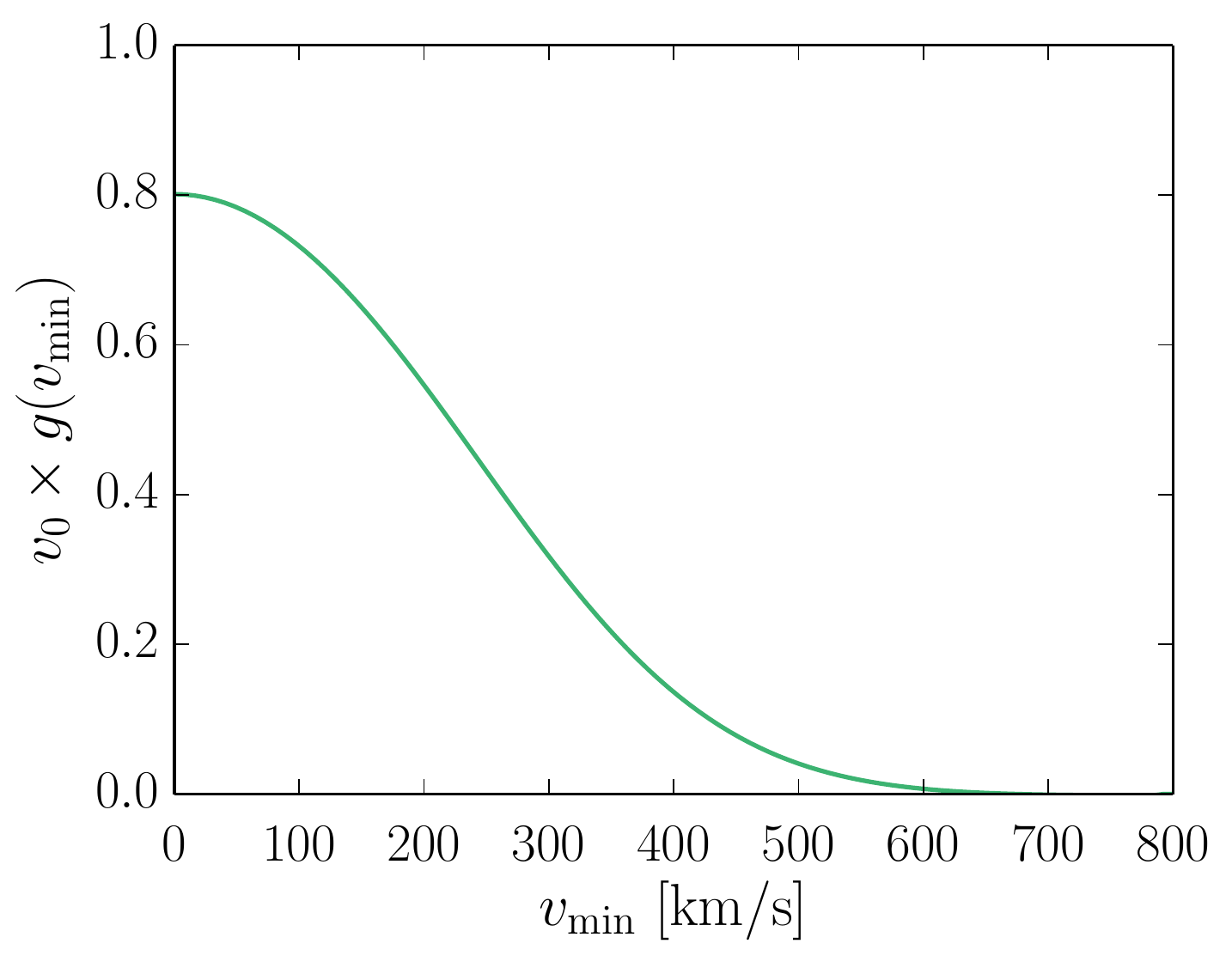}
\includegraphics[width=0.485\textwidth]{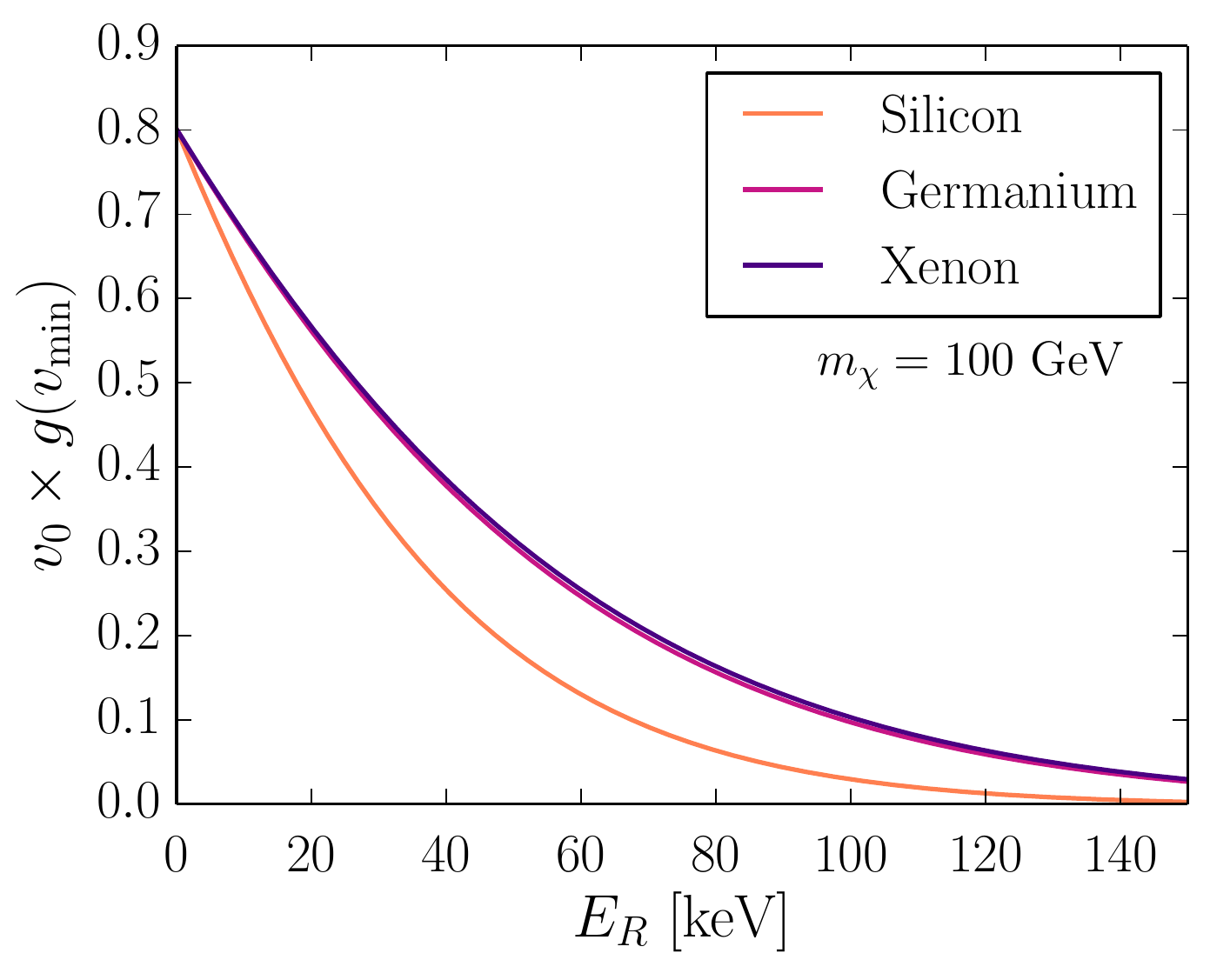}
\caption{ The DM scattering rate falls at large $E_R$ due to two effects. The dominant one is that the available number of DM particles with velocity above $v_{\rm min} = \sqrt{m_N E_R /2 \mu_{\chi N}^2 }$ falls exponentially. This is encapsulated in the function $g(v_{\rm min})$, shown in the left plot for the velocity distribution in Eq.~\ref{eq:fv_earth}. In the right plot, we translate this to $g(v_{\rm min})$ as a function of the recoil energy $E_R$ for several example target nuclei and $m_\chi = 100$ GeV.  In addition, there is a form factor suppression at larger $E_R$, shown in Fig.~\ref{fig:dd_formfactor}. \label{fig:dd_gv} }
\end{figure}

\subsubsection{Astrophysics of direct detection \label{sec:dm_fv}}

The properties of DM in the solar neighborhood enter in the local DM density $\rho_\chi$ and the velocity distribution $f({\bf v})$ of DM particles incident on a target. The local DM density is usually assumed to be $\rho_\chi = 0.4$ GeV/cm$^3$. This is determined from models of the local gravitational potential (usually within 1 kpc), and  different methods lead to variations on this value at the level of $10-50\%$. For a summary of measurements and methods, see Ref.~\cite{Read:2014qva}. 

The DM velocity distribution is more challenging to determine, and this is complicated by the fact that small changes in the velocity distribution can have a large impact on direct detection rates. A simple analytic approximation for the DM velocity distribution is that of a truncated Maxwellian distribution, since we expect the DM in the solar neighborhood to be mostly virialized. In the frame of the Galaxy, this velocity distribution is isotropic with the form:
\begin{align}
	f_{\rm Galaxy}({\bf v}) = \frac{1}{N(v_0)}  \exp \left( -\frac{ {\bf v}^2}{v_0^2} \right) \Theta(v_{\rm esc} - |{\bf v}|) 
\end{align}
where it is usually assumed that $v_0 \approx 220$ km/s. The distribution is cut off at the local escape speed $v_{\rm esc}$, with estimated values around 500-600 km/s~\cite{Piffl:2013mla,2018A&A...616L...9M}. $N(v_0)$ is normalization factor such that $f({\bf v})$ integrates to unity,
\begin{align}
	N(v_0) = \pi^{3/2} v_0^3 \left[ {\rm erf} \left( \tfrac{v_{\rm esc}}{v_0} \right) - \tfrac{2}{\sqrt{\pi}} \tfrac{v_{\rm esc}}{v_0} \exp \left( - \tfrac{v_{\rm esc}^2}{v_0^2}  \right) \right].
\end{align} 
This is known as the Standard Halo Model.
One must then boost this to the Earth (lab) frame
\begin{align}
	f_{\rm Earth}({\bf v}) = \frac{1}{N(v_0)}  \exp \left( -\frac{ ({\bf v} + {\bf V}_e(t))^2}{v_0^2} \right) \Theta(v_{esc} - |{\bf v} + {\bf V}_e(t) |) \label{eq:fv_earth}
\end{align}
where we can decompose the Earth's motion into the Sun's motion in the Galaxy (${\bf V}_{\rm \odot}$) and the orbit of the Earth about the Sun (${\bf V}_{\rm \oplus}$):
\begin{align}
	{\bf V}_e(t) = {\bf V}_{\rm \odot} + {\bf V}_{\rm \oplus}(t)
\end{align}
For calculating rates, we often neglect the orbital motion of the Earth since $|{\bf V}_{\rm \oplus}| \approx 29.8$ km/s, which is much less than the rotational speed of the Sun  $|{\bf V}_{\rm \odot}| \approx 240$ km/s.
The Sun's motion corresponds to a fixed direction with respect to the Milky Way (in the direction of the constellation Cygnus) and the DM velocity distribution appears to be coming at us from $-{\bf V}_\odot$, on average; we therefore often say that there is a DM ``wind'' coming from Cygnus, although of course there is a distribution in the incoming DM direction. The time-dependence in ${\bf V}_{\rm \oplus}(t)$ leads to an additional interesting effect: an annual modulation of the rate as the Earth rotates about the Sun. For a recent review of modulation signals and the astrophysics of direct detection, see Ref.~\cite{Freese:2012xd}.

As defined in Eq.~\ref{eq:gmin}, $g(v_{\rm min})$ is the expectation value of $1/v$ subject to the minimum energy threshold or velocity threshold. Using the velocity distribution given here with $v_{\rm esc} = 550$ km/s, $v_e = 240$ km/s and $v_0 = 220$ km/s, we show in Fig.~\ref{fig:dd_gv}  the resulting  $g(v_{\rm min})$. A given experiment uses a particular target nucleus and has an energy threshold, which means that the experiment is sensitive to particular ranges in the DM velocity distribution. Concentrating on the exponential tail, which gives the highest energy recoils, we can approximate $g(v_{\rm min})$ by
\begin{align}
	g(v_{\rm min}) & \propto \int_{v_{\rm min}}^{v_{\rm max}} dv \, v \, \exp \left( -\frac{v^2}{v_0^2} \right) \\
	 & \propto \exp \left( -\frac{v_{\rm min}^2}{v_0^2} \right) - \exp \left( -\frac{v_{\rm max}^2}{v_0^2} \right)
	 \simeq \exp \left( -\frac{m_N E_R}{2 \mu_{\chi N}^2 v_0^2} \right) \label{eq:gv_approx}
\end{align}
where the maximum possible velocity in the lab frame is given by $v_{\rm max} = v_{\rm esc} + v_e$ and we have used $v_{\rm min} = \sqrt{m_N E_R /2 \mu_{\chi N}^2 }$ in the last equality. Thus the minimum velocity threshold for a given recoil energy leads to an exponential drop in $E_R$ for the differential scattering rate, which can be seen in the right panel of Fig.~\ref{fig:dd_gv}. The typical recoil energy is thus given by $E_R^{\rm typical} = 2 \mu_{\chi N}^2 v_0^2 /m_N$, while the maximum recoil energy is $E_R^{\rm max} = 2 \mu_{\chi N}^2 (v_{\rm esc} + v_e)^2 /m_N$.

Of course, the statements above assume that the tail of the DM velocity distribution is well-described by a Maxwellian. It is well known that the simple analytic result above is not a good approximation to more realistic DM velocity distributions from simulations~\cite{Kuhlen:2009vh,Kuhlen:2012fz,Necib:2018igl}. It is likely that the true distribution is anisotropic within the Galaxy, that there are unvirialized components, and that the hard cutoff at $v_{\rm esc}$ is certainly unphysical. In calculating elastic scattering rates for heavier DM, this affects the recoil spectrum and also has a modest effect on the total rates. The impact is larger for low mass DM, where an experiment is usually sensitive only to DM with the highest speeds, and thus we require knowledge of the tail of the distribution. While we do not have a direct measurement of the velocity distribution, recent work has studied the velocity distribution of the metal-poor stellar halo as a proxy for the DM~\cite{Herzog-Arbeitman:2017fte,Herzog-Arbeitman:2017zbm,Necib:2018iwb}. This gives an empirical way to approximate the DM velocity distribution, and we expect our knowledge to improve with  more data.

\subsection{Scattering rates and experimental status}

\begin{figure}[t]
\includegraphics[width=0.7\textwidth]{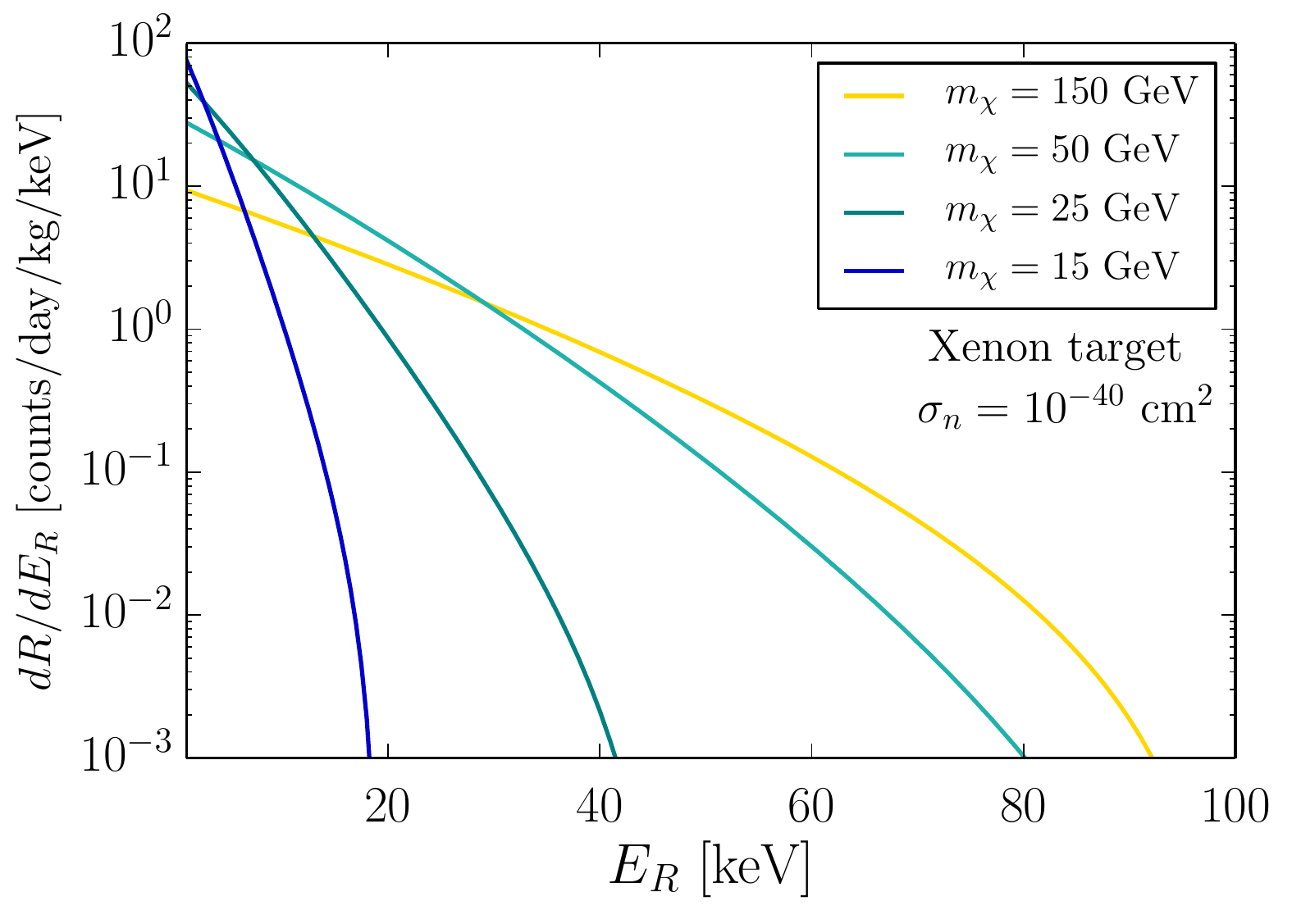}
\caption{ Differential nuclear recoil spectrum, computed using Eq.~\ref{eq:dRdER} with $b_n = b_p$, a Helm form factor, and the truncated Maxwellian DM velocity distribution. We assume a xenon target with $A=131$.\label{fig:dd_dRdER} }
\end{figure}

Finally, we can look at typical recoil spectra and the implication of direct detection constraints on DM models. Fig.~\ref{fig:dd_dRdER} shows the recoil spectra for a xenon target and typical DM masses that can be constrained by liquid xenon experiments. The exponential suppression derived above in Eq.~\ref{eq:gv_approx} can be seen, with most of the signal occurring at low recoil energies. Direct detection experiments thus stand to gain significantly in signal rate with reduced thresholds. Particularly for low mass DM $m_\chi \ll m_N$, the reduced mass $\mu_{\chi N} \approx m_\chi$ and the minimum velocity scales as $v_{\rm min} \approx \sqrt{m_N E_R/(2 m_\chi^2)}$. This results in a steep drop in the rate for low mass DM. Meanwhile, the minimum velocity for $m_\chi \gg m_N$ is $v_{\rm min} \approx \sqrt{E_R/(2m_N)}$, so that the recoil spectra are independent of the DM mass in this limit.

The spectrum shown assumes a spin-independent cross section $\sigma_n = 10^{-40}$ cm$^2$. This is totally excluded by direct detection experiments for masses in the 10 GeV -- $10^5$ TeV range. Fig.~\ref{fig:dd_wimps} shows the landscape of existing constraints for spin-independent scattering. The current strongest limits are from the XENON1T collaboration~\cite{Aprile:2018dbl}, with an exposure of almost $10^6$ kg-day. In the next decade, the sensitivity will improve by another 1-2 orders of magnitude (from experiments such as DarkSide, PandaX, LZ, XENONnT). The sensitivity at low $m_\chi$ drops rapidly due to the energy thresholds in the experiments, $E_R^{\rm th} \gtrsim 5-10$ keV. The sensitivity at high $m_\chi$ drops because the number density of DM drops as $1/m_\chi$. 
The constraints on spin-dependent cross sections are weaker, and at the level of $10^{-41}$ cm$^2$ for coupling to neutron spin and $10^{-40}$ cm$^2$ for coupling to proton spin (see for example Ref.~\cite{Aprile:2019dbj}).

\begin{figure}[t]
\includegraphics[width=0.75\textwidth]{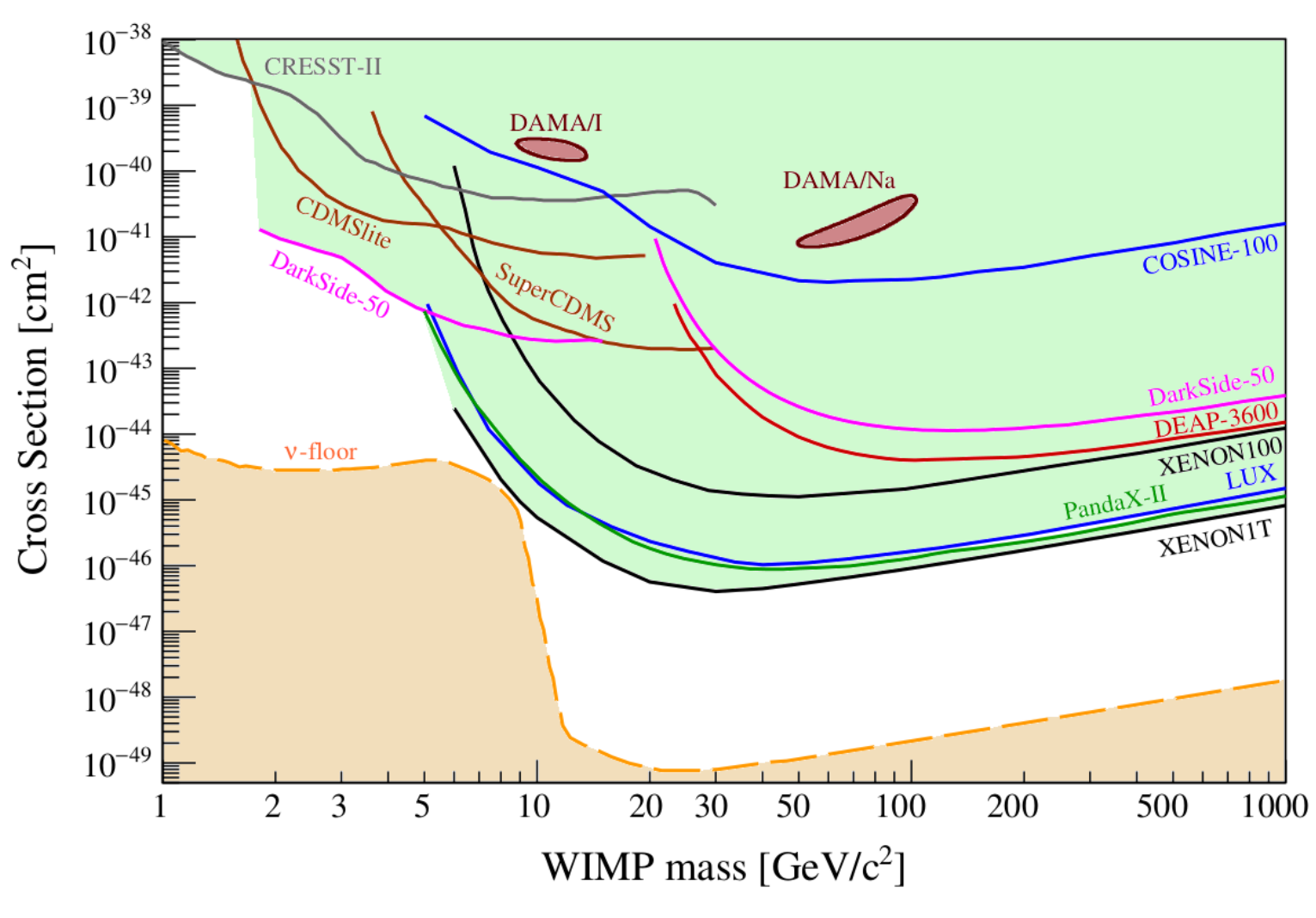}
\caption{Reproduced from Ref.~\cite{Schumann:2019eaa}, constraints on spin-independent DM-nucleus scattering as well as a few best-fit regions from DAMA. The shaded region labeled $\nu$-floor indicates the approximate DM mass and cross section where the sensitivity starts to become limited by the irreducible background from neutrino coherent scattering~\cite{Billard:2013qya}.\label{fig:dd_wimps} 
Note that these exclusion limits do not exclude {\emph{all}} cross sections above the lines, since for sufficiently large $\sigma_n$ the DM scatters too much in the Earth and loses energy~\cite{Kavanagh:2017cru,Emken:2018run,Hooper:2018bfw}.  }
\end{figure}

What are the implications for models of WIMPs? Using Eq.~\ref{eq:sigma_vector} and setting $C_V = g_w^2 \sim 0.1$ where $g_w$ is a weak gauge coupling, the typical cross section for a 100 GeV DM candidate scattering through a vector coupling to the $Z$-boson is
\begin{align}
	\sigma_n^V \simeq 10^{-37} \, {\rm cm}^2 \ .
\end{align}
For Higgs exchange, using Eq.~\ref{eq:sigma_higgs} and setting $C_S = 0.01$, 
\begin{align}
	\sigma_n^S \simeq 6 \times 10^{-47} \, {\rm cm}^2 \ .
\end{align}
For some DM candidates (such as Majorana fermion DM), scattering through the vector coupling to the $Z$ boson is highly suppressed or zero. For $Z$-exchange and an axial vector coupling with $C_A = 0.1$, we use Eq.~\ref{eq:sigma_axial} and obtain
\begin{align}
	\sigma_n^A \simeq 6\times 10^{-38}  \, {\rm cm}^2 \ .
\end{align}
We see that $Z$ exchange with $O(1)$ gauge couplings is strongly excluded, and Higgs exchange with couplings   $C_S = 0.01-0.1$ are being probed in tonne-scale experiments. (Searches for Higgs invisible decays at the LHC provide a complementary constraint on $C_S$ if $m_\chi < m_h/2$, where currently the bounds are comparable for $m_\chi \sim$ 10 GeV~\cite{Hoferichter:2017olk,Aaboud:2018sfi,Sirunyan:2018owy}.) Of course, it is possible to simply lower the couplings to get around the direct detection bounds. The difficulty with this approach is that the annihilation cross sections become correspondingly weaker as well, meaning it is difficult to obtain the relic abundance by thermal freezeout. This makes the simplest WIMP story more and more difficult to achieve. There are important exceptions, including the case of a pseudo-Dirac electroweak doublet fermion. This is also known as a Higgsino, and the thermal relic can be obtained when the mass is about 1.1 TeV. The lightest neutral fermion in the doublet is a good DM candidate, and it does not have diagonal couplings to the $Z$ boson at tree level (which would allow for elastic scattering). The predicted cross section for spin-independent scattering is $\sigma_n \lesssim 10^{-48}$ cm$^2$ due to a cancellation between different contributions~\cite{Hill:2013hoa}. This candidate is extremely challenging to search for in direct detection, indirect detection, and collider experiments. Thus, there are some WIMP candidates that will remain viable for a time to come (see also Refs.~\cite{Beneke:2016jpw,Krall:2017xij,Rinchiuso:2018ajn}). 

There are a number of compelling reasons to go beyond nuclear recoils in the 10--100 GeV regime. First, the current approach rapidly loses sensitivity to low mass DM. As we saw, there are many candidates in the sub-GeV regime which reproduce some of the successes for weak-scale candidates. Furthermore, thermal freezeout for light DM requires the presence of addition of new light mediators. If the mediators are sufficiently light, this further modifies the recoil spectrum to favor lower energy recoils. The factors of $1/m_Z^2$ become replaced with $1/(q^2 - m_V^2) \simeq 1/(2 m_N E_R)$ for $m_V \ll | \bfq|$. Direct detection with lower thresholds opens up the possibility of exploring new classes of low mass candidates.  We turn to the methods and proposals for low-threshold detection next.

\clearpage

\section{New directions in direct detection \label{sec:newdd} }

As highlighted in the previous lecture, conventional direct detection via nuclear recoils becomes much less effective for sub-GeV dark matter due to the much smaller recoil energies. Furthermore, we motivated the presence of light mediators for sub-GeV dark matter, which can further shift the recoil spectra to lower energies. 

Direct detection of sub-GeV dark sectors thus requires a different approach, both theoretically and experimentally. An important difference on the theoretical side is the necessity of incorporating many-body physics. Our idealized picture of a free, nearly-at-rest particle just waiting for an incident DM particle to provide some excitement is hardly reflective of reality. Dark matter direct detection requires significant target mass, and typical targets are liquids or solid state materials. Nuclei and electrons in such materials have enormous interactions with each other! The typical inter-atomic spacing in a material is given by $\sim$\AA $ \sim 1$/keV, and similarly, the size of an atom is also given by $\sim $\AA. For scattering of WIMPs, where the momentum transfer may be as large as $|\bfq| \sim m_\chi v \gtrsim $ MeV, such ``condensed matter'' effects are usually neglected -- but remember that they can still be quite important for experimental detection, and that many-body effects played a role in the nuclear form factors. For light DM, the momentum transfer can be on the keV scale or less, and inter-atom interactions can play an important role in the DM interaction. 

\begin{figure}[t]
\includegraphics[width=0.67\textwidth]{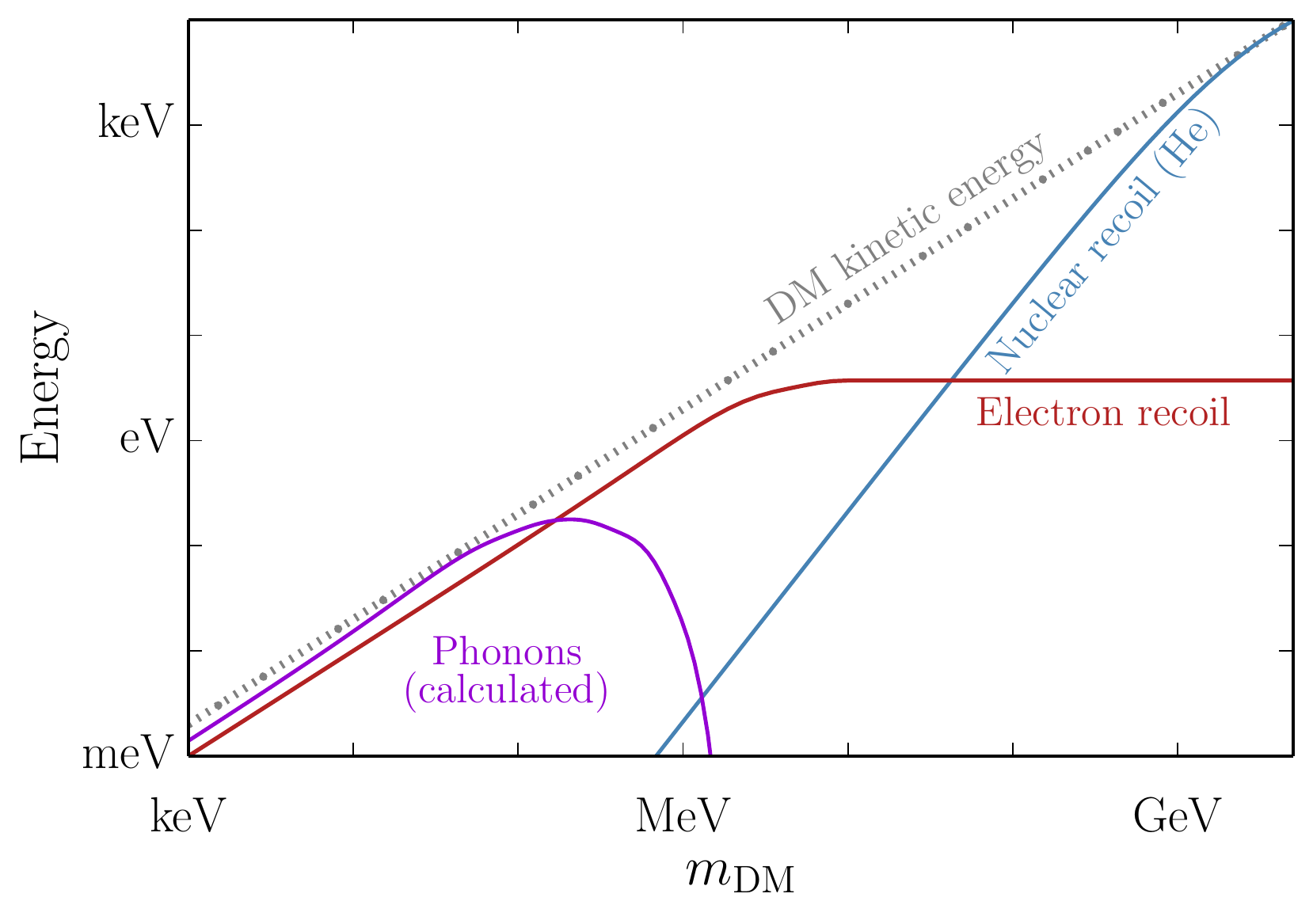}
\caption{A schematic comparison of the total DM kinetic energy (dotted, gray) with the energy deposited in a regular nuclear recoil (blue, taking a helium target), the typical energy deposited in an electron recoil (red), and the typical energy in phonon excitations (purple). Note that the phonon excitation case  cuts off above DM masses above an MeV only because the current theoretical calculations focus on sub-MeV DM; see Section~\ref{sec:phonons} for more details. \label{fig:dd_energies} }
\end{figure}

In this lecture, we will provide an introduction to the theoretical tools used in some proposed detection scenarios. This is a field that is quickly developing, and there will be no attempt to give a complete review. For a recent white paper that provides more details on experimental status, see Ref.~\cite{Battaglieri:2017aum}. Rather, the discussion here will center on toy models to provide intuition for the relevant many-body physics.  Fig.~\ref{fig:dd_energies} illustrates the kinematics of DM scattering on free nuclei, as compared with electron scattering or phonon scattering. One of the goals of this lecture will be to understand this figure, and why these alternate scattering modes allow for more of the DM kinetic energy to be deposited in a scattering. Note that we will use the term `nuclear recoil' (NR) as shorthand for scattering off free nuclei at rest, even though phonon scattering is also sensitive to DM that couples dominantly to nuclei.

\subsection{Electron recoils}

Compared to nuclear recoils, the most important features of electron scattering are that the electron is much lighter than the proton, and its velocity in a typical material is much faster than the DM velocity. Consider an atomic target: the electron is in a bound state with a typical wavefunction size of $R_{\rm Bohr}= 1/(\alpha m_e)$. The typical momentum of this electron is therefore $k_e \simeq 1/R_{\rm Bohr} = \alpha m_e$  and its velocity $v_e \simeq \alpha$. This electron is in an energy eigenstate but not in a momentum eigenstate, and computing scattering requires knowledge of the specific wavefunctions in the material. Before we describe the form factors needed to account for these wavefunctions below, let us repeat the exercise of DM scattering kinematics for free electrons.

As before, the energy deposited $\omega$ in a DM scattering is
\begin{align}
	\omega = \frac{\bfp \cdot \bfq}{m_\chi}  -\frac{\bfq^2}{2 m_\chi} = E_e(\bfk') - E_e(\bfk),
	\label{eq:omega_ER}
\end{align}
but now $E_e(\bfk')$ is an in-medium energy eigenvalue. $\bfq = \bfp-\bfp'$ is again the momentum transfer. As a toy model, we will just consider the situation of scattering off a free electron with initial velocity $v_{e,i} = \alpha$. Then energy conservation gives the condition:
\begin{align}
	\frac{\bfp \cdot \bfq}{m_\chi} = \frac{\bfk \cdot \bfq}{m_e} - \frac{\bfq^2}{2 \mu_{\chi e}},
\end{align}
where $\mu_{\chi e}$ is the DM-electron reduced mass. Since the electron velocity is much larger than the DM velocity, $v_{e,i} = |\bfk|/m_e \gg v_{\chi,i} =  |\bfp|/m_\chi$, let us start by dropping the left-hand-side of the equation above, and estimating the momentum transfer as $\qmod \simeq \mu_{\chi e} v_{e,i}$. We can then estimate the recoil energies in two different DM mass regimes:
\begin{itemize} 
	\item $m_\chi \gtrsim m_e$: for heavier DM, the momentum transfer is approximated by $\qmod \simeq m_e v_{e,i}$. This is the expected result for an electron that recoils off a heavy particle. Then the deposited energy using Eq.~\ref{eq:omega_ER} is 
	\begin{align}
		\omega \simeq m_e v_{e,i} v_{\chi,i} \approx \ \textrm{few eV}, 
	\end{align}
	approximately independent of $m_\chi$. It is worth emphasizing that this is just a typical energy deposition: high $\omega$ is possible in a given material, such as for ionization of atoms in Xe ($\omega > 12$ eV). 
	\item $m_\chi \lesssim m_e$: if we were to use the result $\qmod \simeq \mu_{\chi e} v_{e,i} \approx m_\chi v_{e,i}$, we would actually find in Eq.~\ref{eq:omega_ER} that the energy deposited is negative. The momentum transfer must be bounded from above by $\qmod \simeq m_\chi v_{\chi,i}$ for a positive $\omega$: this effectively tells us that the final state phase space becomes much more restricted. Again, this is the expected result in the limit of a light DM particle scattering off the much more massive electron. The allowed energy deposition is
	\begin{align}
		\omega \simeq \frac{1}{2} m_\chi v_{\chi,i}^2 ,
	\end{align}
	meaning potentially all of the DM kinetic energy can be deposited.
\end{itemize}
These arguments are translated schematically into the red curve in Fig.~\ref{fig:dd_energies}, which shows the typical electron recoil energy as a function of DM mass. Of course, electrons are not actually free in the targets, nor are they in momentum eigenstates, so that arbitrarily high electron velocities are possible (though with suppressed probability). In the remainder of this subsection, we work out the scattering rate explicitly.

As a simple calculation where the result can be obtained analytically, we will consider ionization of a Hydrogen atom from DM-electron scattering. We will take the outgoing states to consist of free electrons; technically, the outgoing electron wavefunction experiences a distortion in the vicinity of the nucleus, but we will neglect this correction for the moment.  
The scattering cross section can be written as~\cite{Essig:2011nj,Essig:2015cda}
\begin{align}
	(\sigma v)_j  =  \int \frac{ d^3\bfp'}{(2\pi)^3} \frac{ d^3 \bfk'}{(2\pi)^3} \, (2\pi) \delta(\Delta E_e - \omega ) \,   \frac{| {\cal M}_{\rm free}|^2}{16 m_\chi^2 m_e^2} \,    | f_{j \to \bfk'}( \bfq ) |^2
\end{align}
with $| {\cal M}_{\rm free}|^2$ is the matrix element squared for scattering off a free electron (e.g. Eq.~\ref{eq:electron_scattering}). The index $j$ labels a possible initial state, while the outgoing states are labelled by a wavevector $\bfk'$. (As before, $\bfp'$ is the outgoing DM momentum.) The form factor accounts for the wavefunctions of initial and final electron states, and is given by
\begin{align}
	 f_{j \to \bfk'}( \bfq ) = \sqrt{V} \int d^3 \bfr \ \psi_j(\bfr) \,  \psi_{\bfk'}^* (\bfr) \, e^{i  \bfq \cdot \bfr},
\end{align}
where $V$ is a volume to account for wavefunction normalization, to be explained below. We are not used to seeing volume factors in our conventional QFT calculations, but including them will be extremely helpful in matching up with quantum mechanics conventions (used more often in other parts of physics!).

Before diving into a calculation for bound electrons, we first use the above result for scattering off free electrons at rest.

\subsubsection{Free electron cross section}

For free electrons, we can use initial and final plane-wave wavefunctions. Then the above result reproduces the expectation for scattering of free particles:
\begin{align}
	 |f_{j \to \bfk'}( \bfq )|^2 &=  \left| \sqrt{V} \int d^3 \bfr \, \frac{e^{i \bfk \cdot \bfr}}{\sqrt{V}} \, \frac{e^{-i \bfk' \cdot \bfr}}{\sqrt{V}} \, e^{i  \bfq \cdot \bfr} \right|^2 \\
	 &= \left[ (2\pi)^3 \delta^{(3)}\left( \bfk + \bfq - \bfk' \right) \right]^2 / V \\
	 &= (2\pi)^3 \delta^{(3)}\left( \bfk + \bfq - \bfk' \right)
\end{align}
where we have included the normalization of the wavefunctions in terms of the volume factor $V$, and take the large volume limit in the end. We have also used that the $(2\pi)^3 \delta^{(3)}({\bf 0})/V \rightarrow 1$ for a momentum-space delta function. This reproduces the standard free particle scattering cross section. 

For scattering via the dark photon mediator for instance, the spin-averaged matrix element is (see definitions and discussion around Eq.~\ref{eq:electron_scattering}):
\begin{align}
	|{\cal M}|^2 = \frac{16 g_\chi^2 \kappa^2 e^2 m_\chi^2 m_e^2}{((q_\mu^2) - m_V^2)^2} \approx  \frac{16 g_\chi^2 \kappa^2 e^2 m_\chi^2 m_e^2}{(\qmod^2 + m_V^2)^2}.
\end{align}
Here we used that for scattering of non-relativistic particles, $q^0 \ll \qmod$.
We will rewrite the matrix element in terms of an average cross section times DM form factor:
\begin{align}
	 \frac{\mu_{\chi e}^2 |{\cal M}|^2}{16 \pi m_\chi^2 m_e^2} \equiv \bar \sigma_e F_{\rm DM}^2(q), \ \ \ F^2_{\rm DM}(q) \equiv  \left( \frac{ (\alpha m_e)^2 + m_V^2}{ \qmod^2 + m_V^2} \right)^2
	 \label{eq:DMformfactor}
\end{align}
where $F^2_{\rm DM}(q)$ absorbs the $q$-dependence of the matrix element. 

For initial electrons at rest $(k=0)$, the final electron has energy $E_R = q^2/(2m_e)$ and the cross section is then given by
\begin{align}
	\sigma v &=  \frac{\bar \sigma_e}{\mu_{\chi e}^2} \int \frac{ d^3\bfq}{4\pi}  \, \delta\left(\tfrac{q^2}{2\mu_{\chi e}}   - \tfrac{p q \cos \theta}{m_\chi} \right) \, F_{\rm DM}^2(q) \\
	&=  \frac{\bar \sigma_e}{\mu_{\chi e}^2} \int d \ln E_R \frac{  E_R m_e}{2v}  \, \theta\left(v - \tfrac{q}{2\mu_{\chi e}} \right) \, F_{\rm DM}^2(q) \label{eq:sigmae_free}
	\end{align}
where we first changed variables $d^3 \bfp' = d^3 \bfq$ and then $q \, dq = dE_R m_e$. The initial relative velocity is just $p/m_\chi$ in this case. We can write the above as
\begin{align}
	 \sigma = \bar \sigma_e \int^{E_R^{\rm max}} \frac{  dE_R }{E_R^{\rm max}} \, F_{\rm DM}^2(q)	 \ ,
\end{align}
which is similar to the result for DM-nucleon scattering except with an additional DM scattering form factor. This motivates the definition above of $\bar \sigma_e$, as when $F_{\rm DM}^2(q)	=1$ we have $\sigma = \bar \sigma_e$. 

\subsubsection{Atomic ionization cross section}

Turning now to bound electrons with a more complicated wavefunction, the scattering form factors are no longer so simple to evaluate. We will start by assuming free unbound wavefunctions for the outgoing states. It is then convenient to consider a spherical harmonic basis for the final states.  Let's take a brief aside to rewrite the sum over plane wave final states appearing in the cross section:
\begin{align}
	\int  \frac{ d^3 \bfk'}{(2\pi)^3} e^{- i {\bfk'} \cdot \bfr} e^{i {\bfk'} \cdot \bfr'} 
	\label{eq:planewavesum}
\end{align}
as a sum instead over final states labelled by  wavenumber $k'$, and angular momentum numbers $\ell', m'$. A plane wave can be expanded in terms of spherical wavefunctions
\begin{align}
	e^{i {\bfk} \cdot \bfr} = 4 \pi \sum_\ell i^\ell j_\ell(k r) \sum_{m=-\ell}^\ell Y_{\ell m}^*(\theta_{\bfk}, \phi_{\bfk})Y_{\ell m}(\theta_{\bfr}, \phi_{\bfr})  
	\label{eq:planewave_expand}
\end{align} 
where we used the normalization that $\int d\Omega \, | Y_{\ell m}(\theta, \phi)|^2 =1$ and $\int dr \, r^2 j_\ell(kr) j_\ell(k' r) = \pi/(2 k^2) \delta(k-k')$. Using these results, we can rewrite Eq.~\ref{eq:planewavesum} as 
\begin{align}
	\sum_{\ell', m'} \int  \frac{ d k'}{(2\pi)^3} (k')^2  \left( 4\pi j_{\ell'} (k' r) Y^*_{\ell' m'}(\theta_{\bfr}, \phi_{\bfr}) \right) \times 
			\left( 4\pi j_{\ell'} (k' r') Y_{\ell' m'}(\theta_{\bfr'}, \phi_{\bfr'}) \right)
\end{align}
In this basis, the cross section to excite state $j$ can be written as
\begin{align}
	(\sigma v)_j  = \int \frac{ d^3\bfp'}{(2\pi)^3} \frac{ (k')^2 \, d k'}{(2\pi)^3} \, (2\pi) \delta(\Delta E_e - \Delta E_\chi ) \,   \frac{| {\cal M}_{\rm free}|^2}{16 m_\chi^2 m_e^2} \,   \sum_{\ell', m'} | f_{j \to k', \ell', m'}( \bfq ) |^2
	\label{eq:sigj_electron}
\end{align}
where the form factor is
\begin{align}
	 |f_{j \to k' ,\ell', m'}( \bfq )|^2 &=  \left|  \int d^3 \bfr \, \psi_j(\bfr) \,  \Psi_{k', \ell', m'}^*  \, e^{i  \bfq \cdot \bfr} \right|^2 \\
	 & =  \left|  \int d^3 \bfr \, \psi_j(\bfr) \, 4\pi j_{\ell'} (k' r) Y^*_{\ell' m'}(\theta_{\bfr}, \phi_{\bfr}) \, e^{i  \bfq \cdot \bfr} \right|^2.
\end{align}

The final state electron energy $E_R = (k')^2/2 m_e$, so that we can turn the above into a differential cross section and sum over all bound initial states $j$:
\begin{align}
	\frac{d (\sigma v)_j}{d \ln E_R}  = \frac{\bar \sigma_e}{16 \pi \mu_{\chi e}^2} \int  d^3\bfq  \, \delta(\Delta E_e - \Delta E_\chi ) \,  F_{\rm DM}^2(q) \,  \underbrace{ \sum_{\ell', m'} \frac{ 2(k')^3}{(2\pi)^3} | f_{j \to k', \ell', m'}( \bfq ) |^2 }_{\text{\small $|f^{\rm ion}_j(k', q)|^2$}} .
	\label{eq:ionformfactor}
\end{align}
where again we used $d^3 \bfp' = d^3 \bfq$. Comparing to Eq.~\ref{eq:sigmae_free}, we find an extra phase space integral (since the initial state is not a momentum eigenstate) as well as the presence of an atomic ionization form factor $|f^{\rm ion}_j(k', q)|^2$ for the initial state $j$.

\begin{figure}[t]
\centering
\begin{tabular}{m{0.3\linewidth}m{0.7\linewidth}}
\includegraphics[width=0.3\textwidth]{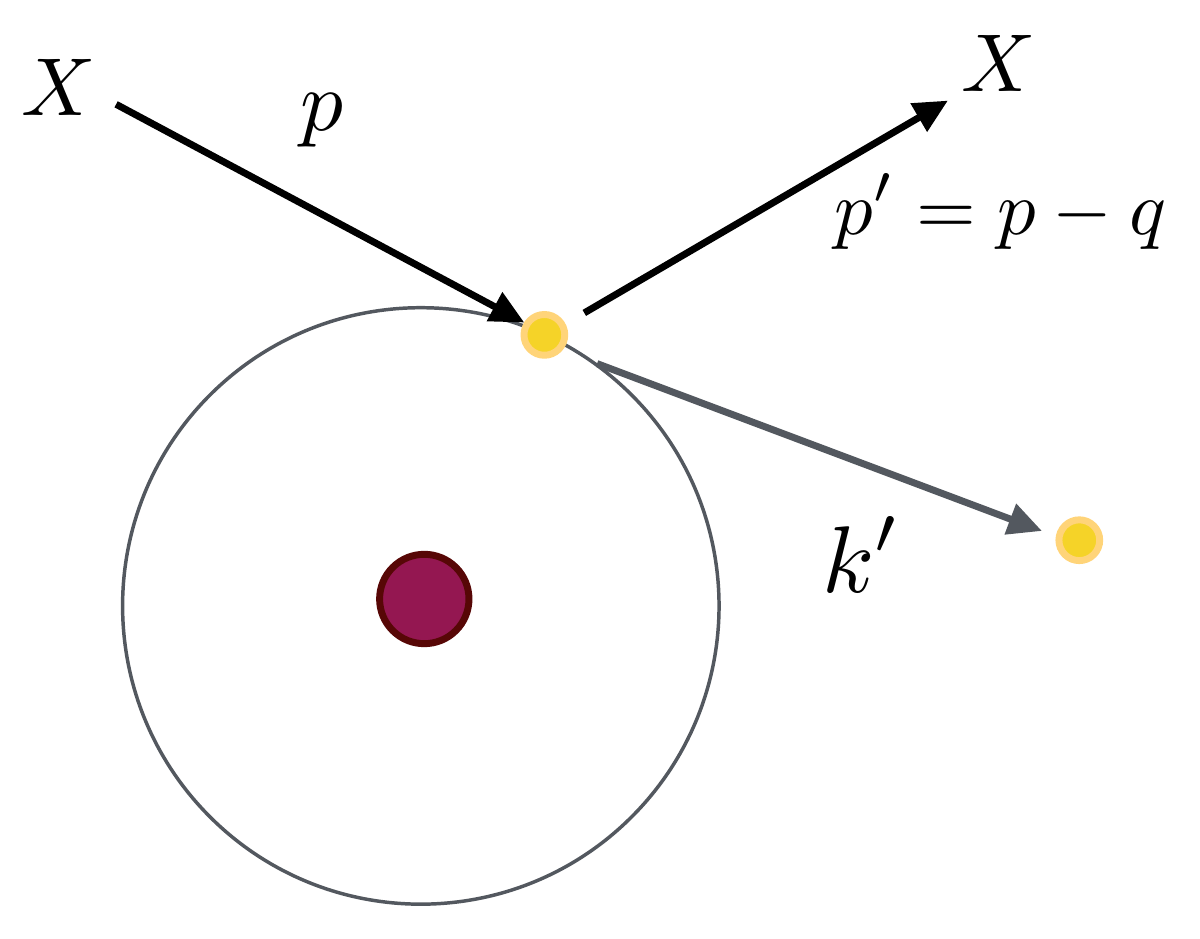} &
\includegraphics[width=0.65\textwidth]{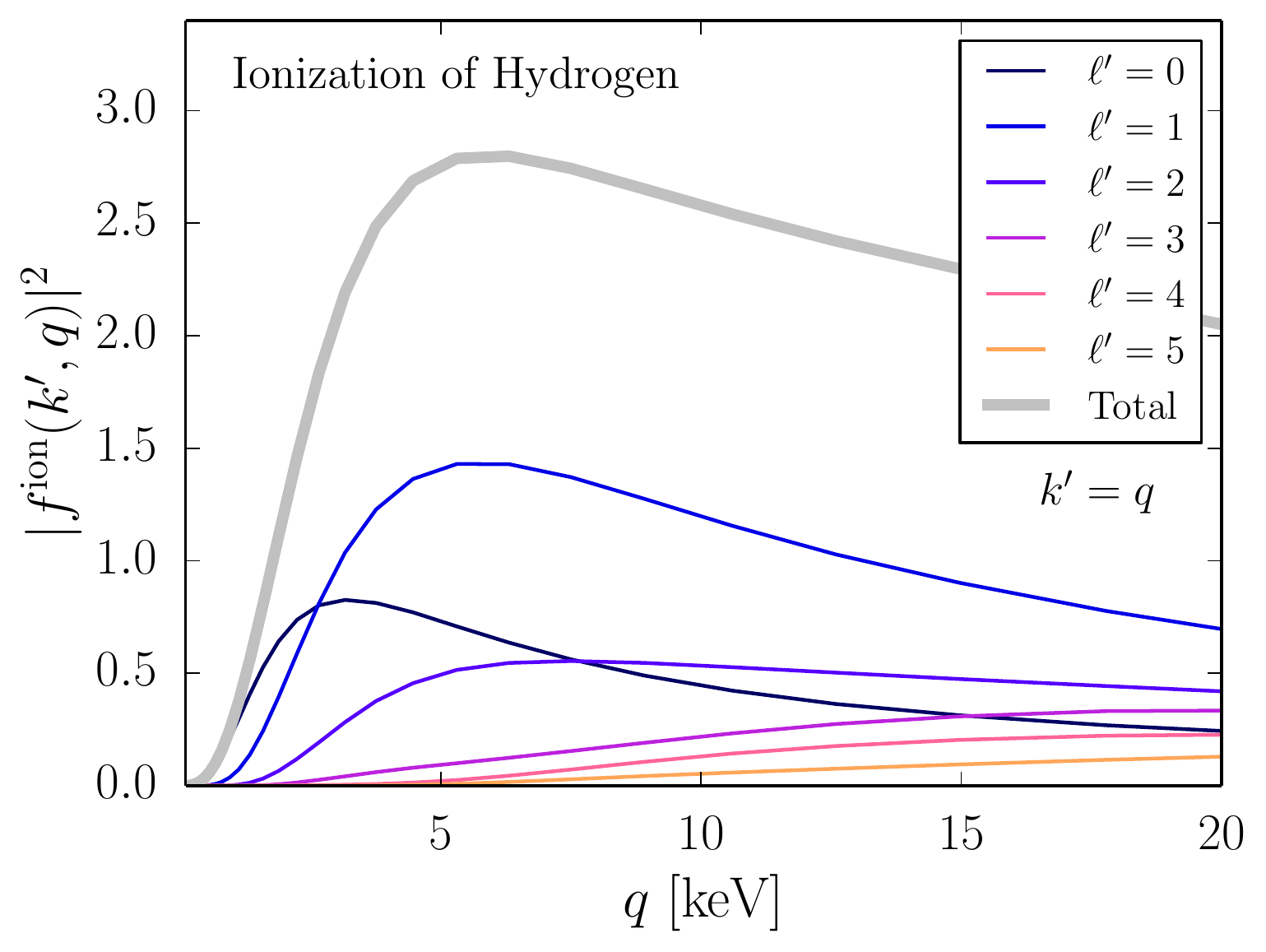}
\end{tabular}
\vspace{-0.5cm}
\caption{  The right plot shows the atomic ionization form factor (Eq.~\ref{eq:ionformfactor}) computed for ionizing an electron in the ground state of Hydrogen, and assuming free outgoing electron states $\Psi_{k', \ell', m'}$. The largest values are for $k' \sim q\sim \alpha m_e =$ 3.7 keV due to the wavefunction of the bound electron. \label{fig:ERformfactor} }  
\end{figure}

As an explicit example, consider scattering from the ground state of Hydrogen into an outgoing $\ell=0, m=0$ state with momentum $k'$. The exercise is left to the reader to compute this. The result is
\begin{align}
	|f^{\rm ion} (k',q)|^2 = \frac{64 (k' r_e)^3}{\pi \left[ r_e^4 ( k'^2 - q^2)^2 + 2 r_e^2 ( k'^2 + q^2) + 1 \right]^2}
\end{align}
where $r_e = 1/(\alpha m_e) \approx 1/3.7$ keV is the Bohr radius. This form factor is peaked at  $q \sim k' \sim 1/r_e$: the outgoing electron wavefunction is correlated with the momentum transfer, while the size of both quantities is set by the initial electron momentum $\sim \alpha m_e$. The total form factor including other $\ell$ states is shown in Fig.~\ref{fig:ERformfactor} for $k' = q$, displaying a similar behavior.\footnote{Note that we still assume free outgoing states here. The presence of the ion distorts the wavefunction of the outgoing state, such that the basis in Eq.~\ref{eq:planewave_expand} needs to be modified. For further details on this, see Refs.~\cite{Essig:2011nj,Essig:2017kqs}.} The typical outgoing electron energy is then $k'^2/(2 m_e) \sim 10$ eV (note that the total deposited energy is the sum of the binding energy plus the outgoing kinetic energy).  

This makes explicit the arguments made at the beginning of this section, where we used free electron kinematics and found that the typical energy deposited is around a few eV.  This also motivates the definition of the scattering cross section $\bar \sigma_e$ in Eq.~\ref{eq:electron_scattering}, since a typical scale for $\qmod \sim \alpha m_e$. To determine the rate in an experiment, there are a few more points to consider. We must also integrate over the DM velocity distribution, discussed already for nuclear recoils in Sec.~\ref{sec:dm_fv}. However, we can obtain a very rough estimate of the rate below. As estimated around the end of Lecture 3, there are thermal relic (and freeze-in) candidates with sizable DM-electron cross sections, $\bar \sigma_e \simeq 10^{-37} {\rm cm}^2$. Then assuming $m_\chi = 10\, \MeV$ and taking a number of atoms per kg of $N_T \simeq 10^{25}$/kg, we have
\begin{align}
	R \sim N_T \frac{\rho_\chi}{m_\chi} \bar \sigma_e \, v_0 \simeq 50-100 \ \textrm{events/kg/day} \, .
\end{align}
Thus, an exposure of around a kg-day is sufficient to observe a DM signal! This can be compared to exposures of $10^5-10^6$ kg-day for experiments targeting heavy DM.  However, there are new difficulties in detecting low mass DM, namely the issue of lower energy thresholds and control of backgrounds at such low energies.

The calculation performed here can be applied to more realistic targets such as liquid Xenon with a similar minimum ionization energy of $\approx 12\, \eV$ (there are differences in ionization from liquid Xenon compared to a free Xenon atom, but so far this effect has been assumed to be small). Experimental sensitivity to events with a single or a few ionization electrons has been demonstrated with XENON10, XENON100~\cite{Essig:2017kqs} and more recently DarkSide~\cite{Agnes:2018oej}, thus allowing for a novel way to probe light DM scattering off of electrons with an existing experimental setup~\cite{Essig:2012yx}. Some of these constraints are shown in Fig.~\ref{fig:sigmae} for both scattering through a heavy vector mediator and a light vector mediator. The maximum energy deposited per scattering is set by the DM kinetic energy in the halo, which means that for an $O(10)$ eV ionization threshold we require DM of mass $m_\chi \gtrsim {\rm few} \, \MeV$ (see Fig.~\ref{fig:dd_energies}).  This is why the XENON10/100 constraints extend at best down to $m_\chi \approx$ 5 MeV. We note also that the XENON10/100 constraints were limited by the large low-energy backgrounds, and that these experiments were not designed to search for DM-electron scattering. A focused effort in this direction could potentially do much better.

\subsubsection{Scattering in a lattice, experimental status, and the sub-MeV frontier}

Detecting lighter DM requires systems with smaller thresholds for detectable electron excitations. Note that we should really think of these as quasiparticle excitations, where the quasiparticle is a nearly free state that behaves much like the electron (but could have a smaller effective mass, for instance). We highlight here some of the proposals and considerations. 

\begin{figure}[t]
\begin{minipage}[h]{\textwidth}
$\vcenter{\hbox{   \includegraphics[width=0.53\textwidth]{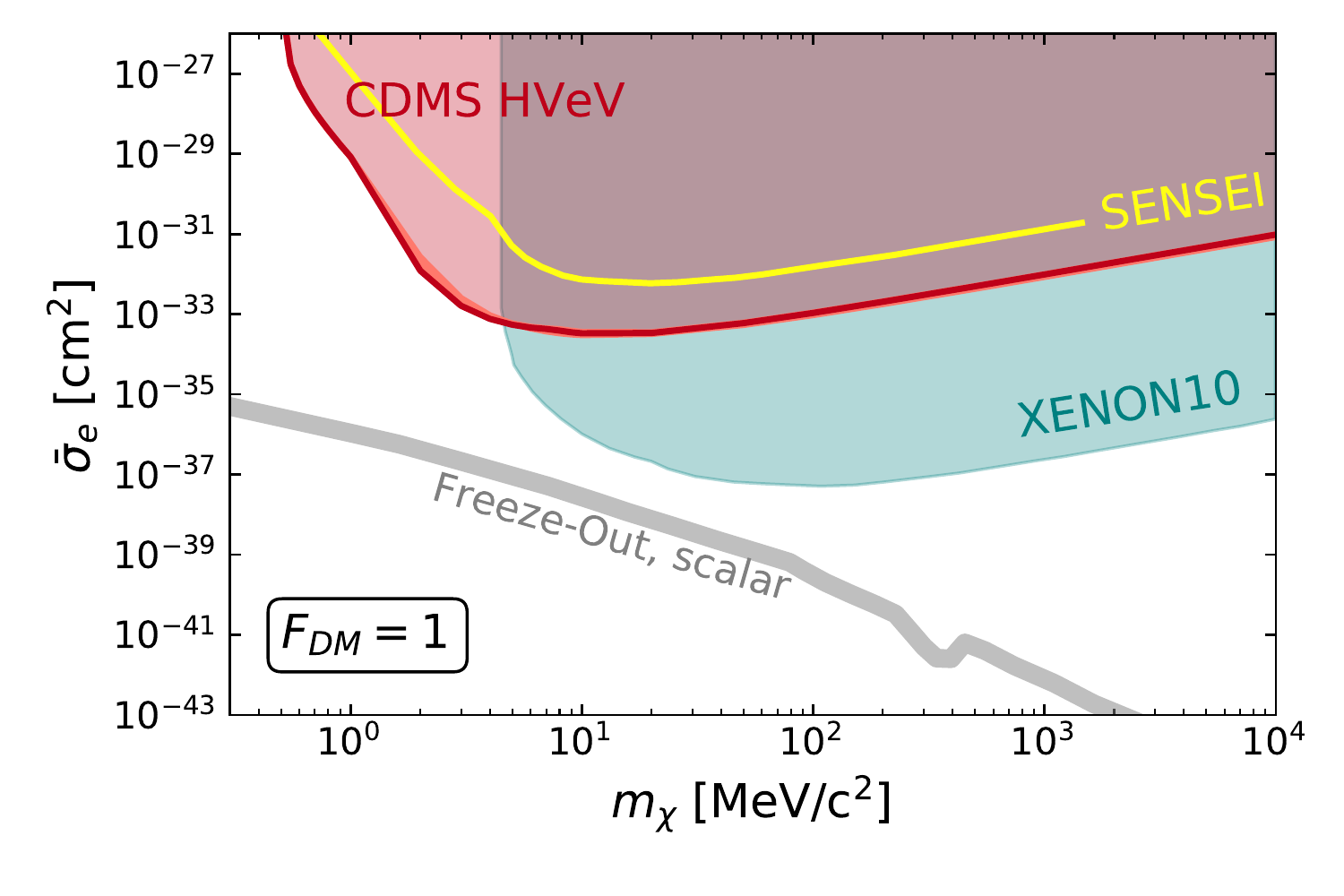} }}$
\hspace{-0.5cm}
$\vcenter{\hbox{   \includegraphics[width=0.45\textwidth]{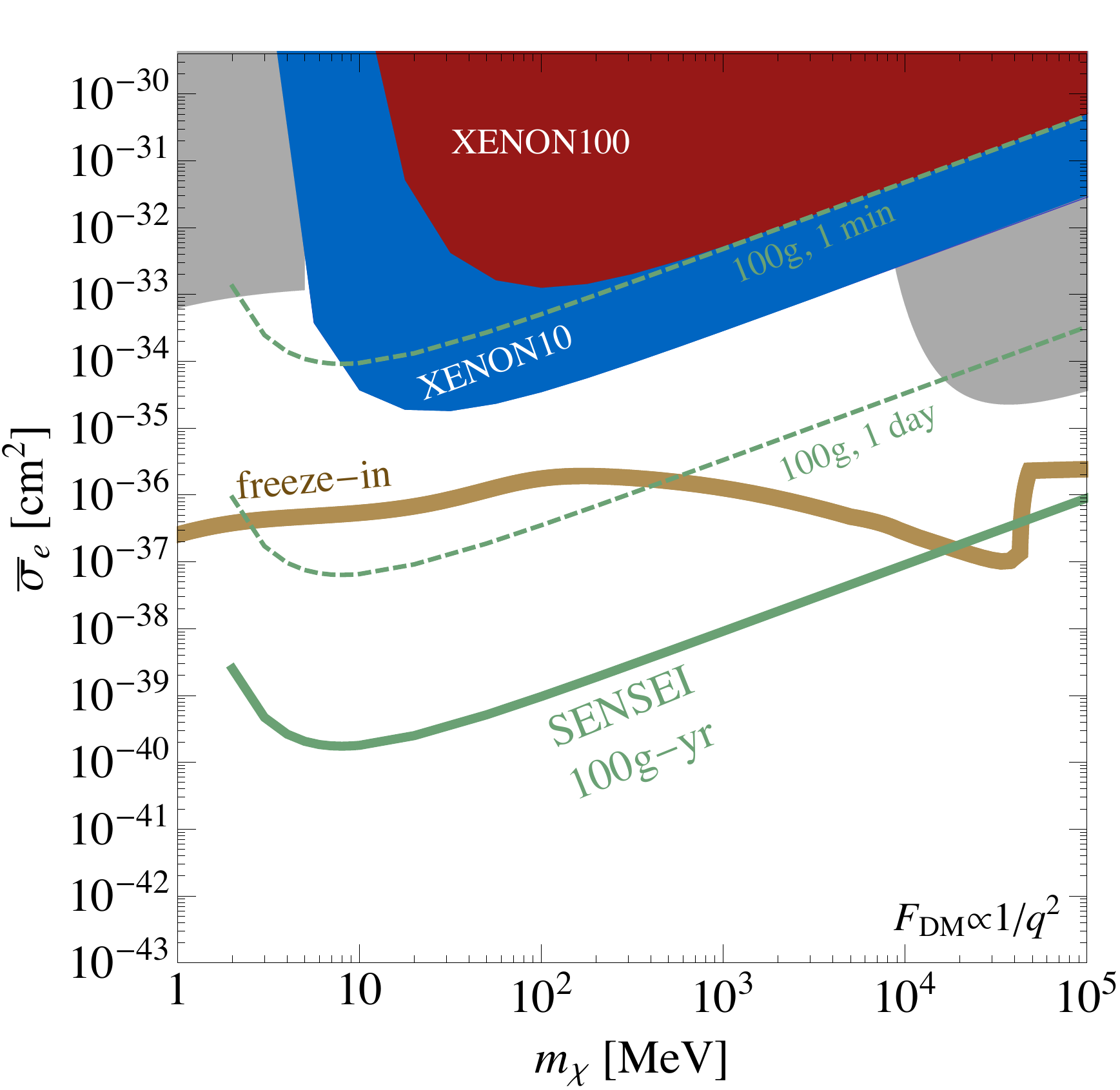} }}$
\end{minipage}
\caption{ \label{fig:sigmae} ({\bf left}) Reproduced from Ref.~\cite{Agnese:2018col}, recent bounds on DM-electron scattering from SuperCDMS~\cite{Agnese:2018col} and SENSEI~\cite{Crisler:2018gci}. The plot assumes scattering through a heavy vector mediator. ({\bf right}) Plot courtesy of Tien-Tien Yu. Projections for DM-electron scattering  through a light vector mediator assuming different exposures in SENSEI~\cite{Tiffenberg:2017aac,Abramoff:2019dfb}. For more discussion of the freeze-out and freeze-in benchmarks in the figures and how the projections were obtained, see Ref.~\cite{Essig:2015cda}.  In both figures, some existing constraints obtained from XENON10 and XENON100 data~\cite{Essig:2012yx,Essig:2017kqs} are shown; there are also constraints from DarkSide~\cite{Agnes:2018oej}, which lie roughly in between those from XENON.
}
\end{figure}

First of all, reducing the threshold to $\sim$ eV is possible by using semiconductor targets such as Ge and Si, and allow sensitivity to DM masses as low as $\sim$ MeV. Semiconductors have typical electron band gaps of around an eV, and the development of first-principles techniques in calculating band structures has made DM-electron scattering a tractable (if computationally intensive) problem. The detailed theoretical treatment of DM-electron scattering in a crystal can be found in Ref.~\cite{Essig:2015cda}. Compared to Eq.~\ref{eq:sigj_electron}, the difference is that we need a crystal form factor summing over initial and final electron states in the lattice. The electron wavefunctions can be obtained the first-principles techniques mentioned above. Relatedly, we also cannot assume that the outgoing electron energy is a simple function of the momentum. 

The SuperCDMS~\cite{Agnese:2018col} and SENSEI~\cite{Tiffenberg:2017aac,Abramoff:2019dfb} collaborations have recently demonstrated experimental sensitivity to single electron excitations in a silicon target. Fig.~\ref{fig:sigmae} shows constraints from surface runs of the two experiments, with only a small exposure. With dedicated underground runs and larger exposure, it is possible to test interesting thermal histories in the near future (see right panel of the figure for some projections). For theoretical studies of the sensitivity of semiconductors to different DM models, see also Refs.~\cite{Essig:2015cda,Hochberg:2016sqx}. 

Going to lower thresholds requires considering materials with smaller band gaps for electron excitations.  There are proposals to use superconductors~\cite{Hochberg:2015pha,Hochberg:2015fth,Hochberg:2016ajh} and Dirac materials~\cite{Hochberg:2017wce}, both of which have a gap of around $\sim $ meV and have long lived quasiparticle excitations. If experimental sensitivity to such low energy excitations can be achieved, this would allow detection of DM as light as $\sim$ keV. Similar to the case of semiconductors, DM-electron scattering is heavily reliant on the in-medium electron band structure.  The intersection of many-body physics with DM physics is a growing area of research, and applying developments in condensed matter or materials science could lead to further improvements in sensitivity to DM mass/models, reduction in backgrounds, or other advantages.

There is one more important consideration for proposals targeting sub-MeV DM scattering through a kinetically-mixed vector mediator. So far we have accounted for the material/many-body effects only in the wavefunctions of initial and final states, while keeping the matrix element $|{\cal M}_{\rm free}|^2$. However, as discussed in Section~\ref{sec:vector} and detailed in Appendix~\ref{sec:inmedium}, the in-medium electromagnetic polarization tensor can have a significant impact on DM and mediator interactions. For instance, in a metal or a superconductor, electromagnetic interactions are screened by the Thomas-Fermi screening length $\lambda_{\rm TF}$, where $1/\lambda_{\rm TF} \simeq$ few keV for a typical materials. These effects are thus most important for DM below an MeV, where the possible momentum transfers are keV and below. The screening leads to a rough suppression in DM-electron scattering rates by $\sim (|\bfq| \, \lambda_{\rm TF})^4 \sim (m_\chi v_0 \, \lambda_{\rm TF})^4$.   The screening is significantly reduced in Dirac materials~\cite{Hochberg:2017wce}, which have a linear dispersion for electrons near the Fermi surface. Another way to obtain sensitivity to vector-mediated scattering is to use a polar materials~\cite{Knapen:2017ekk,Griffin:2018bjn}; since these are semiconductors or insulators, the screening is also fairly mild compared to metals. In polar materials, the DM scattering does not create electron excitations (which have a large gap) but  instead creates phonon excitations of energies $1-100$ meV. We turn to phonon excitations next.

\subsection{Phonon couplings \label{sec:phonons}}

Phonons are collective excitations of the atoms in a solid or liquid, with the simplest examples being sound waves in a medium. There are a number of motivations for considering DM couplings to phonons:
\begin{itemize}
	\item Phonons are the relevant degrees of freedom for sufficiently low momentum transfer (typically $\lesssim$ keV, depending on the material), so we are forced to consider DM-phonon interactions for sub-MeV DM.
	\item The excitation energy of phonons is small: for sound waves there is no energy gap, while there are also gapped phonon modes with typical energies of 10--100 meV. This is well-suited again for low mass DM, since more of the DM kinetic energy can be deposited (Fig.~\ref{fig:dd_energies}).
	\item Since the phonons are excitations of the atoms, a DM-nucleon interaction can create a phonon excitation, allowing an alternate way to probe DM models that does not rely on a DM-electron coupling.
	\item A number of DM experiments exploit phonon collection as one of the channels for energy depositions from DM~\cite{Strauss:2017woq,Agnese:2017njq,Arnaud:2017usi}, with increasing sensitivity to low energy events in recent years. (Since phonons are also thermally produced, these experiments must be operated in cryogenic environments with $T < $ Kelvin.)
\end{itemize}

In this section, we will focus on phonons in a solid state material -- the collective displacements of the atoms in a crystal lattice. We will work through a derivation of the simplest DM-phonon coupling in order to illustrate the basic idea of phonons excitations from DM scattering, and then summarize some of the results in the literature. Before diving in, we clarify one point: what does it really mean when we talk about displacements in a crystal? For a given atom, we can think of the nucleus plus the most tightly bound electrons as being relatively unaffected by the presence of the other atoms. Meanwhile, the outer shell electrons of the atom interact with electrons of neighboring atoms, giving rise to the delocalized electron wavefunctions and complex electron band structure in a material. Therefore, although we usually refer to displacements as atomic displacements, we are really thinking of these as ionic displacements of the nucleus and inner shell electrons.

\subsubsection{Acoustic phonons}

Introductions to phonon modes can be found in standard references~\cite{Kittel}, and for completeness we provide a brief review. The system we will consider is a 1D regular lattice of $N$ atoms of mass $M$, shown in the top left of Fig.~\ref{fig:lattice_1D}. All of the atoms are identical, so the unit cell has a size $a$ (the lattice spacing) and contains one atom. Each atom $i$ has a possible displacement from its equilibrium position, denoted by $u_i$. 
The Hamiltonian for this system is modeled with an effective potential for the relative displacements of neighboring atoms:
\begin{align}
	H = \sum_i \frac{1}{2} M \dot u_i^2  + \frac{1}{2} k_{\rm eff} (u_{i+1} - u_i)^2 + ...
\end{align}
where the $...$ are possible higher order terms in the displacements. Those terms could lead to three-phonon couplings, for instance.

\begin{figure}[t]
\begin{minipage}[h]{\textwidth}
$\vcenter{\hbox{   \includegraphics[width=0.48\textwidth]{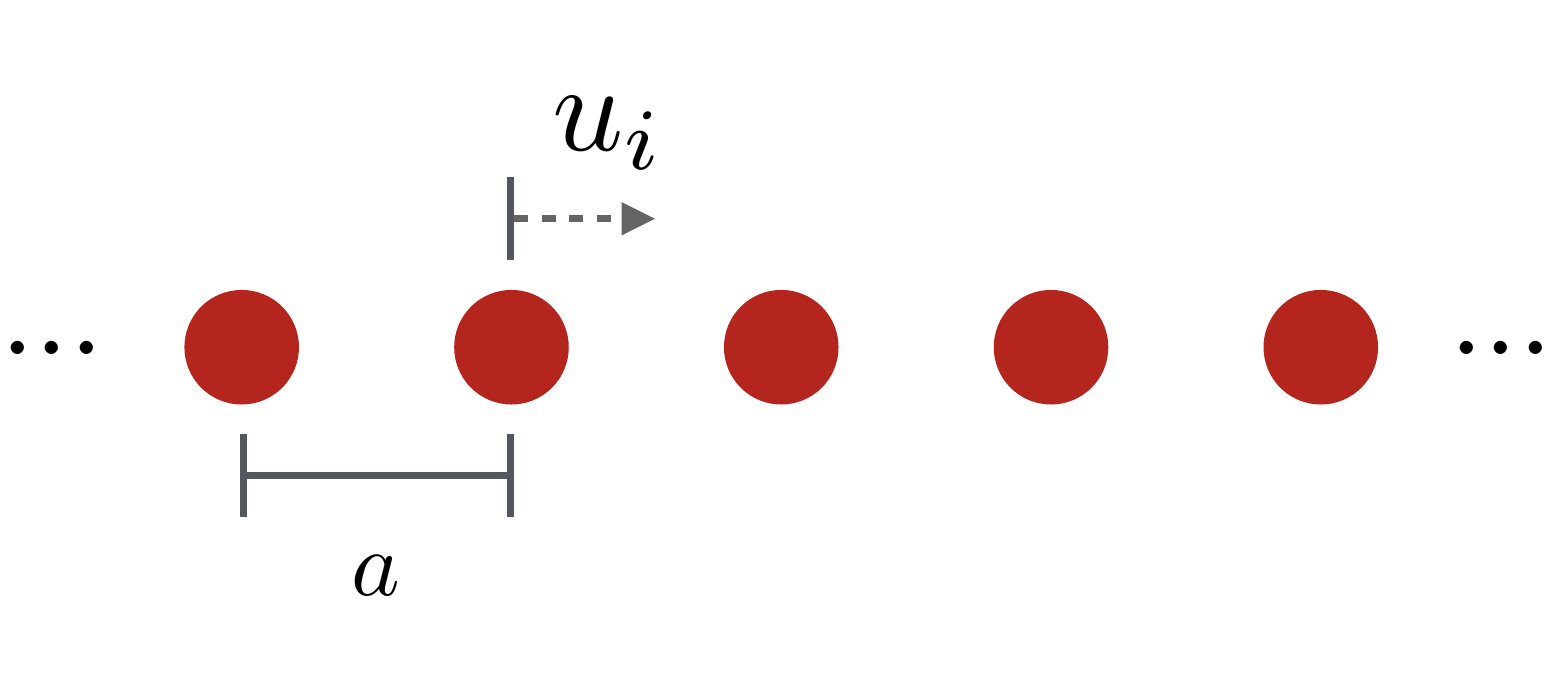} }}$
\hspace{0.2cm}
$\vcenter{\hbox{   \includegraphics[width=0.46\textwidth]{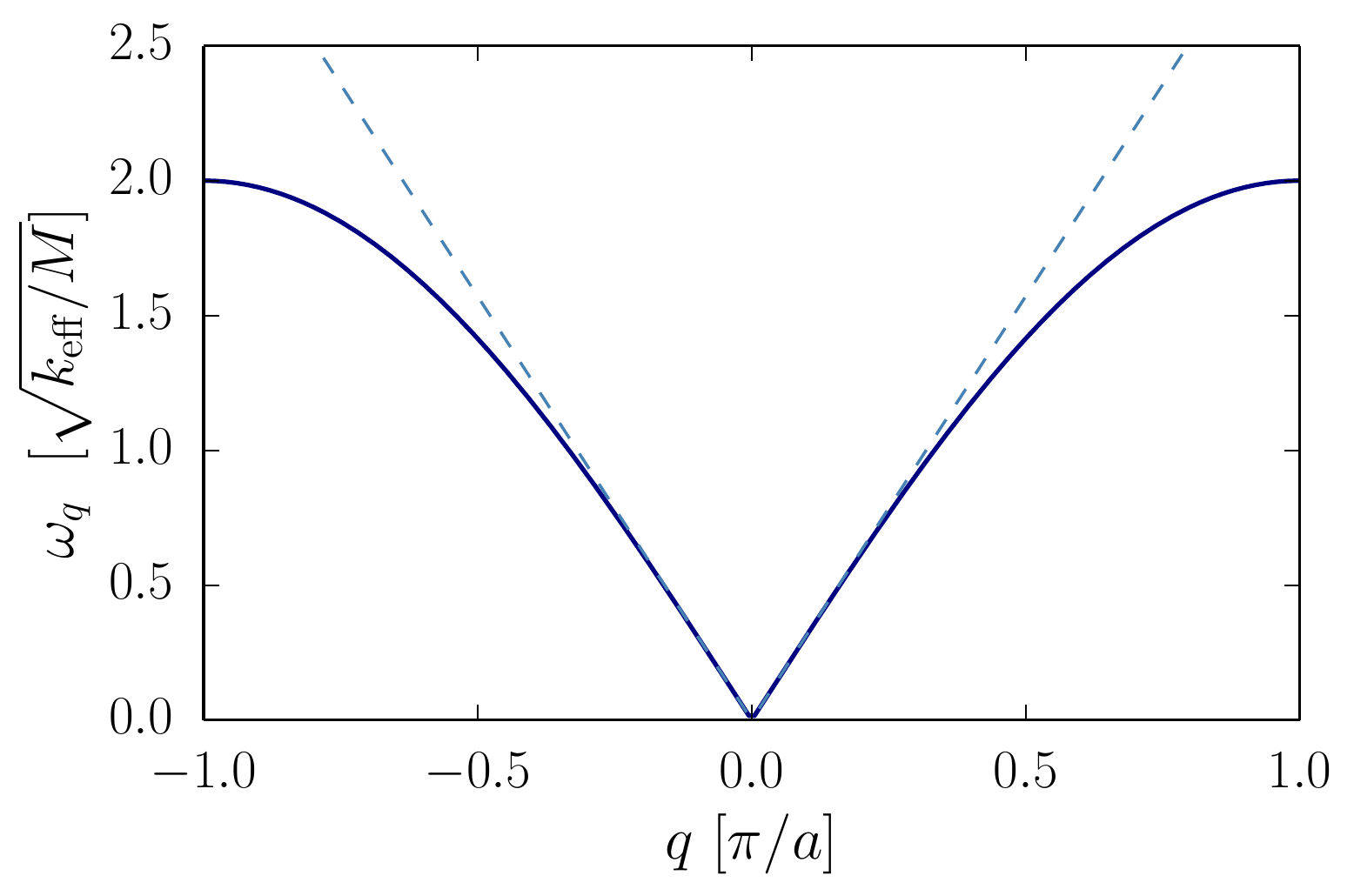} }}$
\end{minipage}
\caption{ \label{fig:lattice_1D} ({\bf left}) A 1D lattice of atoms of mass $M$  has a single longitudinal acoustic phonon branch.  ({\bf right}) The dispersion relation of the acoustic phonon is shown over the first Brillouin Zone; near $q = 0$, the phonon has a linear dispersion with slope given by the speed of sound.}
\end{figure}

We will take the continuum limit for this system, or equivalently consider long-wavelength excitations, so that the displacement field $u(x,t)$ is a function of position and time. Writing $\Delta x = a$, the sum over positions $i$ can be replaced by an integral over $x$:
\begin{align}
	H = \int dx \ \left( \frac{1}{2} \rho \dot u^2 + \frac{1}{2} \tilde k_{\rm eff} (\nabla u )^2 \right)
\end{align}
where $\rho = M/a$ is the mass per unit length, $\tilde k_{\rm eff} = a k_{\rm eff}$, and we have replaced the finite difference with a gradient. Equivalently, introducing a minus sign for the potential, the Lagrangian for the system is:
\begin{align}
	{\cal L} = \int dx \ \left( \frac{1}{2} \rho \dot u^2 - \frac{1}{2} \tilde k_{\rm eff} (\nabla u)^2 \right).
\end{align}
This describes a free, massless particle with linear dispersion, $\omega_q = c_s |\bfq| = c_s q$, and speed of sound $c_s = \sqrt{ \tilde k_{\rm eff}/\rho} = \sqrt{a^2 k_{\rm eff}/M}$. The mode is otherwise known as an acoustic phonon, with an energy that goes to zero in the $q\to 0$ limit. This reflects the fact that the acoustic phonon is a Goldstone boson associated with spontaneous breaking of translational symmetry. As $q\to 0$, all atoms are displaced by the same amount and the arrangement is physically equivalent to the original ground state.

The quantized displacement field (phonon) is written in a standard way,  in terms of a mode expansion in the interaction picture:
\begin{align}
	u(x,t) = \sum_j   \frac{1}{\sqrt{2 N a \rho \omega_{q_j}}}  \left( \hat a_{q_j} \ e^{i {q_j} x - i \omega_{q_j} t} + \textrm{h.c.} \right) \\
	 =  \frac{\sqrt{a N}}{\sqrt{\rho}} \int \frac{dq}{(2\pi)} \frac{1}{\sqrt{2\omega_q}}   \left( \hat a_q \ e^{i q x - i \omega_q t} + \textrm{h.c.} \right)
	 \label{eq:1D_modedecomp}
\end{align}
with creation and annihilation operators $\hat a_q, \hat a^\dagger_q$ satisfying the commutation relations 
$[ \hat a_q, \hat a^\dagger_{q'} ] = \delta_{q,q'}$. In the first line, we have written the expansion in terms of a discrete sum -- this reflects the actual discrete lattice, with $q_j = 2 \pi j/(a N)$ for a lattice of length $aN$. In the second line, we have given the continuum limit result. (If the factor of $\sqrt{aN}$ looks funny in the continuum limit, it is because in the QFT convention we typically normalize the creation and annihilation operators differently, $[ \hat a_q, \hat a^\dagger_{q'} ] = 2 \pi \delta(q - q') \to V = aN$ when $q=q'$, see Appendix~\ref{sec:convention}. This would remove the factor.) As an exercise, you can check that plugging the above expansion into the Hamiltonian gives, in the discrete limit:
\begin{align}
	H = \sum_j \omega_{q_j} \left( \hat a^\dagger_{q_j}  \hat a_{q_j} + \frac{1}{2} \right).
\end{align}
We have found that the excitations are described by phonons created by the $\hat a^\dagger_{q}$ operator.

The energy eigenvalues $\omega_{q_j}$ can be solved for exactly, see Ref.~\cite{Kittel}. The right panel of Fig.~\ref{fig:lattice_1D} shows the exact phonon band structure over the first Brillouin zone (BZ), accounting for the lattice periodicity. In the long-wavelength limit  $q \ll \pi /a$, we see the linear dispersion expected for Goldstone modes. The size of the first BZ is set by $\pi/a$;  for a typical material $a \sim {\rm few}$ \AA, and so $ q \lesssim $ keV in the first BZ.

Going to three spatial dimensions, the displacement field becomes a vector field ${\bf u}({\bf x}, t)$ and there are more phonon branches corresponding to the additional degrees of freedom. 
The band structure for a possible direct detection material (GaAs) is shown in the right panel of Fig.~\ref{fig:GaAsphonons}, with the phonon energies plotted against a specific path within the first BZ. Here the $q\to0$ limit is labelled by the $\Gamma$ point, and there are three acoustic modes for a 3-dimensional lattice  -- two transverse acoustic (TA) branches where the oscillation of the atoms is perpendicular to ${\bf q}$ and one longitudinal acoustic (LA) branch. The acoustic modes have linear dispersions for sufficiently small $q$, although the sound speed is different for transverse vs. longitudinal phonons.  Because the unit cell for GaAs contains more than one atom, it is seen that there are also optical phonon branches (LO and TO), which we discuss further below. Before introducing this feature, we turn to the coupling of a DM candidate with acoustic phonons.

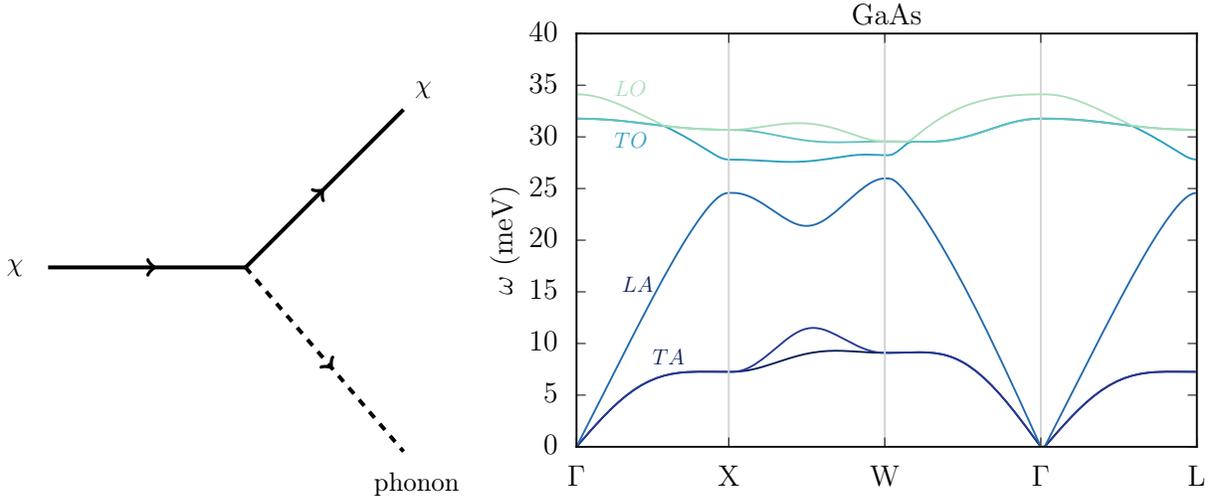
\begin{figure}[t]
\begin{tikzpicture}[line width=1.5 pt, scale=1.75]
	\draw[scalar] (0,0)--(1.2,-1.4);
	\draw[fermionbar] (1.2,1.2)--(0,0);
	\draw[fermion] (-1.5,0)--(0,0);
	\node at (1.3,-1.65) {phonon};
	\node at (1.35,1.35) {$\chi$};
	\node at (-1.75,0) {$\chi$};	
\end{tikzpicture}
\includegraphics[width=0.6\textwidth]{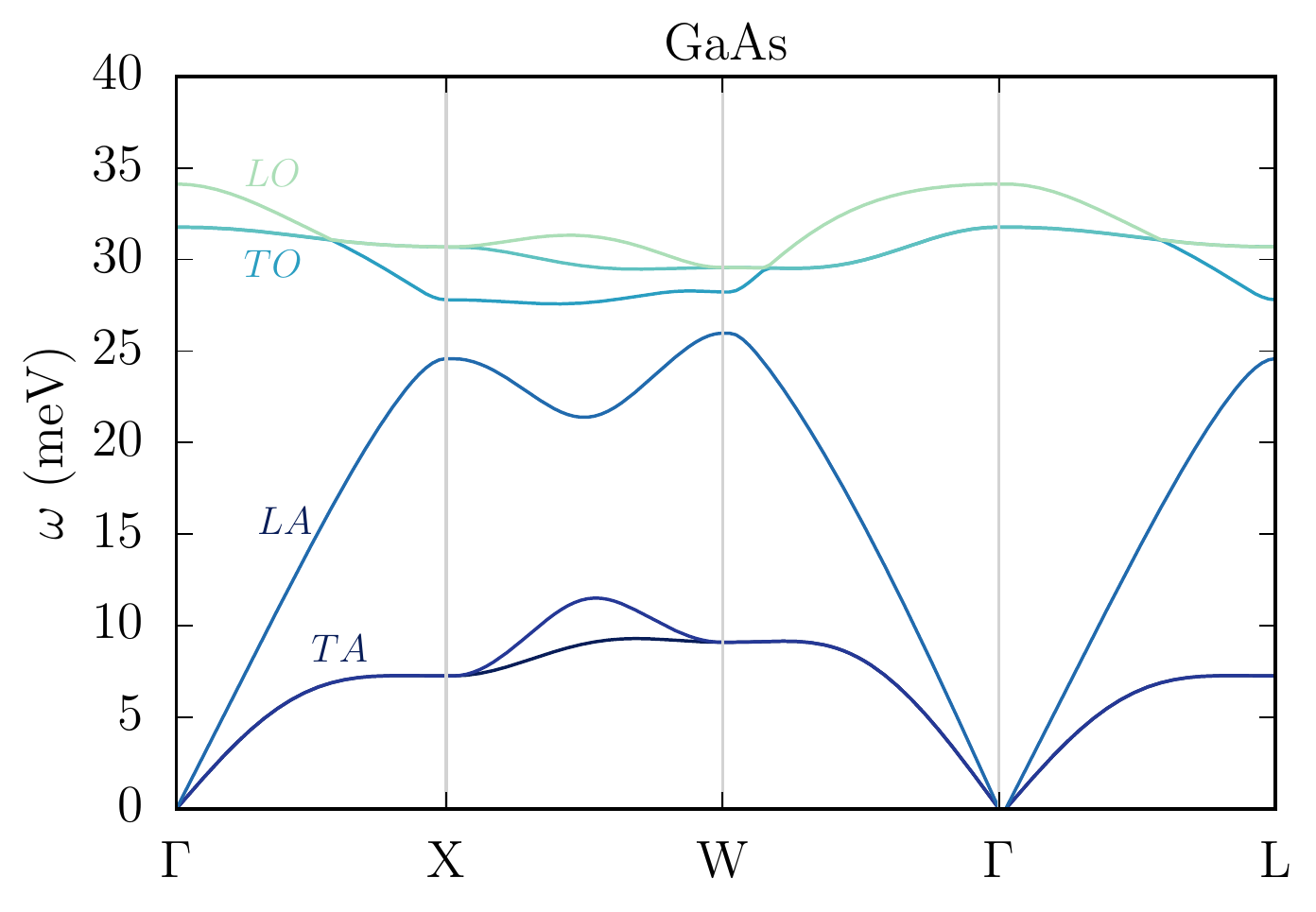}
\caption{({\bf left}) A typical diagram for exciting a single phonon. ({\bf right}) A representative phonon band structure. The $\Gamma$ point is where $\bfq=0$, and the acoustic modes are indicated by TA (transverse acoustic) and LA (longitudinal acoustic). The TO and LO branches are the optical phonon modes. Reproduced from Ref.~\cite{Knapen:2017ekk}. \label{fig:GaAsphonons} } 
\end{figure}

\subsubsection{Dark matter coupling to acoustic phonons}

To describe the interaction of a DM candidate with a phonon, we must first determine the underlying DM interactions with electrons and nuclei. To start, a simple case is fermion DM with a scalar mediator coupling to nucleons:
\begin{align}
	{\cal L} \supset - m_\chi \bar \chi \chi -\frac{1}{2} m_\phi^2 \phi^2 - y_n (\bar n n + \bar p p) \phi - y_\chi \phi \bar \chi \chi
\end{align}
for both protons and neutrons. For this model, DM scattering into acoustic phonons has been studied in polar materials \cite{Knapen:2017ekk,Griffin:2018bjn} and superfluid helium \cite{Schutz:2016tid,Knapen:2016cue,Acanfora:2019con}. Of course, superfluid helium is not a crystal, but acoustic phonons are also present and a number of techniques are similar.

Since we are working in the non-relativistic limit, we can use the interaction Hamiltonian to compute the scattering rate. This will also make it straightforward to write the operators in terms of phonon creation and annihilation operators $\hat a^\dagger_{\bf q}, \hat a_{\bf q}$ for 3D momentum vector ${\bf q}$. For a massive mediator $\phi$, the Hamiltonian for the DM-nucleon interaction can be written in the non-relativistic limit as
\begin{align}
	\frac{y_\chi y_n}{m_\phi^2}  \int d^3 \bfr \ \bar \chi \chi (\bar n n + \bar p p).
\end{align}
The operator $\bar n n + \bar p p$ becomes the number density of nucleons in the non-relativistic limit.
Since we are dealing with low momentum transfer $q$, we can sum coherently over all of the nucleons in a given nucleus $N$ with mass number $A$. The interaction can then be written in terms the number density operator $\rho(\bfr)$ for the nuclei:
\begin{align}
	{\cal H}_I = \frac{A y_\chi y_n}{m_\phi^2}  \int d^3 \bfr \ \bar \chi \chi \, \bar N N   \to  \frac{A y_\chi y_n}{m_\phi^2}  \int d^3 \bfr \ \bar \chi \chi \, \rho(\bfr),
	\label{eq:Hint0}
\end{align}
with additional terms if multiple types of nuclei are present. We consider just one type of atom here. 

In order to make a connection with the QM language used in the condensed matter literature, we perform a mode expansion for the fermionic $\chi$ field into creation and annihilation operators $\hat b^\dagger_{{\bf p},s}, \hat b_{{\bf p},s}$ for the DM, as well as analogous operators $\hat c^\dagger_{{\bf p},s}, \hat c_{{\bf p},s}$ for antiparticles. Explicitly, we use the expansion in the interaction picture:
\begin{align}
	\chi({\bf r}, t) =  \sqrt{V} \int \frac{d^3 \bfp}{(2\pi)^3} \frac{1}{\sqrt{2 \omega_{\bf p} } } \sum_s \left( \hat b_{{\bf p},s} \, u^s(p) \, e^{- i \omega_{\bf p} t + i {\bf p } \cdot {\bf r} } \, + \, \hat c^\dagger_{{\bf p},s} \,v^s(p) \, e^{i \omega_{\bf p} t - i {\bf p } \cdot {\bf r} }  \right)
	\label{eq:Xmode}
\end{align}
and similar for $\bar \chi({\bf r},t)$. The factor of $\sqrt{V}$ is because we are using the non-relativistic normalization for the operators where $\{ \hat b_{{\bf p},s}, \hat b_{{\bf p'},s'}^\dagger \} = \delta_{\bf p, \bfp'}  \delta_{s,s'} $. For convenience, we consider only the scattering of DM particles in the discussion below; similar interaction terms are present for DM antiparticles.

The next step is to take the nonrelativistic limit and rewrite the interaction Hamiltonian in terms of a discrete sum for the finite volume $V$. The convention to go between continuum limit and discrete sum is given in Appendix~\ref{sec:convention} for convenience. This is roughly the reverse of the procedure in the acoustic phonon toy model above. (Our approach of starting from the relativistic Lorentz invariant theory may seem backwards, but we are assuming the typical reader is coming from a particle physics background.) Inserting Eq.~\ref{eq:Xmode} into Eq.~\ref{eq:Hint0} then gives 
\begin{align}
	{\cal H}_I = \frac{A y_\chi y_n}{m_\phi^2}   \frac{1}{V} \sum_{s} \sum_{\bf p, \bf p'} \hat b^\dagger_{{\bf p'},s} \, \hat b_{{\bf p},s} \, \int d^3 \bfr \, e^{\left[ i ({\bf p} - {\bf p}') \cdot \bfr  - i  (\omega_{\bf p} - \omega_{{\bf p}'}) t \right]} \, \rho(\bfr).
	\label{eq:HI_int}
\end{align}
The integral is proportional to a Fourier transform of the density field. Thus a DM scattering creates a density excitation with momentum ${\bfp - \bfp'} = \bfq$.  In the limit of a dilute gas of atoms, then the density excitation would be just a single atom that has been given a momentum kick. However, in treating a lattice or a liquid, we must translate the density excitation into the creation of a phonon.

Since the size of a typical nucleus is a few fm, much smaller than the interparticle spacing, we can approximate the density field as a sum over delta functions:
\begin{align}
	\rho(\bfr) = \sum_J \delta\left( \bfr - \bfr_J \right), \ \ \bfr_J \equiv \bfr_J^0 + \bfu_J
\end{align}
where we have decomposed $\bfr_J$, the position of ion $J$, into its equilibrium position $\bfr_J^0$ and displacement $\bfu_J(t)$. Inserting this density field into Eq.~\ref{eq:HI_int}, we can now write the Hamiltonian in terms of ${\bfu_J}$ and brazenly take the limit of small ${\bfu}_J$:
\begin{align}
	{\cal H}_I &= \frac{A y_\chi y_n}{m_\phi^2}  \frac{1}{V} \sum_{s} \sum_{\bf p, \bf p'}  \hat b^\dagger_{{\bf p'},s} \, \hat b_{{\bf p},s} \, e^{ - i  (\omega_{\bf p} - \omega_{{\bf p}'}) t }  \, \sum_J \exp \left[ i ({\bf p} - {\bf p}') \cdot (\bfr_J^0 + \bfu_J)   \right] \\
	& \to \frac{A y_\chi y_n}{m_\phi^2}   \frac{1}{V}  \sum_{s} \sum_{\bf p, \bf p'} \hat b^\dagger_{{\bf p'},s} \, \hat b_{{\bf p},s} \, e^{  - i  (\omega_{\bf p} - \omega_{{\bf p}'}) t }  \, \sum_J e^{  i ({\bf p} - {\bf p}') \cdot \bfr_J^0  } \, ({\bf p} - {\bf p}') \cdot  \bfu_J(t) \ .
	\label{eq:DMphonon_approx}
\end{align} 
In addition, we have dropped the piece independent of $\bfu_J$, since this doesn't lead to any scattering. 

For the next step, we will gloss over some of the details and instead refer the interested reader to Ref.~\cite{Griffin:2018bjn}. (A warning that the notation is not always identical, in order to simplify the presentation here.) The basic idea is the following: similar to the mode decomposition given in Eq.~\ref{eq:1D_modedecomp}, the displacement ${\bfu_J}$ can be expanded as a sum over creation and annihilation operators for phonons of momentum ${\bfq}$. In a 3D lattice with $N$ atoms, and including a phonon eigenvector ${\bf e}_{{\bf q},\nu}$ to account for the phonon polarization,
\begin{align}
	{\bf u}_J(t) =  \sum_{\bfq, \nu}  \frac{1}{\sqrt{2 N M \omega_{\bfq, \nu}}} \left( \hat a_{\bfq}  \,{\bf e}_{\bfq, \nu} \ e^{i {\bfq} \cdot {\bfr}^0_J - i \omega_{\bfq, \nu} t} + \textrm{h.c.} \right).
\end{align}
Here $\nu$ is an index labeling the different phonon mode branches and $\omega_{\bfq, \nu}$ is the energy of the phonon in branch $\nu$ with momentum $\bfq$.

Inserting the mode expansion of ${\bfu_J}$ into Eq.~\ref{eq:DMphonon_approx} and performing the sum over all $J = 1...N$ enforces the momentum conservation condition. We see that the interaction Hamiltonian includes a term where a phonon of momentum ${\bfq}$ is excited by the DM:
\begin{align}
	{\cal H}_I = \frac{A y_\chi y_n}{m_\phi^2}  \frac{ \sqrt{N}}{V}   \sum_{s} \sum_{\bf p, \bf q, \nu} \hat b^\dagger_{{\bf p}-{\bf q},s} \, \hat b_{{\bf p},s} \,  \hat a^\dagger_{\bf q} \  \frac{ {\bf q}\cdot  {\bf e}^*_{{\bf q},\nu} }{\sqrt{2 M \omega_{\bf q, \nu}} } \ e^{ - i  (\omega_{\bf p} - \omega_{{\bf p} - \bfq} - \omega_{\bfq, \nu}) t } 
\end{align}
where ${\bf q} = {\bf p} - {\bf p}'$ and ${\bf e}_{\bf q}$ is a normalized eigenvector for the phonon with momentum ${\bf q}$. At this point, you may be concerned about the factors of $\sqrt{N}/V$ where we will take the $V, N \to \infty$ limit. This will drop out in the end;  in computing the DM scattering, we will sum the matrix element squared over all excited phonon states, where the density of states scales as $N$ in the infinite volume limit.

For a given DM particle with momentum ${\bf p} = m_\chi \bfv$,  the above interaction Hamiltonian gives rise to a (spin-averaged) transition rate of 
\begin{align}
	\Gamma(\bfv)  &= 2\pi \, \frac{A^2 y_\chi^2 y_n^2}{m_\phi^4} \frac{N}{V^2}  \sum_{\bfq, \nu}  \frac{|{\bf q}\cdot  {\bf e}^*_{{\bf q},\nu}|^2}{2 M \omega_{\bfq, \nu}}   \,  \delta(\omega_{\bf p} - \omega_{{\bf p} - \bfq} - \omega_{\bfq, \nu}), \nonumber \\
	&= 2\pi \, \frac{A^2 y_\chi^2 y_n^2}{m_\phi^4} \frac{1}{\Omega}  \sum_{\nu}  \int \frac{d^3 {\bf q}}{(2\pi)^3}  \frac{|{\bf q}\cdot  {\bf e}^*_{{\bf q},\nu}|^2}{2 M \omega_{\bfq, \nu}}  \,  \delta(\omega_{\bf p} - \omega_{{\bf p} - \bfq} - \omega_{\bfq, \nu})
\end{align}
(i.e., Fermi's Golden Rule for a time-dependent interaction). Note that in the second line, we took the continuum limit and defined the unit cell volume $\Omega \equiv V/N \sim a^3$ for lattice spacing $a$.

The above result captures the qualitative form of the scattering rate, but is only approximate. Aside from the simplification in Eq.~\ref{eq:DMphonon_approx}, we did not distinguish between different atoms inside a primitive cell of a crystal lattice. The full derivation and expansion in small ${\bf u}_J(t)$ is more lengthy. It accounts for the fact that there is also a typical `zero-point' motion of the ions in the lattice, which is encapsulated in the Debye Waller factors $W_\alpha(\bfq)$. More generally, scattering off nuclei is typically described in terms of the {\emph{dynamic structure factor}} $S(\bfq, \omega)$. For scattering into an acoustic phonon, this quantity can be written in the long-wavelength limit as
\begin{align}
	S(\bfq, \omega) \approx \frac{1}{2 \omega_{\bfq,\nu} }  \left| \sum_\alpha \frac{A_\alpha}{\sqrt{M_{\alpha}} } {\bf q}\cdot  {\bf e}^\alpha_{\bfq, \nu} \, e^{-W_\alpha(\bfq)} \right|^2 \delta(\omega_{\bfq,\nu} - \omega)
\end{align}
where $\alpha$ runs over all atoms within a primitive unit cell. The complete results and derivation can be found in Ref.~\cite{Griffin:2018bjn}, which was based on the review by Schober~\cite{Schober2014}.
 
Summarizing our efforts here, the DM scattering rate for a general lattice is given in terms of the dynamic structure factor:
\begin{align}
	\Gamma = 2\pi \frac{(2 \pi b_\chi)^2}{m_\chi^2} \frac{1}{\Omega} \int_{\rm BZ} \frac{d^3 {\bf q}}{(2\pi)^3} \, S( {\bf q}, \omega),
\end{align}
where we have defined a DM scattering length $b_\chi$, such that the single nucleon cross section is given by $\sigma_n = 4 \pi b_\chi^2 = y_\chi^2 y_n^2 m_\chi^2/(\pi m_\phi^4)$ for sub-GeV fermion DM scattering through a massive mediator. In the equation above, it is implicit that $\bfq = \bfp - \bfp'$ is the momentum transfer and $\omega = \omega_{\bf p} - \omega_{{\bf p} - \bfq}$ is the energy deposited.

The total scattering rate per unit target mass is obtained by integrating over the DM velocity distribution,\begin{align}
	R &=  \frac{1}{\rho_T} \frac{\rho_\chi}{m_\chi} \int d^3 {\bf v} \, f({\bf v}) \,  \Gamma(\bfv) \\
	&= \frac{\rho_\chi}{m_\chi} \frac{ \sigma_n }{4\pi m_\chi^2} \int d^3 {\bf v} \, f({\bf v}) \left( \frac{1}{\rho_T \Omega}  \int_{\rm BZ}  d^3 {\bf q}  \, S( {\bf q}, \omega) \right)
	\label{eq:rate_Sqw}
\end{align}
where $\rho_T$ is the target mass density. 
This is the basis of the approach to DM-phonon scattering in a lattice used in Ref.~\cite{Griffin:2018bjn}. Eq.~\ref{eq:rate_Sqw} can also be applied to superfluid He, even though there is no lattice. A number of works have used analytic or numerical results for the dynamic structure factor to show that multiphonon creation in superfluid He could be exploited for sub-MeV DM scattering~\cite{Schutz:2016tid,Knapen:2016cue,Acanfora:2019con}.

Let's make a heuristic estimate of the total rate for scattering into a long wavelength acoustic phonon with energy $\omega = {\rm meV}$. The acoustic phonon dispersion is given by $\omega_{\bfq} = c_s q$ with $c_s \sim 10^{-5}$ in a typical material. Assuming each unit cell consists of a single atom of mass $M_N$, the integral over the dynamic structure factor can be simplified to
\begin{align}
	\int_{\rm BZ}  d^3 {\bf q}  \, S( {\bf q}, \omega) \simeq 4 \pi \int dq \, q^2 \frac{q A^2}{2 c_s M_N} \delta( c_s q - \omega) = \frac{4 \pi A^2}{M_N} \frac{\omega^3}{2 c_s^5} \ .
\end{align}
With this, we can get a rough idea of the rate for MeV DM scattering off a target with $A =100$ and $M_N = A m_n$:
\begin{align}
	R(\omega = {\rm meV}) &\simeq \frac{\rho_\chi}{m_\chi} \frac{ \sigma_n }{m_\chi^2}  \frac{1}{\rho_T \Omega}  \frac{A^2}{M_N} \frac{\omega^3}{2 c_s^5} \\
	& = 2 \,  \frac{{\rm events}}{{\rm kg}\cdot{\rm day}} \left( \frac{\sigma_n}{10^{-41} {\rm cm}^2} \right) \left( \frac{\MeV}{m_\chi} \right)^3 \left( \frac{10^{-5}}{c_s} \right)^5 \left( \frac{ 3 \, {\rm g/cm}^3 \times \textrm{\AA}^3}{\rho_T \Omega} \right) .
	\label{eq:rate_phonon}
\end{align}
We have neglected the DM velocity integral for this estimate. Remarkably, we find that a cross section of $\sigma_n \simeq 10^{-41} {\rm cm}^2$ could be probed with a kg-day exposure in a low-threshold experiment. Of course, this assumes zero background and requires detectability of $\sim$ meV phonon excitations, which is an enormous experimental challenge. However, it is promising to see that $O$(kg) size targets could be sensitive to relatively small DM scattering cross sections.

\exercise{Suppose DM couples to nucleons via a light mediator. Taking the massless mediator limit, there is an additional momentum-transfer dependence which can be parameterized by including a DM form factor $F^2_{\rm DM}(\bfq) = (m_\chi v_0)^4/|\bfq|^4$ with $v_0 = 220$ km/s. (See for example the analogous definition for electron scattering in Eq.~\ref{eq:DMformfactor}.) Estimate the acoustic phonon excitation rate for $\sigma_n \simeq 10^{-41}$ cm$^2$, where now $\sigma_n$ is a reference cross section at $|\bfq|_{\rm ref} = m_\chi v_0$.  }  

\begin{figure}[t]
\begin{minipage}[h]{\textwidth}
$\vcenter{\hbox{   \includegraphics[width=0.48\textwidth]{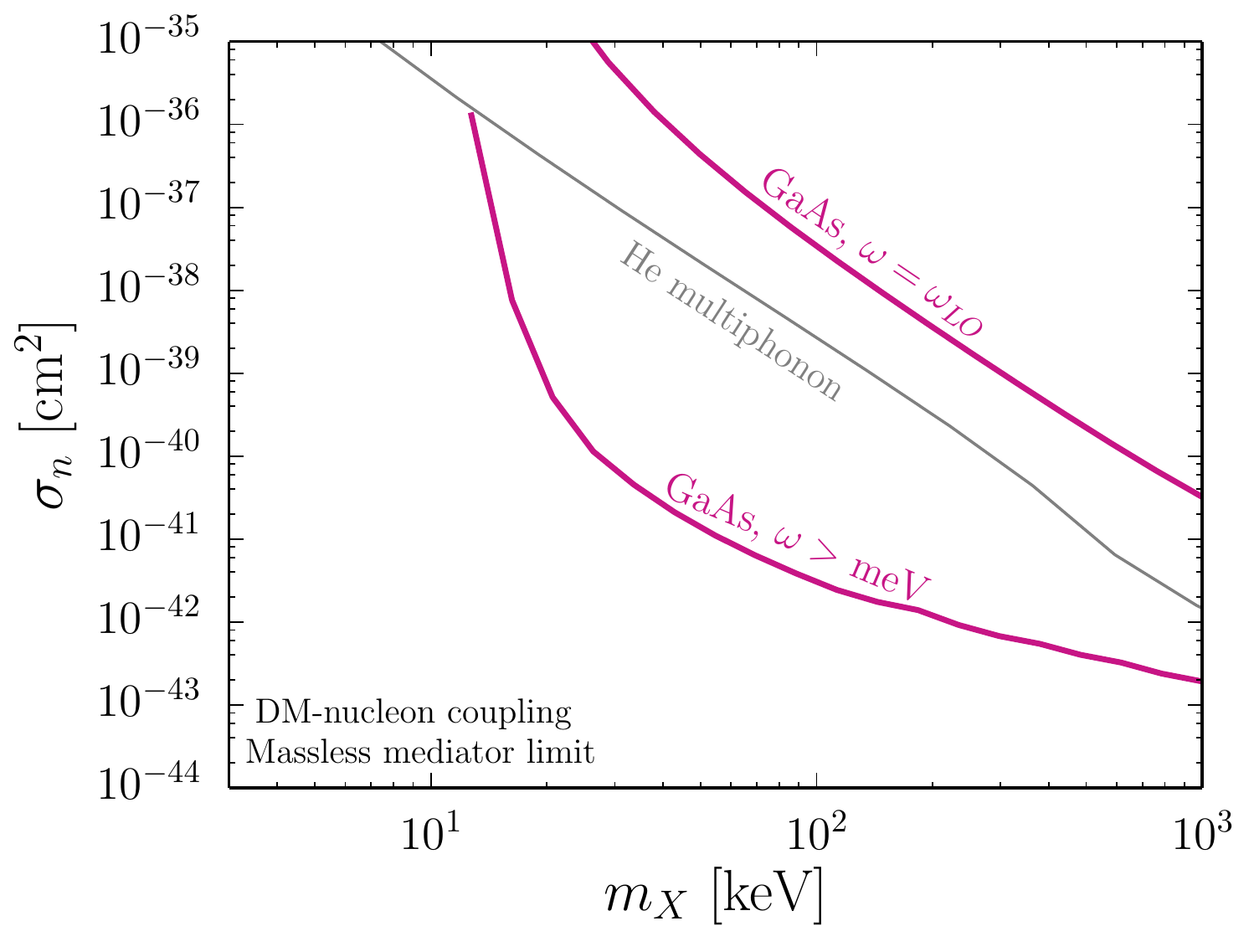} }}$
$\vcenter{\hbox{   \includegraphics[width=0.48\textwidth]{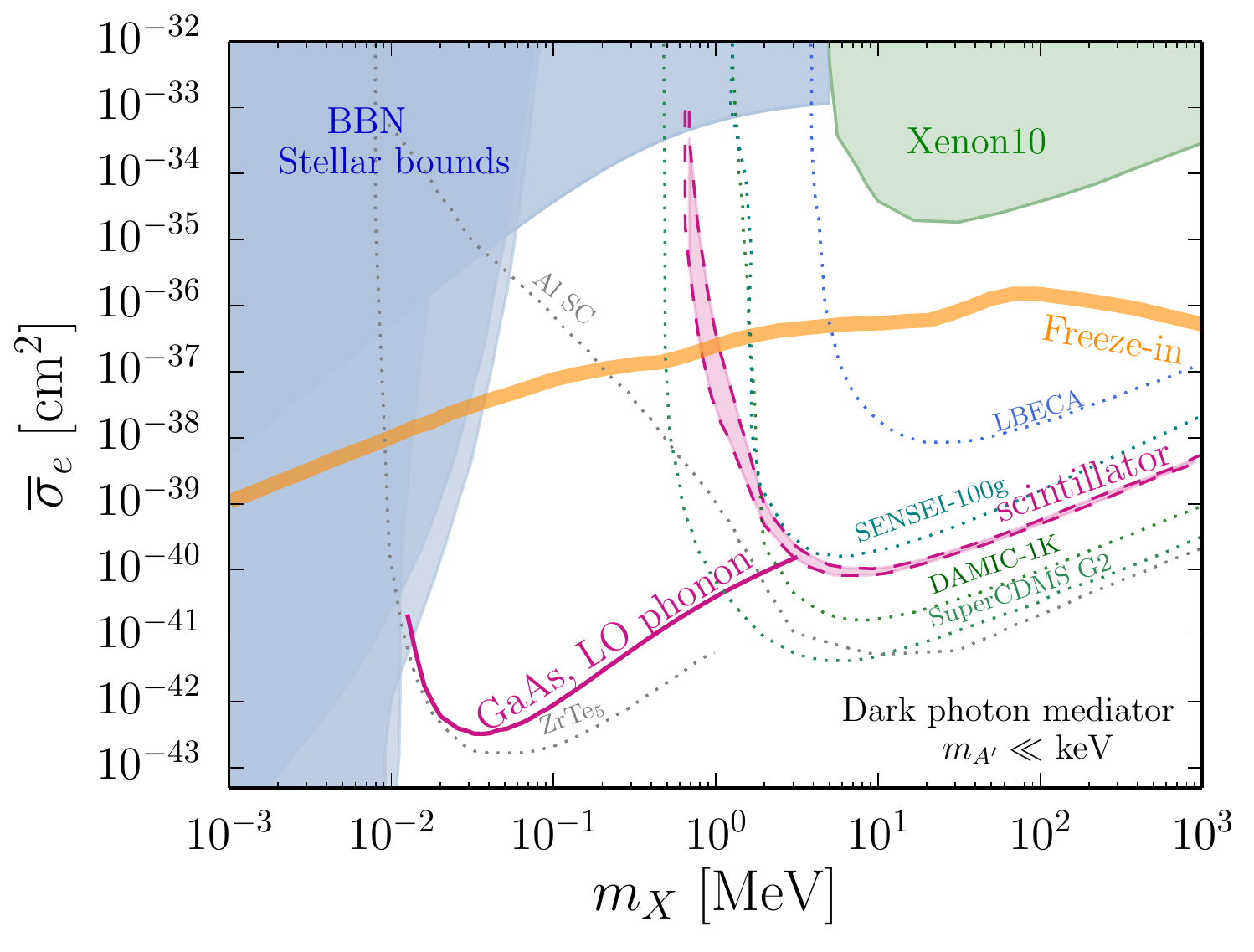} }}$
\end{minipage}
\caption{ \label{fig:sigma_phonon} Reproduced from Ref.~\cite{Knapen:2017ekk}. ({\bf left}) Sensitivity of a GaAs target to DM-nucleon interactions mediated by an massless scalar. Results are shown assuming kg-year exposure and zero background and for scattering into acoustic phonons $(\omega >$ meV) and for scattering into optical phonons $(\omega = \omega_{\rm LO} \approx$ 35 meV). Also shown is the sensitivity for multiphonon production superfluid He~\cite{Schutz:2016tid,Knapen:2016cue}. ({\bf right}) Sensitivity to DM-ion scattering mediated by an ultralight kinetically-mixed vector. The reach for $m_\chi < $MeV is from DM excitation of a single longitudinal optical phonon, and assumes kg-year exposure. The orange line denotes the couplings where the DM abundance can be entirely explained by the freeze-in mechanism. Dashed and dotted lines show sensitivities of other proposed experiments. }
\end{figure}

From  Eq.~\ref{eq:rate_phonon}, one might think the sensitivity improves dramatically with smaller $m_\chi$ and $c_s$. However, this is not really case once we integrate over the DM velocity distribution. For smaller $m_\chi$, the available phase space for creating an excitation of $\omega = {\rm meV}$ also becomes smaller. In particular, the maximum energy deposition from DM scattering can be estimated as
\begin{align}
	\omega_{\rm max} = c_s q_{\rm max} \lesssim 2 c_s m_\chi v  \sim {\rm few}\, {\rm meV} \times \frac{ m_\chi}{100 \, {\rm keV}} \times \frac{ c_s}{10^{-5}}
\end{align}
which scales as $m_\chi$ and $c_s$. Similarly, for smaller $c_s$ the phonon energies are smaller for a given momentum transfer. This again leads to a phase space suppression since we typically have to impose a minimum recoil energy, which is very optimistically $\omega \simeq $ meV. One can also expect some correlation of the target nucleus mass with the speed of sound, so it is not obvious that increasing $A$ leads to a larger rate. The yield-limited sensitivity from more detailed calculations for DM scattering via a light mediator is shown in Fig.~\ref{fig:sigma_phonon}. This result is for a GaAs target which has $c_s \approx 1.3 \times 10^{-5}$. For a sapphire (Al$_2$O$_3$) target, which has $c_s \approx 3.3 \times 10^{-5}$, the reach is quite similar~\cite{Griffin:2018bjn}.

\subsubsection{Optical phonons}

A realistic 3D lattice has more than one atom per unit cell, resulting in additional phonon branches associated with the relative motions of the atoms within the cell. We will illustrate this briefly with the toy 1D lattice, shown in Fig.~\ref{fig:lattice_1D_optical}. Taking the same effective spring constant everywhere for simplicity, the Hamiltonian for this lattice then has the form
\begin{align}
	H = \sum_i \frac{1}{2} M_1 \dot u_{2i}^2 +  \frac{1}{2} M_2 \dot w_{2i+1}^2 + \frac{1}{2} k_{\rm eff} ( u_{2 i} - w_{2i+1})^2 + \frac{1}{2} k_{\rm eff} ( u_{2i +2} - w_{2i+1})^2 \, 
\end{align}
where $i=0..N-1$ is an index running over all unit cells. As before, the exact dispersions can be derived in this toy model, see Ref.~\cite{Kittel}. In order to put things in a form more familiar to field theorists, we can again turn this into a Lagrangian by taking a continuum limit. Define fields $u(x,t)$ and $w(x,t)$ for the displacements of the two atoms in the unit cell. Then the Lagrangian can be written as
\begin{align}
	&\int dx  \left( \tfrac{ \rho_1}{2} \dot u(x,t)^2 +\tfrac{ \rho_2}{2}  \dot w(x,t)^2 - \tfrac{k_{\rm eff}}{2a} \left[ u(x,t) - w(x+ \tfrac{a}{2},t) \right]^2 - \tfrac{k_{\rm eff}}{2a} \left[ u(x,t) - w(x- \tfrac{a}{2},t) \right]^2 \right) \nonumber \\
	&= \int dx  \left( \tfrac{ \rho_1}{2}\dot u(x,t)^2 +\tfrac{ \rho_2}{2}  \dot w(x,t)^2 - \tfrac{k_{\rm eff}}{a} \left[ u(x,t)^2 + w(x,t)^2 \right] + \tfrac{k_{\rm eff}}{a} u(x,t)  \left[ w(x+ \tfrac{a}{2},t) + w(x- \tfrac{a}{2},t) \right] \right) \nonumber
\end{align}
where we have taken care that the displacements of the atoms  $u_{2i}$ and $w_{2i+1}$ are the displacements of the fields at positions $x$ and $x+a/2$, respectively. The densities are $\rho_1 = M_1/a$ and $\rho_2 = M_2/a$.

\begin{figure}[t]
\begin{minipage}[h]{\textwidth}
$\vcenter{\hbox{   \includegraphics[width=0.48\textwidth]{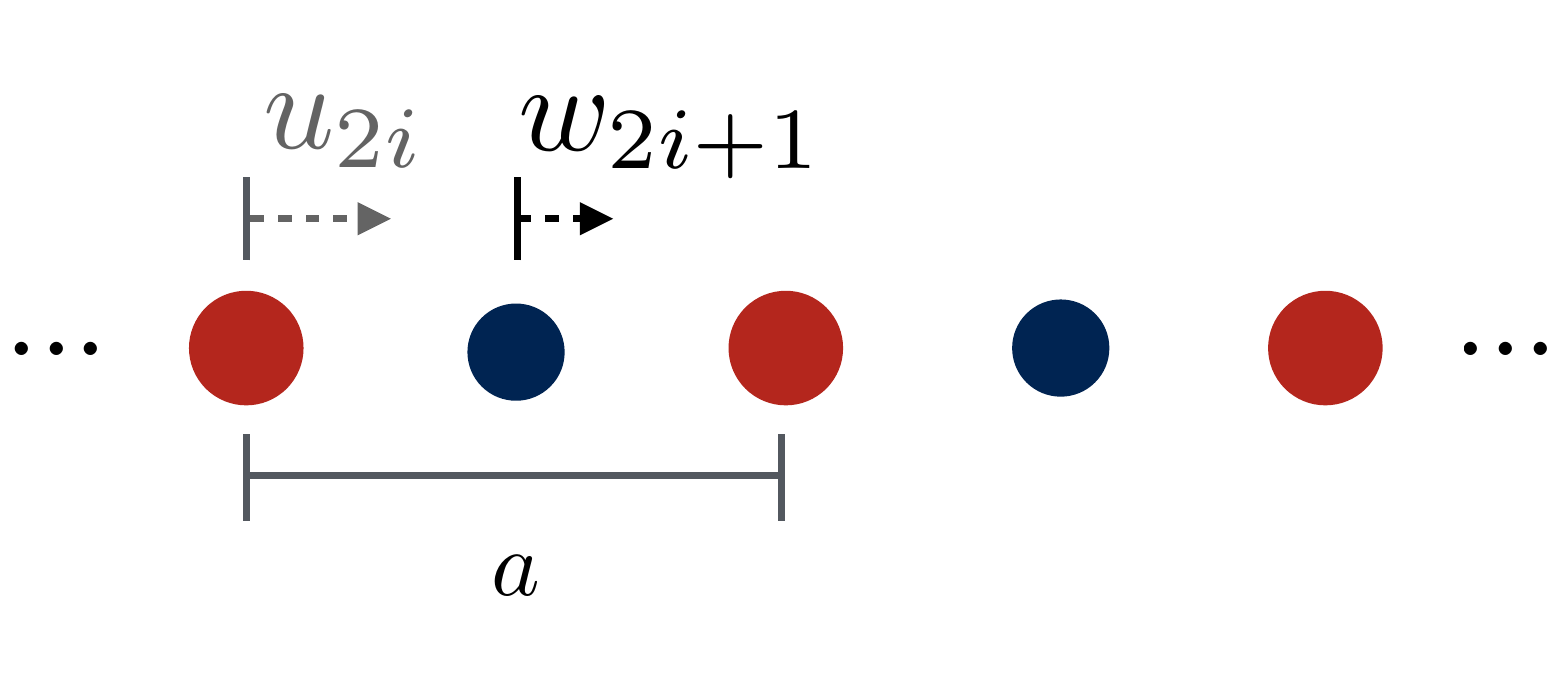} }}$
\hspace{0.2cm}
$\vcenter{\hbox{   \includegraphics[width=0.46\textwidth]{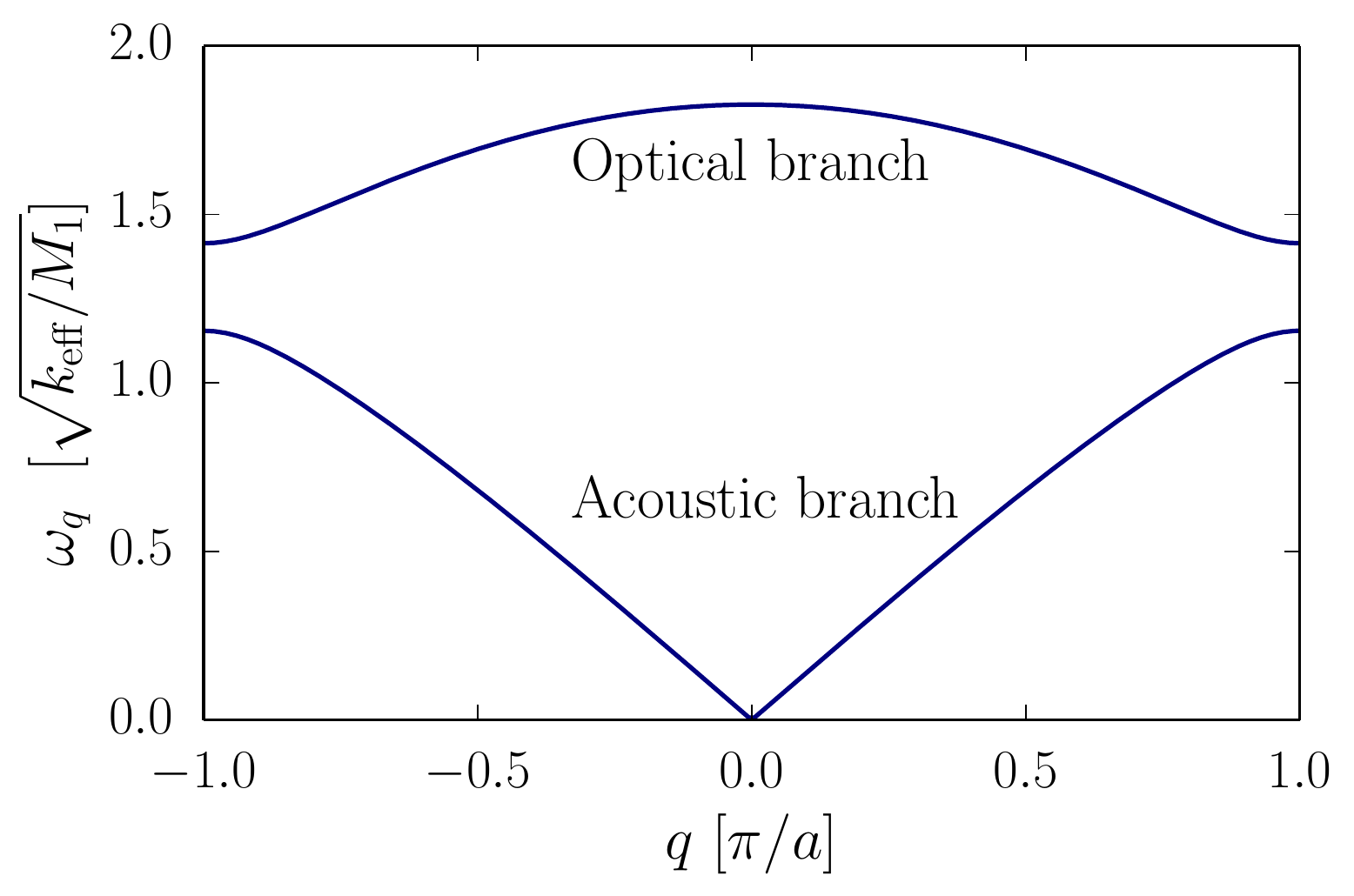} }}$
\end{minipage}
\caption{ \label{fig:lattice_1D_optical} ({\bf left}) A 1D lattice of atoms of mass $M_1$ and $M_2$, with respective displacements from equilibrium $u_i$ and $w_i$. This lattice has both a longitudinal acoustic and longitudinal optical phonon branch. ({\bf right}) The dispersion relations are shown over the first Brillouin Zone for $M_2/M_1 = 1.5$.}
\end{figure}

We substitute the following expansion for the $w$ field in the long wavelength limit:
\begin{align}
	w(x \pm \tfrac{a}{2},t) \approx w(x,t) \pm \tfrac{a}{2} \nabla w(x,t) + \tfrac{a^2}{8} \nabla^2 w(x,t) .
\end{align}
Then this Lagrangian can be rewritten as
\begin{align}
	\int dx  \left( \frac{1}{2} \rho_1 \dot u^2 +\frac{1}{2}  \rho_2 \dot w^2 - \frac{\tilde k_{\rm eff}}{a^2} ( u - w )^2 - \frac{\tilde k_{\rm eff}}{4} ( \nabla u \nabla w)\right) 
\end{align}
where $\tilde k_{\rm eff} = k_{\rm eff} a$ as before. This time we find an apparent mass mixing for the $u(x), w(x)$ fields.  In the limit $\rho_1 = \rho_2$, there is one massless mode $\propto (u +w)/\sqrt{2}$ (the Goldstone mode where all the atoms in a unit cell move in phase) and one massive/gapped mode $\propto (u-w)/\sqrt{2}$ (where the atoms in a unit cell move out of the phase). 
\exercise{Show that for $\rho_1 \neq \rho_2$, the dispersion relations and eigenmodes in the $q \to 0$ limit are given by:
\begin{align}
	\omega_{\rm acoustic} &\approx |\bfq| \sqrt{ \frac{\tilde k_{\rm eff}}{2(\rho_1 + \rho_2)} } \quad , \quad u(x) =  w(x) \\
	\omega_{\rm optical}  &\approx \sqrt{ \frac{2 \tilde k_{\rm eff} (\rho_1 + \rho_2)}{a^2 \rho_1 \rho_2 }} \quad, \quad u(x) =  -\frac{\rho_2}{\rho_1}  \, w(x) .
\end{align}
}
The exact dispersion relations for the toy model are shown in the right panel Fig.~\ref{fig:lattice_1D_optical}.  As before, the slope of the acoustic phonon dispersion is typically given by the speed of sound, around $10^{-5}-10^{-6}$.  The gapped optical phonon mode typically has energies of 10--100 meV in crystals. 

Consider the kinematics of DM scattering, it can be even more advantageous to search for excitations into optical phonons. The reason is that the optical phonon has an approximately constant energy as $q \to 0$. For sub-MeV DM, $q \ll $ keV and so more energy can be deposited by exciting an optical phonon mode compared to an acoustic phonon mode. This is helpful since the experimental threshold does not have to be quite so low.

In addition, for compound materials such as GaAs the different atoms within the unit cell generally have different effective charges. These materials are also known as polar materials. The Born effective charges $Z$ describe the induced polarization due to the displacement of an atom. For GaAs, these are equal and opposite with $|Z| \approx 2.27$. As a result, the out-of-phase oscillations in an optical phonon can be viewed as oscillating dipoles. DM that interacts via a kinetically-mixed dark photon can thus create these dipole excitations, whereas it may not have enough energy to create an electron excitation (the gap for electron excitations in polar materials is usually at least eV). The sensitivity from optical phonon excitations in GaAs is shown in Fig.~\ref{fig:sigma_phonon}, where it is particularly promising for the vector portal model.

DM-phonon excitations have only been explored for a few materials so far, and it is an interesting question as to how to optimize for a target material sensitive to different types of phonon scattering. The directional dependence of phonon modes in a crystal also means it may be possible to observe a directional signal from DM scattering~\cite{Griffin:2018bjn}. Finally, the phonon properties are important in determining the experimental prospects of detecting low energy excitations. See Ref.~\cite{Knapen:2017ekk} for a brief discussion of the detector concept and some of the relevant backgrounds.

\subsection{The ultralight frontier and conclusions}

We have focused on summarizing the basic physics of sub-GeV DM scattering off electrons and phonons.  Other quasiparticle excitations may also be promising for different types of DM interactions! In the case of electrons and phonons, the same experimental proposals are also sensitive to much lighter DM through an absorption process. The idea is similar to the photoelectric effect, where bosonic DM is absorbed to create electron and phonon excitations.  In  many cases the DM absorption rate can thus be related to the optical absorption properties of the material. This idea has been explored in a range of targets~\cite{An:2013yua,Hochberg:2016ajh,Hochberg:2016sqx,Bloch:2016sjj,Knapen:2017ekk,Hochberg:2017wce,Arvanitaki:2017nhi,Griffin:2018bjn} for DM down to the $\sim$ meV scale, and experiments have recently set direct detection limits on eV-scale DM~\cite{Crisler:2018gci,Aguilar-Arevalo:2016zop,Agnese:2018col}.

The quest to detect even lower mass DM has also proceeded with great fervor recently. There are methods to search for DM as light as the fuzzy DM limit ($\sim 10^{-22}$ eV), as well as those that target the QCD axion as a DM candidate specifically~\cite{Hook:2018dlk,Graham:2015ouw,Battaglieri:2017aum}. As with sub-GeV DM scattering, some of these take advantage of technological developments in neighboring fields of physics. The landscape of experimental and observational probes of DM is just as varied as the landscape of models. It is an exciting time in the search for DM, and with some luck and thoughtfulness, we may look forward to some of our efforts bearing fruit.

\clearpage

\begin{acknowledgements}
These lectures were originally given at TASI 2018: Theory in an Era of Data. I would like to thank the organizers (Tom DeGrand, Tilman Plehn, and Tracy Slatyer) for the opportunity to speak and for putting together this event. It was a great pleasure to give these lectures and to get to know the students of TASI 2018. I am grateful to Simon Knapen, Jung-Tsung Li, Aneesh Manohar, and Dipan Sengupta for valuable feedback on these notes, and I'd especially like to thank Nikita Blinov, Brian Campbell-Deem, and Katelin Schutz for providing in-depth comments on a working draft. 
Parts of this document were prepared while at the KITP, supported by the National Science Foundation under Grant No. NSF PHY-1748958. This work was supported in part by an Alfred P. Sloan foundation fellowship and the Department of Energy under grant DE-SC0019195.
\end{acknowledgements}

\appendix

\section{Units and Conventions \label{sec:convention}}

\def\arraystretch{1.4}\setlength{\tabcolsep}{10pt}
\begin{table}[th]
\begin{tabular}{|c|c|} 
\hline
1 Kelvin & $8.62 \times 10^{-5}$ eV \\ \hline	
$M_\odot$ & $10^{66}$ eV \\ \hline
Mpc & $( 6.4 \times 10^{-30} \ \eV )^{-1}$ \\  \hline
$G_N$ & $1/\Mpl^2 = 1/( 1.22 \times 10^{19}\ \GeV )^{2}$ \\   \hline	
1 cm & ($1.97 \times 10^{-14}$ GeV)$^{-1}$ \\
\hline
\end{tabular}
\caption{Some conversions from units used in astrophysics into natural units where $\hbar = c =1$. \label{tab:units}}
\end{table}

In this appendix, we summarize some conventions and other facts that may be useful in reading these notes or exploring some of the topics in more depth. Table~\ref{tab:units} gives some unit conversions to help in translating between cosmology, condensed matter, and particle physics.

Sub-GeV direct detection often requires referring to the condensed matter literature, and we note that the units in condensed matter often assume cgs-Gaussian units. However, these notes (and most papers in the direct detection literature) follow the convention in the particle physics community in using Lorentz-Heaviside units:
\begin{align}
	e &= \sqrt{4\pi \alpha \hbar c}\ , \ \ \ {\rm Lorentz-Heaviside} \nonumber \\
	e &= \sqrt{\alpha \hbar c}\ , \ \ \ \ \ \ \ {\rm cgs-Gaussian} \nonumber
\end{align}

In lecture 5, we switch between discrete sums in finite volume and integrating over continuous variables in the infinite volume limit. For reference, the relations are:
\begin{align}
	\label{eq:discretization1}
	\sum_{\bfk} &\to V \int \frac{d^3 \bfk}{(2\pi)^3} \quad , \\
	\delta_{\bfk, \bfk'} &\to \frac{(2\pi)^3}{V} \delta^{(3)}( \bfk - \bfk') 
	\label{eq:discretization2} \\
	\hat a_{\bfq} &\to \frac{\hat a_{\bfq}}{\sqrt{V} }
\end{align}
In the last line, we relate the creation and annihilation operators with normalization $[ \hat a_{\bfq},  \hat a^\dagger_{\bfq'}] = \delta_{\bfq, \bfq'}$ in the discrete limit to the operators in the continuum limit where $[ \hat a_{\bfq},  \hat a^\dagger_{\bfq'}] = (2\pi)^3 \delta^{(3)}(\bfq - \bfq')$.

\section{In-medium dark photon couplings and photon polarization \label{sec:inmedium}}

A kinetically mixed dark photon is a nearly ubiquitous feature in models of dark sectors. As discussed in Sec.~\ref{sec:vector}, the appropriate couplings in a given scenario depend sensitively on the photon polarization tensor $\Pi^{\mu \nu}$. In this appendix, we examine the effective in-medium couplings of the dark photon and also provide a summary of the behavior of $\Pi^{\mu \nu}$ in several limits and materials.

Starting from the basis in Eq.~\ref{eq:massiveV_basis}, 
we include additional terms in the Lagrangian which account for the in-medium polarization tensor, here defined as $\Pi^{\mu \nu}(q) \equiv i e^2 \langle J^\mu_{\rm EM} J^\nu_{\rm EM} \rangle$, 
\begin{align}
	{\cal L} \supset & -\frac{1}{4} \tilde F_{\mu \nu} \tilde F^{\mu \nu} -\frac{1}{4} V_{\mu \nu} V^{\mu \nu} + \frac{1}{2} m_V^2 V_\mu V^\mu + e (\tilde A_\mu + \kappa V_\mu) J^\mu_{\rm EM} + g_\chi V_\mu J^\mu_{D} \nonumber \\
	&- \frac{1}{2} \tilde A_\mu \Pi^{\mu \nu} \tilde A_\nu -  \kappa \tilde A_\mu \Pi^{\mu \nu} V_\nu \, .
	\label{eq:inmedium_basis}
\end{align}
In the limit that $\kappa \ll 1$ and $g_\chi \ll 1$, we will drop terms in the Lagrangian of $O(\kappa^2)$ or $O(g_\chi^2)$ and neglect the effect of any in-medium polarization tensor quadratic in $V$. 
The polarization tensor $\Pi^{\mu \nu}(q)$ can be decomposed into longitudinal and transverse pieces:
\begin{align}
	\Pi^{\mu \nu}(q) = \Pi_L(q) \eta_L^\mu \eta_L^\nu + \Pi_T(q) \, (\eta_+^\mu)^* \, \eta_+^\nu+ \Pi_T(q) \, (\eta_-^\mu)^* \, \eta_-^\nu
\end{align}
where $q = (\omega, \bfq)$. Taking $\bfq = |\bfq|  \hat z$, then the vectors above are defined as\footnote{Note that the $\eta_{L,T}$ here  are a useful shorthand for decomposing the polarization tensor and satisfy $q_\mu \Pi^{\mu \nu} = 0$ as required for current conservation. They are not the same as the renormalized external photon polarization vectors.}
\begin{align}
	\eta_L^\mu &= \left(1, 0,0, \frac{\omega}{|\bfq|} \right) \\	\eta_{\pm}^\mu &= \frac{1}{\sqrt{2} }  (0, 1, \pm i, 0)
\end{align}
This general form of $\Pi^{\mu \nu}(q)$ thus satisfies the requirement of current conservation, and we have picked sign conventions for $\Pi^{\mu \nu}$ such that $\Pi_T$ will correspond to a positive mass-squared contribution. Note that some references use a definition of $\Pi_L$ which differs from this one by a factor of $\Pi_L =q^2/|\bfq|^2 \Pi_L^{\rm here}$ and $\Pi^{\mu \nu}$ is instead written in terms of a normalized longitudinal vector $\tfrac{|\bfq|}{\sqrt{q^2}}(1,  0,0,\frac{\omega}{|\bfq|} )$. We follow here the convention of Ref.~\cite{Braaten:1993jw}.

From Eq.~\ref{eq:inmedium_basis}, we can see that in-medium effects generate an effective mass for the photon, where the on-shell condition is defined by $q^2 - \Pi_{T}(q) = 0$ for transverse modes and $|\bfq|^2 - \Pi_{L}(q) = 0$ for  longitudinal modes. The solutions to these equations as well as the residues near the poles determine the in-medium dispersion relations $\omega_{L,T}({\bf q})$ as well as wavefunction renormalization for each polarization. The effective mass is ${\bfq}$-dependent, but is approximately given by the plasma mass.  Note that the polarization tensor in various thermal plasmas has been discussed in works such as Refs.~\cite{Braaten:1993jw,Raffelt:1996wa}, and we briefly review this along with properties of the polarization tensor relevant for direct detection in the Appendix. 

Next, we turn to the $q$-dependent and polarization-dependent mixing of $V$ and the photon which arises from the in-medium polarization. For a given vector, we can decompose it into polarization states as:
\begin{align}
	V^\mu &= V^+ \epsilon_+^\mu + V^- \epsilon_-^\mu + V^L \epsilon_L^\mu \\
		&\equiv (V^+)^\mu + (V^-)^\mu + (V^L)^\mu
\end{align}
where we neglect the gauge dependent piece proportional to $q^\mu$. 
We start with Eq.~\ref{eq:inmedium_basis}  and first just focus on a single transverse polarization and specific value of $q$, which gives the Lagrangian
\begin{align}
	{\cal L} \supset & -\frac{1}{4}  F^+_{\mu \nu}  F_+^{\mu \nu} -\frac{1}{4} V^+_{\mu \nu} V_+^{\mu \nu} + e (A^+_\mu + \kappa V^+_\mu) J^\mu_{\rm EM} + g_\chi V^+_\mu J^\mu_{D} \nonumber \\
	& +  \frac{1}{2} \Pi_T(q) A^+_\mu   A_+^\mu +  \kappa \, \Pi_T(q) A^+_\mu  V_+^\mu + \frac{1}{2} m_V^2 V^+_\mu V_+^\mu \, .
\end{align}
(The sign in front of the quadratic $A^2, AV$ terms changed because the polarization vectors satisfy $\epsilon_\mu \epsilon^\mu = -1$.)

We define a change of basis to the in-medium states $\bar A^{+}_\mu, \bar V^{+}_\mu$ which diagonalize the mass and kinetic terms: 
\begin{align}
	A^+_\mu &\equiv \bar A^+_\mu + \frac{\kappa \,  \Pi_T(q)}{m_{V}^2 -  \Pi_T(q)} \bar  V^+_\mu \\
	V^+_\mu &\equiv \bar V^+_\mu - \frac{\kappa \,  \Pi_T(q)}{m_{V}^2 - \Pi_T(q)}  \bar A^+_\mu \quad .
\end{align}
Then dropping terms of $O(\kappa^2)$ or smaller, the above Lagrangian can be rewritten in the terms of the in-medium states as
\begin{align}
	{\cal L} \supset & -\frac{1}{4}  \bar F^+_{\mu \nu}  \bar F_+^{\mu \nu} -\frac{1}{4} \bar V^+_{\mu \nu} \bar V_+^{\mu \nu} +  \frac{1}{2} \Pi_T(q) \bar A^+_\mu  \bar A_+^\mu +  \frac{1}{2} m_V^2 \bar V^+_\mu \bar V_+^\mu \nonumber \\
	& + e \left(\bar A^+_\mu + \frac{ \kappa m_V^2}{m_V^2 - \Pi_T(q)} \bar V^+_\mu \right) J^\mu_{\rm EM} + g_\chi \left(\bar V^+_\mu - \frac{\kappa \, \Pi_T(q)}{m_{V}^2 -  \Pi_T(q)}  \bar A^+_\mu \right)J^\mu_{D}  \quad.
	\label{eq:inmedium_massdependent}
\end{align}
We see that the couplings of the dark photon and dark matter are highly dependent on the properties of the medium and the process considered, since this determines the $q$ values. For example, for a nonrelativistic plasma with $|\bfq| \sim \omega$, then $ \Pi_T(q) \approx \omega_p^2$. If $\omega_p^2 \gg m_V^2$, the coupling of the dark photon with $J_{\rm EM}$ is suppressed from the vacuum value by a ratio of $\sim m_V^2/\omega_p^2$.  Meanwhile, the dark matter current $J_D^\mu$ couples to the in-medium photon with an effective charge parameter $\kappa g_\chi$. Note that we could also have started with the basis of Eq.~\ref{eq:vacuumL_darkphoton}, introduced in-medium effects in that basis, and arrived at the same conclusion. See for example Ref.~\cite{Hardy:2016kme} where identical results are obtained in the two bases.

The calculation for the longitudinal polarization proceeds similarly, except that some care must be taken with the normalization since $\eta_L^\mu \eta_{L,\mu} = - q^2/|\bfq|^2$. The effective mass terms for longitudinal vector polarizations  $A_L, V_L$ are instead given by
\begin{align}
	{\cal L} \supset   \frac{1}{2} \frac{q^2}{|\bfq|^2} \Pi_L(q) A^L_\mu   A_L^\mu +  \kappa \, \frac{q^2}{|\bfq|^2} \Pi_L(q) A^L_\mu  V_L^\mu + \frac{1}{2} m_V^2 V^L_\mu V_L^\mu \, .
\end{align}
Performing a similar basis rotation as above to in-medium fields $\bar A_L, \bar V_L$, we obtain diagonal mass terms and  the effective couplings are given by
\begin{align}
	{\cal L} \supset   e \left(\bar A^L_\mu + \frac{ \kappa m_V^2}{m_V^2 - \Pi_L(q) q^2/|\bfq|^2} \bar V^L_\mu \right) J^\mu_{\rm EM} + g_\chi \left(\bar V^L_\mu - \frac{\kappa \, \Pi_L(q)}{m_{V}^2 |\bfq|^2/q^2 -  \Pi_L(q) }  \bar A^L_\mu \right)J^\mu_{D}  \quad.
	\label{eq:inmediumcouplings_long}
\end{align}
For a typical nonrelativistic plasma $|\bfq| \sim \omega$, $ \Pi_L(q) \approx \omega_p^2 |\bfq|^2/\omega^2$. However for $|\bfq| \gg \omega$, relevant for scattering processes, $ \Pi_L(q) \approx m_D^2 |\bfq|^2/q^2 \approx - m_D^2$ where $m_D$ is the Debye mass. This is connected to the Debye screening of Coulomb scattering in a plasma; for an explicit expression of the Debye screening length $\lambda_D = 1/m_D$, see Eq.~\ref{eq:Pidebye} below.

The in-medium mixing effects are important in determining constraints or searches for the dark photon portal. 
One set of important constraints on dark photons comes from stellar emission considerations. The dark photon can be emitted in the core of the star through the in-medium interactions above and escape due to the small couplings, leading to anomalous energy loss. This was studied recently in Refs.~\cite{An:2013yfc,An:2013yua,Redondo:2013lna,Hardy:2016kme,Chang:2016ntp} for the Sun, red giants, horizontal branch stars, and SN1987a. Here the plasma mass in the core of the star ranges from $\sim 0.1-1$ keV (for the Sun or horizontal branch stars) up to $\sim 10$ MeV (for core collapse supernovae such as SN1987A). 

We can use the in-medium couplings above to estimate the production of on-shell dark photons. As mentioned above, the real parts of the polarization tensors are $ {\rm Re}\, \Pi_T(q) \approx \omega_p^2$ and $ {\rm Re}\,  \Pi_L(q) \approx \omega_p^2 |\bfq|^2/\omega^2$ for the $q$ values relevant for this process. Meanwhile, the imaginary parts of the polarization tensors are proportional to the emission (and absorption) of photon modes in the medium, ${\rm Im}\, \Pi_T(q) \equiv - \omega \Gamma_T$ and ${\rm Im}\, \Pi_L(q) \equiv - (|\bfq|^2/\omega^2) \omega \Gamma_L $. Since $\Pi^{\mu \nu} = i e^2 \langle J^\mu_{\rm EM}J^\nu_{\rm EM} \rangle$, we can write the emission rate $Q$ of each {\emph{dark}} photon polarization as
\begin{align}
	Q_T  &\propto -  {\rm Im}\, \Pi_T(q) \frac{\kappa^2 m_V^4 }{(m_V^2 -\omega_p^2)^2 + ({\rm Im}\, \Pi_T(q))^2 } = \frac{\kappa^2 m_V^4 \,  \omega \Gamma_T}{(m_V^2 - \omega_p^2)^2 + \omega^2 \Gamma_T^2} \\
	Q_L &\propto -{\rm Im}\, \Pi_L(q) \frac{q^2}{|\bfq|^2} \frac{\kappa^2 m_V^4}{(m_V^2 - \omega_p^2 q^2/\omega^2)^2 + ({\rm Im}\, \Pi_L(q) \, q^2/|\bfq|^2)^2 }  = \frac{\kappa^2 m_V^2\, \omega^3 \Gamma_L}{(\omega^2 - \omega_p^2)^2 + \omega^2 \Gamma_L^2} 
\end{align}
where for the longitudinal polarization we used that $q^2 = m_V^2$ for on-shell production and the overall factor of $q^2/|\bfq|^2$ comes from contracting $\Pi^{\mu \nu}$ with external polarization vectors $\epsilon_L^\mu = \tfrac{|\bfq|}{\sqrt{q^2}}(1,  \frac{\omega}{|\bfq|} \hat \bfq)$. While the detailed calculation of the production rate can be found in works such as Refs.~\cite{An:2013yfc,Redondo:2013lna,Hardy:2016kme}, we highlight just a few things. First, there is resonant transverse polarization when $m_V = \omega_p$, but for $m_V \ll \omega_p$ the production rate scales as $\kappa^2 m_V^4$. For the longitudinal modes, there is resonant production for any $m_V$ where $\omega = \omega_p$ is kinematically possible. Furthermore, the process scales as $\kappa^2 m_V^2$ and so this gives the dominant emission mechanism for $m_V \ll \omega_p$. This explains why stellar emission constraints on $\kappa$ scale as $1/m_V$ in the low mass limit. This $m_V^2$ suppression also means that ultralight vectors would {\emph{not}} be efficiently produced in the early universe by interactions of SM charged particles, so there are no $N_{\rm eff}$ constraints on extra contributions to relativistic degrees of freedom.

\subsection{Behavior of polarization tensor in various limits}

The polarization tensor determines the in-medium dispersion relations as well as the optical properties of the medium. It is related to the dielectric function $\epsilon(\bfq, \omega)$ and conductivity $\sigma(\bfq, \omega)$ by:
\begin{align}
	\epsilon_T(\bfq, \omega) &= 1 - \frac{\Pi_T(q)}{\omega^2} = 1 + i \frac{\sigma_T(\bfq, \omega)}{\omega} \\
	\epsilon_L(\bfq, \omega) &= 1 - \frac{\Pi_L(q)}{|\bfq|^2} = 1 + i \frac{\sigma_L(\bfq, \omega)}{\omega}
\end{align}
Below we summarize the behavior of the real part of the polarization tensor. Meanwhile, the imaginary part determines absorption and emission rates via the optical theorem, ${\rm Im} \, \Pi = - \omega \Gamma(\omega)$~\cite{Weldon:1983jn}. Ref.~\cite{Raffelt:1996wa} provides more in-depth discussion, especially in the context of stellar environments.

For accurate in-medium dispersion relations, see Ref.~\cite{Braaten:1993jw}, which explicitly gives $\Pi^{\mu \nu}$ in the regime where $|\bfq| \simeq \omega$. For a non-degenerate medium, a rough approximation when $|\bfq| \simeq \omega$ is to take
\begin{align}
	\Pi_L(q) &\simeq \omega_p^2 - \bfq^2 \\
	\Pi_T(q) &\simeq \omega_p^2
\end{align}
where $\omega_p$ is the plasma frequency. For a nonrelativistic electron gas, $\omega_p^2 = n_e e^2/m_e$. For a relativistic and non-degenerate electron gas at temperature $T$, $\omega_p^2 = (e T/3)^2$. (A simple model for the real and imaginary parts of $\Pi_{T,L}$ in a metal is also discussed in Ref.~\cite{Hochberg:2016ajh}.)

The limit of $|\bfq| \gg \omega$ is relevant for nonrelativistic scattering processes. In this limit, Coulomb scattering via the longitudinal polarization dominates. In the classical limit (non-degenerate, non-relativistic gas), 
\begin{align}
	\Pi_L(q) \simeq -\lambda_D^2 = - \frac{n_e e^2}{T}  \simeq - \frac{m_e}{T} \omega_p^2
	\label{eq:Pidebye}
\end{align} 
where $\lambda_D$ is the Debye length and we have assumed an electron gas. In the degenerate limit, 
\begin{align}
	\Pi_L(q) \simeq -\lambda_{\rm TF}^2 \simeq - \frac{4 \alpha E_F p_F}{\pi}
\end{align} 
where $\lambda_{\rm TF}$ is the Thomas-Fermi screening length and $E_F$ ($p_F$) is the energy (momentum) at the Fermi surface. For the complete dielectric function in the degenerate case, see Ref.~\cite{Hochberg:2015fth} which reproduces the result from Ref.~\cite{dresselgruner}. Ref.~\cite{dresselgruner} provides useful derivations of $\Pi_{L,T}$ from a materials point of view.

\bibliographystyle{apsrev4-1}
\bibliography{tasi}

\end{document}